\documentclass[12pt]{book}
\usepackage{aastex2}
\usepackage{subfigure,lscape,color}
\input{psfig.sty}

\def \beq {\begin{equation}}
\def \eeq {\end{equation}}

\begin{document}

\pagenumbering{}
\begin{titlepage}
\begin{figure*}[h!]\centerline{\psfig{figure=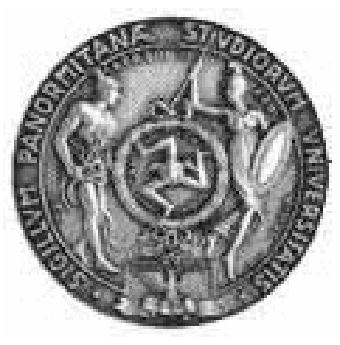,angle=0,width=0.15\textwidth}\hspace{9cm}\psfig{figure=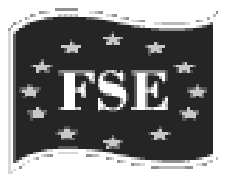,angle=0,width=0.2\textwidth}}\end{figure*}
\begin{center} 
\LARGE Universit\`a di Palermo\\
\Large Dipartimento di Scienze Fisiche ed Astronomiche\\
\vspace*{0.5cm}

Dottorato di Ricerca in Fisica\\
XIV ciclo\\
A.A. 2000/2001
\vspace*{2.5cm}

\noindent
{\huge\bf X-ray Emission from Early-Type Galaxies in Groups and Poor Clusters.}

\vspace*{4cm}

\noindent
\Large
~~Author\hspace*{6cm}~~~~~Supervisor\\ 
Paolillo Maurizio\hspace*{5cm}Prof. G. Peres\\
\end{center}
\end{titlepage}

\thispagestyle{empty}
\centerline{\bf \Large ABSTRACT}
\vspace{1cm}
{\small
We present the study of three bright Early-Type galaxies hosted in groups or poor clusters: NGC 1399, NGC 1404 and NGC 507. We used ROSAT HRI and PSPC data to study the morphological structure of the X-ray halo and, when available, the higher resolution {\it Chandra} data.
Our study revealed a complex halo structure for the two dominant cluster galaxies, NGC 1399 and NGC 507. The halo center was found to be dominated by a bright X-ray peak coincident with the position of the optical galaxy. This central peak is not explained by the classical homogeneous or inhomogeneous cooling flow models. Instead our data suggest that these features are produced by stellar ejected material, kinetically heated by stellar mass losses. The total mass distribution shows that, within the effective radius, the hot halo distribution is essentially tracing the galactic potential due to luminous matter. At larger distances the halo dynamics is dominated by a large amount of dark matter extending on group and cluster scales.
The NGC 1399 X-ray halo possess a more complex morphology than NGC 507. This finding requires either that the dark matter distribution has a  hierarchical structure, or that environmental effects (ram stripping from ICM, tidal interaction with nearby galaxies) are producing departures from hydrostatic equilibrium.
We found significant density fluctuations in the hot gas distribution of both NGC 1399 and NGC 507. Some of these features are explained by the interaction of the radio-emitting plasma, filling the radio jets and lobes, with the surrounding ISM. This evidence indicates that Radio/X-ray interactions are a more widespread phenomena than observed before. The nature of the remaining structures is more uncertain: we speculate that they can be the result of galaxy-galaxy encounters or wakes produced by the motion of the galaxy through the ICM. Alternatively they may reflect the inhomogeneity of the cooling process invoked by many authors as an explanation for the failure of the standard cooling-flow models.

NGC 1404 represents a puzzling case. In fact, despite a large X-ray luminosity and halo temperature, which suggest a similarity with dominant cluster members, significant differences are found in the regular halo profile, small velocity dispersion and low metallicity. These conflicting evidences can be explained assuming that environmental effects, such as stifling of galactic winds by the ICM, are affecting the physical status of the halo.

Finally, we studied the population of discrete sources found in proximity of the two dominant galaxies. We were able to assess that the observed excess of sources observed in the Fornax cluster is associated with NGC 1399. This result was later confirmed by the higher resolution {\it Chandra} data. 
}

\pagenumbering{Roman}
\tableofcontents

\addcontentsline{toc}{chapter}{\protect\numberline{}{INTRODUCTION}}
\pagenumbering{roman}\setcounter{page}{1}

\chapter*{INTRODUCTION\markboth{INTRODUCTION}{INTRODUCTION}}

This thesis discusses the X-ray properties of Early-Type (i.e. Elliptical and S0) galaxies in groups and poor clusters. We focus on the brightest Early-Type (BETG) whose X-ray luminosity is dominated by the thermal emission from the gaseous halo. The study of these hot halos is of fundamental importance for understanding BETG, since its properties are connected with the evolution of the stellar population and to the formation history of the galaxy; moreover they are very good tracers of the gravitational structure and allow us to study the amount of dark matter hosted by these systems. 

Despite the fast development of scientific instrumentation and analysis methods, the X-ray properties of BETG are still a matter of debate. 
The X-ray satellites of the last generation ({\it Chandra} and XMM) permit to clearly separate, for the first time, the contribution of different components (i.e. accreting binaries, hot gas, cooling flow); however their discoveries are also seriously challenging many of the classical models used to describe the physical status of the halos.
For instance the large scatter of the X-ray properties of BETG, which show instead a high degree of homogeneity in optical bands, is not fully understood. The cooling flow models, which were assumed to be responsible for the structure of the inner halo, are not able to describe the temperature structure of the hot gas; the presence of multiphase medium, that they predict, is still uncertain and the fate of the accreted material is still unknown. Studies of large samples of Early-Type galaxies are suggesting that the X-ray structure of these systems is mostly determined by the gravitational potential well produced by large amounts of dark matter rather than by thermal mechanisms. 

These problems are enhanced in dominant Early-Types at the center of groups or clusters of galaxies since these objects lie at bottom of large-scale potential structures which must affect the hot gas properties. Moreover
they are likely to form and evolve through interactions with nearby galaxies  and with the Intra Cluster Medium, and may thus reveal how important these mechanisms are in the galaxy formation framework. 

In the present work we study BETG through pointed and archival data of the ROSAT and {\it Chandra} X-ray satellites. We choose galaxies at the center of two nearby systems: the Fornax cluster, a poor cluster located located $\sim 20$ Mpc from our galaxy, and the NGC 507 group, which is part of the Pisces cluster located at $\sim 66$ Mpc. These systems host very bright dominant galaxies which, due to their proximity, permit a detailed study of their X-ray properties. The choice to study BETG in poor clusters and groups is justified by the fact that these environments allow us to better disentangle the properties of individual galaxies from those of the ICM: the intergalactic gas is less dense than in rich clusters and there are no large cluster cooling flows which may affect the formation and evolution of the dominant galaxies. 

We will focus on the analysis of the morphology and of the dynamical structure of the X-ray halo in order to separate small and large scale structures and to determine to which extent the galactic halo properties are influenced by the local environment. We will then investigate the presence of interaction between the radio sources, hosted in the core of some galaxies, and the surrounding interstellar medium.
When possible we will also study the properties of the discrete source population.

The comparison between different galaxies will help to understand which mechanisms play a fundamental role in determining the X-ray properties of bright Early-Type galaxies and, more in general, affect the formation and evolution of these massive systems.\\

\vspace{.5cm}\noindent
The thesis is organized as follows:\\
- in chapter 1 we introduce the properties of Early-Type galaxies in the X-ray band and discuss the main sources of X-ray emission. We then focus on the informations that can be inferred from the study of the X-ray halos and on the effect of the local environment, paying particular attention to the latest developments;\\
- in chapters 2, 3 and 4 we present the data analysis concerning the individual galaxies and  discuss the results on each object separately. A preliminary version of the work discussed in Chapter 2 is presented in \cite{paolillo00,paolillo01}; the final results of Chapters 2 and 3 have been published in \cite{paolillo02}. Chapter 4 will be included in a separate paper that is currently in preparation;\\
- in chapter 5 we compare the results on the individual galaxies obtained in previous chapters and discuss global properties of bright Early-Type X-ray emission in the context of the different halo formation and evolution scenarios.

\pagenumbering{arabic}\setcounter{page}{1}
  \chapter{X-RAY EMISSION FROM EARLY-TYPE GALAXIES}
\section{THE EARLY-TYPE GALAXIES}
\label{earlytype}
\begin{figure}[t]
\centerline{\psfig{figure=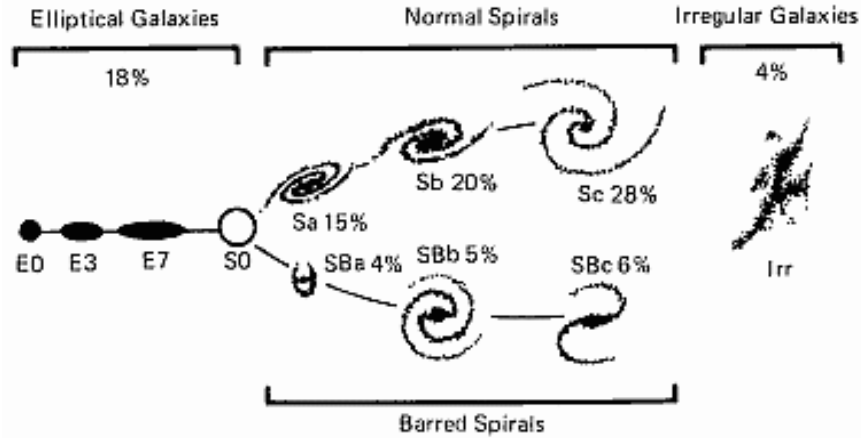,angle=0,width=0.7\textwidth}}
\caption{The \cite{Hubble} classification of galaxies. The numbers represent the relative abundance of each type.}
\label{Hclass}
\end{figure}

Early-type galaxies are large, bright stellar systems with considerable structural regularity. According to the \cite{Hubble} criteria (Figure \ref{Hclass}), which classifies galaxies based on their morphology and light concentration toward the nucleus, the Early-Type galaxies represent a group which includes both Elliptical (E0-E7) and Lenticular (S0) galaxies. These stellar systems are characterized by approximately elliptical and concentric isophotes with axial ratios ranging from 1 (E0) to 0.3 (E7). In the case of S0 galaxies the elliptical bulge is integrated by a disk component, missing the `arms' typical of spiral (S) galaxies. 

The Early-Type galaxies represent a class of well studied objects both because of their homogeneous properties and because of their high luminosity (absolute magnitudes in the range $-24 < M_B < -20$) which allows to detect them at large distances. These objects are usually composed by an `old' stellar population, since the last major star formation event occurred several Gyrs ago. Their optical emission is thus centered at longer wavelengths with respect to the `bluer' spiral galaxies. Photometric studies \citep{Cap89,Caon93} revealed that the surface brightness of Ellipticals is well represented by a {\it de Vaucouleurs} profile \citep{deVau48}:
\beq
\label{deVau_eq}
\log\left[I(r)\over I(r_c)\right]=-3.331\left[\left(r\over r_c\right)^{1/4}-1\right]
\eeq
where $I$ represents the surface brightness, $r$ is the projected distance from the galactic center and $r_c$ is the {\it effective radius}, i.e. the radius which includes half of the total light of the galaxy. S0 galaxies possess an additional disk component with an exponential profile:
\beq
\label{expo_prof}
I(r)=I_0\exp(-r/r_d)
\eeq
where $I_0$ is the central brightness and $r_d$ is the disc scale length.

Unlike Late-Type galaxies (Spirals and Irregulars), Early-Types have depleted already their hydrogen supply either by forming stars or ejecting it into the Intra Cluster Medium (ICM) \citep{Sar88}. However while the content of cold gas, emitting at optical wavelengths, is usually small, we will see in $\S$ \ref{Xprop} that this situation is reversed when we take into account the hot, fully-ionized Inter Stellar Medium (ISM).

The stellar dynamic of these systems is characterized by a high triaxial velocity dispersion and small or no rotation \citep[ and references therein]{deZee91}. For this reason Early-Type galaxies are often called `hot systems' with a dynamical temperature $T_{dyn}=\mu m_p\sigma^2/k \sim 1.4-5.4\times
10^6$ K, where $\mu$ is the mean molecular mass, $m_p$ the proton mass, $k$ the Boltzmann constant and $\sigma\sim 150-300$ km s$^{-1}$ the stellar velocity dispersion \citep[e.g.][]{Whit85,Gra98}.
The homogeneity of Elliptical galaxies is supported by the fact that stellar velocity dispersion, surface brightness and effective radius are constrained to a `fundamental plane' \citep{Djorg87,Dressler87}.

Early-Type galaxies are found to be more numerous in high-density environments, such as groups or clusters of galaxies. This suggests that their formation is related, to some extent, to galaxy-galaxy interactions: Elliptical galaxies may accrete either by means of tidal interactions, capturing smaller galaxies or part of them in their deeper potential well, or by merging of smaller systems, maybe Spirals which loose part of their angular momentum, creating a spheroidal system with high velocity dispersion \citep{Bar88,Bar92,Hern93}.

\subsection{cD Galaxies}
Even though Early-Type galaxies seem to be a nearly homogeneous class, the data acquired in the last 20 years has shown that significant departures from the general properties are often present in the form of large quantities of dust and gas \citep{Cap94}, anomalous or counter-rotating nuclei, polar rings \citep{Ber78,Ben88,Whit87}. Moreover Elliptical galaxies are often strong X-ray and radio emitters.

Within the peculiar Early-Type galaxies there are the largest known stellar systems: the giant cD Ellipticals. cD galaxies were defined by \cite{Math64} as galaxies with a nucleus of a very luminous Elliptical (represented by the {\it de Vaucouleur} brightness profile) embedded in an extended amorphous halo of low surface brightness. 
They are usually found at the center of regular, compact clusters of galaxies \citep{Bau70} but several cDs have been also found in poor clusters and groups \citep{Mor75,Alb77,kill88}.
Their peculiar nature is confirmed by the fact that these galaxies are too bright ($M_V\geq -24$) to belong to the same galaxy luminosity function that fits the other bright Ellipticals distribution \citep{Sche76}. They also often have double or multiple nuclei \citep{Min61,Hoe80,Sch82} and/or intense X-ray and radio emission. 

Their special structural and kinematic properties suggest that they have been formed or modified by dynamical processes in clusters. \cite{Gal72} and \cite{Rich75,Rich76} have suggested that cDs consist of the debris from galaxy collisions in which the outer envelopes of galaxies are stripped by tidal effects and then settle at the cluster center. An alternative scenario \citep{Ost75,Gunn76} is one in which the orbits of massive cluster galaxies decay due to dynamical friction and subsequently merge in the cluster center forming a single massive galaxy. This galaxy would then swallow any smaller galaxy passing through the cluster center (the so called `galactic cannibalism', \citealp{Ost77}). 
However the core of a rich cluster is a very active physical environment in which many other processes may be important.

  \section{X-RAY PROPERTIES OF NORMAL EARLY--TYPE GALAXIES}
\label{Xprop}
The first identification of extragalactic X-ray sources and their association with galaxy clusters were made with balloon and rocket-borne detectors \citep[e.g.][]{Byr66,Fri71} while the study of their physical properties begun with the {\it Uhuru}, {\it Ariel} and HEAO satellites \citep[e.g.][]{Kel73,Kel74,Kel75,Elvis75,Ulm81}. However, before the launch of the {\it Einstein} satellite \citep{Giacc79}, just four individual galaxies had been detected in X-rays\footnote{Excluding sources associated with Seyfert nuclei}: the Milky Way, M31 and the Magellanic Clouds. In general the so called `normal' galaxies were thought to be of little interest for X-ray astronomy because they were missing an active nucleus or star forming regions.

The {\it Einstein} satellite showed that normal Early-Type galaxies are strong X-ray emitters with luminosities ranging from $10^{39}$ erg s$^{-1}$ to $10^{43}$ erg s$^{-1}$, the upper limits being reached by cD galaxies in the center of rich clusters. 
{\it Einstein} (see \citealp{Fab89} for a review of {\it Einstein} results) and the following generation of X-ray satellites with improved spatial and spectral resolution (ROSAT: \citealp{pfeff87}, \citealp{Dav96}; ASCA: \citealp{tanaka94}; Beppo-SAX: \citealp{boella97}) allowed to study in detail the properties of these systems.

Different physical mechanism were suggested to be responsible for the X-ray luminosity of normal Early-Type galaxies: i) integrated stellar coronal emission; ii) thermal emission due to the hot ISM; iii) end products of stellar evolution (compact binary systems and supernovae remnants); iv) X-ray diffuse emission due to non-thermal processes.

Stellar coronae have X-ray luminosities reaching up to $10^{33}$ erg s$^{-1}$ \citep{Vai81} but their integrated emission does not contribute significantly to the total X-ray emission of normal Early-Type galaxies, except for the fainter ones or at low energies ($<0.3$ keV, e.g. \citealp{Pel&Fab94}).
Instead, the emission coming from hot ISM and end products of stellar evolution must be taken both into account to explain the Early-Type X-ray emission.
\cite{Fab84} and \cite{For85} have shown that Elliptical galaxies must be divided in two groups depending on their total luminosity. Objects with $L_X<10^{41}$ erg s$^{-1}$ seems to be dominated by the integrated emission of compact accreting systems. This is demonstrated by the `hard' spectra (kT$ > 2$ keV) and by the correlation between optical and X-ray luminosity similar to the one seen for Spiral galaxies, whose emission is mostly due to discrete sources \citep[ and references therein]{KFT92,Irw98b,Tri00}. 

\begin{figure}[t]
\centerline{\psfig{figure=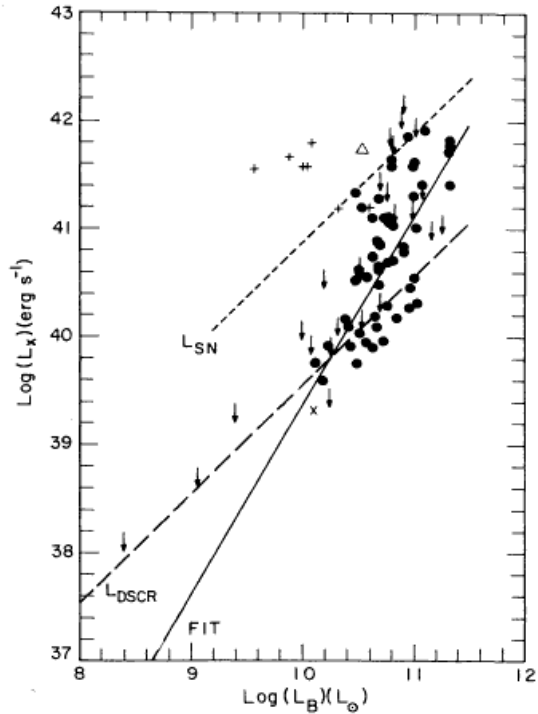}}	
\caption{Early-Type galaxies with high X-ray luminosity (filled circles)
show a steeper correlation (continuous line), in the L$_X$-L$_B$ plane, different from the one expected if the emission was due only to discrete sources (long-dashed line), as happens for Spirals and fainter Ellipticals.
 L$_{SN}$ represent the expected correlation if the hot gas is heated by Supernovae explosions (from \citealp{Can87}).}
\label{Xcorr}
\end{figure}

Early-Type galaxies with $L_X>10^{41}$ erg s$^{-1}$ show a correlation between $L_B$ and $L_x$ steeper than the one observed for Spirals and fainter Ellipticals (Figure \ref{Xcorr}, \citealp{Can87}).
This result implies the presence of an additional component in the bright  galaxies. This components is represented by hot ($\sim 1$ keV) ISM
trapped by the galaxy potential well \citep{Esk95} whose emission is due to thermal processes (see $\S$ \ref{thermal}).

The origin of the hot ISM is still unclear. The gas may be the remainings of the primordial gas from which the galaxy formed or may be the result of the stellar mass losses and supernovae ejections accumulated during the galaxy life. 
The possibility to ascertain which of these theories is true relies on our ability to measure the ISM metallicity: low metallicities would confirm the primordial origin while near-solar ones would indicate metal enrichment after the galaxy formation. In fact primordial nucleosyntesis models predict that during the formation of the universe (the Big Bang) no elements than hydrogen, helium and lithium were formed, because of the lack of any stable isotopes with atomic weights of 5 or 8 (\citealp{Wei72}; see also \citealp{Long98}). All the mechanisms which have been suggested for the formation of heavier elements involve processing in stars.

The low spectral resolution of {\it Einstein} and ROSAT allowed both single-temperatures spectral models with subsolar metallicities or multi-temperature fits with solar metallicities \citep[e.g.][]{Kim95}. However extensive ASCA studies \citep{Mat00,Buo00} have now shown that the ISM metallicity is nearly solar up to great distances ($\sim 100$ kpc) from the galaxy centers, thus supporting the stellar origin of the ISM. This result seems to be further confirmed by the latest {\it Chandra} results  \citep[e.g.][]{Dav00}.
   
~\\
In X-ray faint Early-Type the hot interstellar gas, that dominates the emission of their bright counterparts, may have been lost in galactic winds \citep{loew87,David91}, by ram-pressure stripping by ambient intracluster or intragroup gas \citep{White91,toni01}. The bulk of the X-ray emission in X-ray faint Early-Type is uncertain: it might be due to low-mass X-ray binaries (LMXBs), to an active galactic nucleus (AGN, \citealp{allen00}), to interstellar gas \citep{Pel&Fab94} or to fainter stellar sources.

In general X-ray faint galaxies have significantly different X-ray spectral properties than their X-ray bright counterparts. In fact X-ray faint galaxies exhibit two distinct spectral components. They have a hard $\sim 5-10$ keV component, most easily seen in ASCA spectra \citep{Matsumo97}, which is roughly proportional to the optical luminosity of the galaxy. Actually, both X-ray faint and X-ray bright Early-Type galaxies appear to have this hard X-ray component, thus suggesting that it is due to LMXBs like those seen in the bulge of our Galaxy \citep[e.g.][]{White95}. However, the ASCA observations cannot resolve this component into discrete sources because of its low angular resolution, nor they provide much detailed information on its spectrum.

X-ray faint galaxies also have a very soft ($\sim 0.2$ keV) component, whose origin is uncertain \citep{Fab94,Pel94,Kim96}. Suggested stellar sources for this soft emission include M stars and RS CVn binaries or supersoft sources \citep{Kaha97}, but none of these appears to work quantitatively \citep{Pel&Fab94,Irw98a}. It was also proposed that the soft X-rays may be due to warm (0.2 keV) ISM \citep{Pel&Fab94} or to the same LMXBs population responsible for the hard emission \citep{Irw98a,Irw98b}. 
Recent ROSAT and {\it Chandra} \citep{weiss96} studies have now identified  these discrete sources in the faint Early-Type galaxy NGC 4697 and demonstrated that they can account for $>70\%$ of the total X-ray emission \citep{Irw00,Sar01}. They also showed that, in this galaxy, the soft emission is mainly produced by warm ISM.

~\\
The presence of diffuse non-thermal X-ray emission in Early-Type galaxies requires the presence of a population of high energy electrons. The existence of such a population is now well established through radio measurements. In fact radio observations have shown collimated radio jets emanating from the nucleus itself, with velocities approaching the speed of light. The jets are believed to be made up of very energetic particles expelled from the nucleus and fueling diffuse structures called `radio lobes'. Bright spots (hotspots) are often observed at the extremities of the lobes. The radio emission is invariably polarized and is characterized by large brightness temperatures ($10^{10}-10^{12}$ K) so that it is assumed to be generated by the synchrotron process. 

The two primary mechanisms normally considered in order to explain X-ray emission in radio sources are synchrotron emission itself and inverse Compton (IC) emission \citep{Har99}. 
Synchrotron emission is often considered to explain X-ray emission from knots and hotspots associated with radio jets. The gas flowing in the jet at supersonic velocity, decelerates suddenly near the hot spot and this causes a shock wave to form across the jet. Before reaching the shock wave, most of the energy is in the form of ordered kinetic energy. The passage through the shock converts this into relativistic electron energy and magnetic field energy. The relativistic electrons with $\gamma\sim 10^7$ are thus able to radiate synchrotron emission in the X-ray band. An example is knot A in the M87 jet \citep{Bir91}.

Alternatively the relativistic electrons, responsible for the radio emission, may scatter low energy photons into the X-ray energy band through inverse Compton. The IC process can work on any photon distribution: in particular with synchrotron emission and 3 K background photons. The synchrotron self Compton (SSC) emission requires compact, high-brightness radio structures in order to be detectable by current X-ray systems. For instance it has been established to be responsible for the hotspots of Cygnus A  \citep{Har94}. IC scattering of background photons, instead, may produce diffuse, low-brightness X-ray emission corresponding with the position of the radio lobes, such as has been observed in the strong radio galaxy Fornax A \citep{Fei95,Kan95}. This process 
may be important also in faint radio galaxies and will be discussed in greater detail in $\S$ \ref{IC}.

    \section{X-RAY EMISSION MECHANISMS}
	\subsection{Thermal Emission}
	\label{thermal}
The ISM responsible for the thermal component of the X-ray emission in normal Early-Type galaxies is in the form of a hot ($T\sim 10^7$ K) and low density ($n < 10^{-1}$ cm$^{-1}$) plasma. In these conditions it is possible to make several simplifications \citep{Sar88}: i) the time scale for elastic Coulomb collisions between particles in the plasma is much shorter than the age, or cooling time (see $\S$ \ref{cooling}), of the plasma, and thus the free particles can be assumed to have a Maxwell-Boltzmann distribution at temperature $T_g$; this is the kinetic temperature of electrons which determines the rate of all excitation and ionization processes; ii) due to the low density, collisional excitation and de-excitation are much slower than radiative decays and thus any ionization or excitation process can be assumed to be initiated from the ground state of an ion; three-body (or more) collisional processes can be ignored; iii) the radiation field in a cluster is sufficiently dilute that stimulated radiative transition are not important, and the effect of the radiation field on the gas is insignificant; iv) at these low densities the gas is optically thin and the absorption processes do not affect the radiation field.
These assumptions constitute the 'coronal limit', under which ionization and emission result primarily from collisions of ions with electrons (collisions with ions can be ignored). Finally, v) the time scales for ionization and recombination are generally considerably less than the age of the galaxy or any other relevant hydrodynamic time scale, so that the plasma can be assumed to be in ionization equilibrium.

In the ISM (and almost in all astrophysical plasma) hydrogen and helium are the most common elements with all the heavier elements being considerably less abundant. It is conventional to use solar abundances as a standard when studying many astrophysical systems. Since most of the electrons originate in hydrogen and helium atoms, and they are fully ionized under the conditions considered here, the electron number density is nearly independent of the state of ionization and is roughly given by $n_e=1.21~n_p$ where $n_p$ is the hydrogen number density. This is the assumption that we will use throughout this work.

The X-ray continuum emission of a hot diffuse plasma is due primarily to three processes: thermal bremsstrahlung (free-free emission), recombination (free-bound) emission and two-photon decay of metastable levels. Processes that contribute to the X-ray line emission include collisional excitation of valence or inner shell electrons, radiative recombination, inner shell collisional excitation and radiative cascades following any of these processes. Compilations of emissivities (emitted energy per unit time, frequency and volume) for X-ray lines and continua have been given by various authors \citep[e.g.][]{Mewe85,Landini90}; however throughout this work we adopted the Raymond-Smith model \citep[ and more recent updates]{Ray77} which is one of the most used in literature and is included in many spectral fitting packages (e.g. Xspec, CIAO).

\begin{figure}[t]
\centerline{\psfig{figure=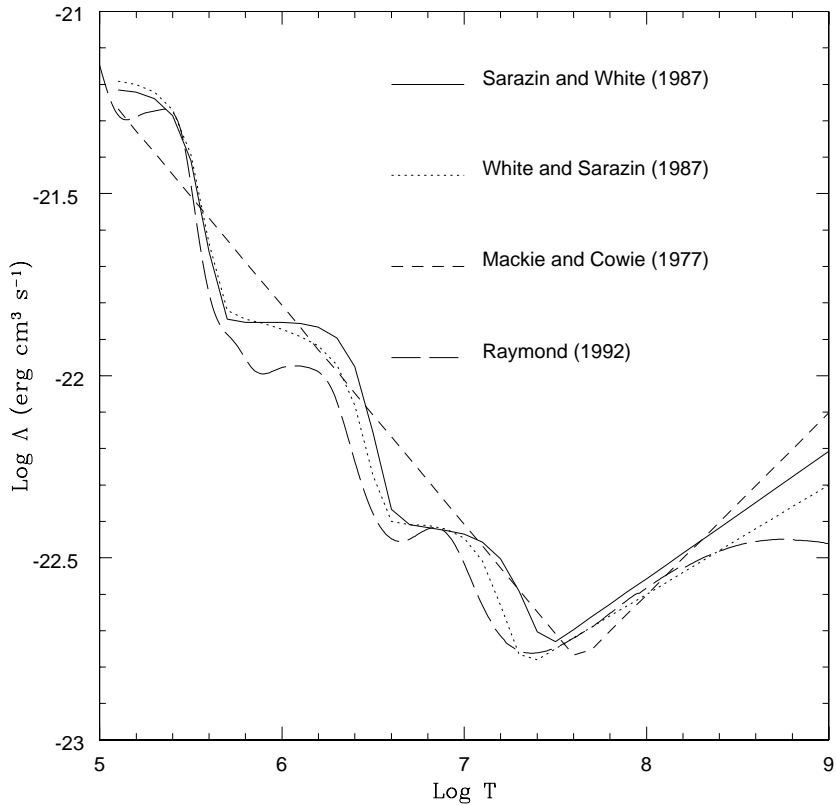,width=0.7\textwidth}}	
\caption{Bolometric cooling functions from literature. }
\label{cfunc}
\end{figure}

Since all the emission processes just mentioned have emissivities proportional to the product of the ion and of the electron densities, and otherwise depend only on the temperature, we can write:
\beq
\label{emiss1}
\epsilon(\nu,T_g)=\sum_{X,i}~\Lambda_{X^i}(\nu,T_g)~n_{X^i}~n_e
\eeq
where $\epsilon(\nu,T_g)$ is the emissivity at frequency $\nu$ and gas temperature $T_g$, $\Lambda_{X^i}$ is the emission per ion at unit electron density, $n_{X^i}$ is the number density of the $i$-th ionization stage of the element $X$ and $n_e$ is the electron number density. If $n_X$ is the total density of the element $X$, in equilibrium the ionization fractions $f(X^i)=n_{X^i}/n_X$ depend only on the temperature and equation (\ref{emiss1}) becomes:
\beq
\label{emiss2}
\epsilon(\nu,T_g)=n_p n_e\sum_{X,i}\frac{n_X}{n_p}[f(X^i,T_g)~\Lambda_{X^i}(\nu,T_g)]
\eeq
If we assume solar abundances the ratio $n_X/n_p$ is known and the quantity  $[f(X^i,T_g)~\Lambda_{X^i}(\nu,T_g)]$ depends only on the temperature. Thus equation (\ref{emiss2}) can be written as:
\beq
\label{emiss3}
\epsilon(\nu,T_g)=n_p n_e~\Lambda(\nu,T_g)= \frac{n_e^2}{1.21}\Lambda(\nu,T_g)
\eeq
where 
\beq
\Lambda(\nu,T_g)= \sum_{X,i}\frac{n_X}{n_p}[f(X^i,T_g)~\Lambda_{X^i}(\nu,T_g)]
\eeq
is usually called 'cooling function'. Examples of cooling functions from literature, integrated over the whole spectrum, are shown in Figure \ref{cfunc}. The differences arise from the different approximations and atomic models adopted by the authors to calculate the relative contribution of each physical process involved in the radiative process.

In discussing which properties can be derived by X-ray measurements we must consider that X-ray detectors allow us to measure the X-ray flux and spectrum, both of which are the result of the ISM emissivity integrated along the line of sight. $\Lambda(\nu,T_g)$, instead, depends only on the local gas properties. The shape of the spectrum thus reflects the distribution of temperatures, abundances $n(X)/n(H)$ and density through the gas. Deriving these quantities from the measured X-ray spectrum through thermal models fitting can be a very difficult task, unless we make some assumptions on how these parameters vary within the ISM.
In the simplest case we can assume that the gas temperature and the chemical composition are constant along the line of sight, so that the total luminosity $L_\nu$ of a galaxy is given by:
\beq
\label{totlum}
L(\nu,T_g)=\int \epsilon(\nu,T_g) {\rm d}V=\int n_p n_e\Lambda(\nu,T_g) {\rm d}V= EM\cdot\Lambda(\nu,T_g)  
\eeq
where
\beq
\label{EM}
EM=\int n_p n_e {\rm d}V
\eeq
is the {\it emission measure} of the plasma and $V$ is the emitting volume. Thus knowing $T_g$ from the spectral fit we can derive $EM$. 

To define completely the status of the gas we need to find $n_e$, and thus the gas density. This implies the knowledge of the observed volume and of the gas distribution through this volume. This problem can be solved in part assuming some simple distribution of the hot gas, such as those discussed in $\S$ \ref{halo}. We will also see that the use of `deprojection' techniques allows to consider more complex situations in which the ISM is characterized by different temperatures and abundances at each radius from the galaxy center. 

	\subsection{Inverse Compton scattering}
	\label{IC}
Among the physical processes discussed in $\S$ \ref{Xprop}, which produce non-thermal X-ray emission, we are mostly concerned with Inverse Compton  scattering from the 3 K background photons (3 K IC). This is because the galaxies considered here are only weak radio sources. The radio emitting plasma is most probably moving at subsonic speed so that the shock produced at the lobe leading edge is not strong enough to produce `hotspots'. Similarly the radio brightness is too low to produce detectable SSC emission. 3 K IC instead may be an important process also in radio faint sources and will be discussed in greater detail to obtain some important relations that will be useful later.

Nonthermal synchrotron radio emission generally has a spectrum in which the flux $S_r(\nu)$ is well represented as a power-law over a wide range of frequencies $\nu$ \citep{Rob83}:
\beq
S_r\propto \nu^{-\alpha_r}
\eeq
where $\alpha_r$ is the radio spectral index, with typical value of $\sim 0.8$.
It is generally assumed that this spectral shape is due to the power-law energy distribution of the relativistic electrons:
\beq
\label{pow_distr}
N_e(E){\rm d}E=N_0E^{-p}{\rm d}E~~~~~~~E_l<E<E_u
\eeq
where $N_e{\rm d}E$ is the number of electrons with energies between $E$ and $E+{\rm d}E$, $E=\gamma m_ec^2$ is the electron energy, $E_l$ and $E_u$ define the lower and upper energy cutoff of the distribution. In this case the number of electrons with $\gamma< 10^3$, which contribute to IC scattering of background photons to energies $\sim 1$ keV \citep{Har79}, will be larger than those in the range $\gamma\sim 10^3-10^5$ which are responsible for the synchrotron emission, so that every radio galaxy must be a source of IC emission as well. 

These extremely relativistic electrons ($\gamma >> 1$) will scatter photons with initial frequency $\nu_b$ up to an average frequency \citep{Lon92}:
\beq
<\nu_x>=\frac{4\gamma^2\nu_b}{3}.
\eeq  
The resulting IC radiation has a spectrum given by \citep{Sar88}:
\beq
\frac{{\rm d}L_x}{{\rm d}\nu_x}=3\sigma_T ch2^pN_0\frac{p^2+4p+11}{(p+3)^2(p+1)(p+5)}\left [\int \nu^{\alpha_x}_{b}n_b(\nu_b){\rm d}\nu_b\right ] \nu^{\alpha_x}_x
\eeq
where $L_x$ is the X-ray luminosity emitted at frequency $\nu_x$, $n_b$ is the number density of background photons as a function of frequency $\nu_b$, $\sigma_T$ is the Thompson cross section, and the X-ray spectral index is related to the electron distribution index by:
\beq
\alpha_x=\frac{p-1}{2}
\eeq
If the source of low energy photons is the 3 K cosmic background radiation, taken to be a blackbody at temperature $T_r$, then:
\beq
\frac{{\rm d}L_x}{{\rm d}\nu_x}=\frac{3\pi\sigma_T}{h^2c^2}b(p)N(kT_r)^3\left ({kT_r\over h\nu_x}\right)^{\alpha_x}
\eeq
for frequencies in the range $\gamma_l^2<<(kT_r/h\nu_x)<<\gamma_u^2$. The function $b(p)$ is given by \cite{Sar88}.

The synchrotron radio luminosity $L_r$ produced by the same electron population is given by
\beq
\frac{{\rm d}L_r}{{\rm d}\nu_r}=\frac{8\pi^2e^2\nu_B N_0}{c}~a(p)\left ({3\nu_B\over 2\nu_r}\right )^{\alpha_r}
\eeq  
where $\nu_B\equiv eB/2\pi m_ec$ is the gyrofrequency in the magnetic field $B$ and $a(p)$ is an analytic function of $p$ \citep{Sar88}.

The radio spectral index $\alpha_r$ is equal to the X-ray index $\alpha_x$.
The synchrotron radio and IC X-ray emission from the same relativistic electrons will thus have the same spectral shape, reflecting the underlying electron energy distribution. The luminosities in the two bands are simply related:
\beq
{L_x\over L_r}={U_b\over U_B}
\eeq  
where $U_b$ and $U_B=B^2/8\pi$ are the energy density of the background radiation and of the magnetic field, respectively.
This result shows that the X-ray luminosity expected from IC scattering in a radio source can be easily determined if one knows the radio luminosity and the magnetic field energy density, since the cosmic background energy density is well known. Equivalently we can determine $U_B$ if we can measure both $L_x$ and $L_r$.

In practice, fluxes at a single frequency are easier to measure than integrated luminosities. Harris \& Grindlay (\citeyear{Har79}; also see \citealp{Fei95}) obtained an expression for the magnetic field $B_{IC}$ required to produce a measured ratio of radio to X-ray flux:
\beq
\label{B_IC}
B_{IC}=\left [\frac{(4790)^{\alpha_r}C(\alpha_r)G(\alpha_r)(1+z)^{3-\alpha_r}S_rE_x^{\alpha_r}}{10^{70}S_x\nu^{\alpha_r}\sin\phi}\right ]^{1/(1-\alpha_r)}
\eeq
where $z$ is the source redshift, $\phi$ is the angle between the (assumed) uniform magnetic field and the line of sight, $S_r$ is the radio flux density measured in Jansky at $\nu_r$ (in GHz) and $S_x$ is the X-ray flux density (in ergs s$^{-1}$ cm$^{-2}$ Hz$^{-1}$) at energy $E_x$ (in keV). The functions $C(\alpha_r)$ and $G(\alpha_r)$ are given in \cite{Har79}. 

The problem in estimating the X-ray flux expected from a given radio source is to estimate the magnetic field intensity in the radio lobes, where the bulk of the IC emission originates. In fact the nature of synchrotron emission is such that it is difficult to separate the number density of relativistic electrons from the magnetic field strength. The general approach is to invoke equipartition to allow order of magnitude estimates.

The total energy contained in the lobes is given by the sum of the magnetic field energy density $U_B=B^2/8\pi$ and the particles energy density $U_p\propto B^{-3/2}$ (obtained assuming a power law distribution of the particles and a high energy cutoff, as in equation (\ref{pow_distr}) ) so that:
\beq
\label{tot_energy}
U_{tot}=aB^{-3/2}+bB^2
\eeq
Here $a$ and $b$ are two constants which depend on the power-law index $p$, the total radio luminosity and the energy cutoffs (see for instance \citealp{Pach70}). The equipartition approach consists simply in determining the value of $B$ that yields the minimum energy density:
\beq
\label{min_energy}
{\partial U_{tot}\over\partial B}=-{3\over 2}aB^{-5/2}+2bB=0
\eeq
The full expression for the minimum energy field $B_{ME}$ is given by \cite{Mil80}:
\beq
\label{B_ME}
B_{ME}=5.69\times 10^{-5}\left [{(1+k)\over\eta}(1+z)^{3-\alpha_r}{1\over \theta^2 s \sin^{3/2}\phi}~{S_r\over \nu_r^{\alpha_r}}~{\nu_2^{\alpha_r+(1/2)}-\nu_1^{\alpha_r+(1/2)}\over \alpha_r+{1\over 2}}\right ]
\eeq
where $\eta$ is the filling factor of the synchrotron emitting region, $\theta$ is the lobe diameter (in arcseconds), $s$ is the path length through the source (in kpc) and $\nu_1$ and $\nu_2$ (in GHz) are the upper and lower cutoff frequencies for the radio synchrotron spectrum. Since the synchrotron emission is essentially due to electrons, the energy density contained in heavier particles can not be inferred directly from observations. Thus the factor $k$, defined as the ratio of the energy in heavy particles to the energy in electrons, is introduced in equation (\ref{B_ME}) to take into account this contribution.

Equation (\ref{min_energy}) also yields $b=3/4~aB^{-7/2}$ and thus: 
\beq
{U_p \over U_B} = {aB^{-3/2}\over bB^2}\simeq 1
\eeq
which explains why the minimum energy condition is also called `equipartition'. Of course the equipartition condition is not necessarily valid but allows to obtain lower limits on the expected X-ray emission from IC scattering in radio sources.

 	\subsection{Discrete Sources Emission}
	\label{discrete}
We have seen in $\S$ \ref{Xprop} that discrete sources contribute significantly to the X-ray emission of faint Early-Type galaxies and are present in bright galaxies too. These sources are basically represented by binary stellar systems emitting in X-rays \citep{Tri85,Can87,Irw98a,Irw98b,Irw00,Sar01}.
The knowledge of the properties of these objects is very important to understand the star formation history and the evolution of the stellar population in galaxies. An exhaustive treatment of the physical processes responsible for the X-ray emission in these systems is a very complex task and is beyond the aim of this work. Instead, I will qualitatively discuss the scenario in which the X-ray emission is produced and mention the main properties of this emission. More detailed informations can be found in the references included below.

An X-ray binary is composed either by a neutron star or a black hole accreting material from a companion star (see \citealp{Lewin95} for an extensive review). The primary factors that determine the emission properties of an accreting compact object are i) whether the central object is a black hole (BH) or a neutron star (NS), ii) if it is a neutron star, the strength and geometry of its magnetic field, and iii) the geometry of the accretion flow from the companion (disk vs spherical accretion). These determine whether the emission region is the small magnetic polar cap of a neutron star, a hot accretion disk surrounding a black hole, a shock heated region in a spherical inflow, or the boundary layer between an accretion disk and a neutron star. Two more factors are the mass of the central object and the mass accretion rate; these influence the overall luminosity, spectral shape and time variability of the emission.

A neutron star with a strong magnetic field ($\sim 10^{12}$ G) will disrupt the accretion flow at several hundred neutron-star-radii and funnel material onto the magnetic poles \citep{Prin72,Dav73,Lamb73}. If the magnetic and rotation axes are misaligned, X-ray pulsations will be observed if the beamed emission from the magnetic poles rotates through the line of sight \citep{Mesz80,Nagel81a,Nagel81b}. When the magnetic field of the neutron star is relatively weak ($< 10^{10}$ G), the disk may touch or come close to the neutron star surface. The energy released from the inner accretion disk and the boundary layer between the disk and the neutron star will dominate the emission \citep[e.g.][]{Mits84}. If the central object is a black hole the X-rays come from the inner disk and are the results of viscous heating \citep{Shak73}.

Instabilities in the emission region, or its influence on the nearby accretion flow, can give rise to rapid fluctuations, or quasi-periodic oscillations. The material, as it accumulates on the neutron star, may reach a critical size and undergo a thermonuclear flash, resulting in an X-ray burst. Instabilities in the accretion flow can also give rise to X-ray bursts or flares \citep{Taam88}.

The spectral type of the companion determines the mode of the mass transfer to the compact object and the overall environment in it's vicinity. In the low-mass X-ray binaries (LMXBs) the companion is later than type A and can, in some very evolved systems, be a white dwarf. A late type or degenerate star does not have a natural wind strong enough to power the observed X-ray source. Significant mass transfer will occur only if the companion fills its critical gravitational lobe: the Roche lobe. 

In high-mass X-ray binaries (HMXBs) the companion is an O or B star whose optical/UV luminosity may be comparable to, or greater than, that of the X-ray source \citep{Conti78,Pett78}. X-ray heating is minimal, with the optical properties dominated by the companion star. The companion star has substantial stellar wind, removing $10^{-6}-10^{-10} M_\odot$ yr$^{-1}$ with a terminal velocity up to 2000 km s$^{-1}$. A NS or BH in a relatively close orbit will capture a significant fraction of the wind, sufficient to power the X-ray source. Roche lobe overflow can be a supplement to the mass transfer rate. 

Many X-ray binaries are transient sources that appear on a timescale of a few days, and then decay over tens or hundreds of days \citep{White84,Par84}. The transient episodes may result from an instability in the accretion disk, or a mass ejection episode from the companion.

The flow geometry is determined by the angular momentum per specific mass of the accretion flow. If the companion star fills its critical Roche lobe, then a stream of material will be driven through the inner Lagrangian point. This stream will orbit the compact object at a radius determined by its specific angular momentum \citep{Lubow75}. Viscous interactions and angular momentum conservation will cause the ring to expand into a disk. The disk's outer radius is limited by tidal forces, which will transfer angular momentum back to the binary orbit.

Our knowledge of discrete sources comes basically from objects in our galaxy or in very nearby ones (e.g. M31). Only recently the use of XMM and {\it Chandra} satellites is revealing the properties of these sources in more distant galaxies. In particular the emission due to discrete emission in Early-Type galaxies is thought to be mainly due to low-mass X-ray binaries. This is because in HMXBs 
the time taken by the massive companion of the neutron star to evolve from the  neutron-star-forming supernova event to the end of the X-ray emitting phase 
 is relatively short ($\sim 5\times 10^6$ yr , \citealp{vdH92}). In LMXBs, on the contrary, this interval lasts $\sim 1-2\times 10^9$ yr \citep{Ghosh01}.
HMXBs are thus expected to fade away shortly after the last relevant star formation event, while LMXBs survive much longer.
In Early-Type galaxies star formation ended several Gyrs ago ($\S$ \ref{earlytype}) and very few (if any) HMXB are expected to be still present.

The spectral properties of the LMXBs population in Early-Type galaxies is still a matter of debate. The assumed spectra is usually a thermal bremsstrahlung
with characteristic temperature $\geq 5$ keV \citep[e.g.][]{Irw00}. However
studies of galactic LMXBs revealed that these objects may exhibit different spectral features. The physics of the accretion process, which is very complex and not fully understood, allows the presence of other components such as a multi-temperature blackbody spectrum, attributed to the accreting disk, or a power-law spectrum, due to IC scattering of the emitted photons in the dense environment surrounding the compact accreting object (\citealp[e.g.][]{laparola01}; also see \citealp{White95} for a review of the LMXBs spectral properties). 
 
\cite{Sar01} found that a bremsstrahlung model with $kT\sim8$ keV fits well the average spectra of X-ray sources in faint elliptical galaxies. However the 
X-ray luminosity function of LMXBs has a `knee' at $L_x=3.2\times 10^{38}$ erg s$^{-1}$, and sources fainter than this limit have additional soft emission ($kT\sim 0.15$). The different spectra and the change in slope of the luminosity function may reflect difference in the emission mechanism due to the presence of a BH instead of a NS in the binary system. 
In fact, stellar evolution theories predict that the upper mass limit of a NS is $\sim 1.4$ M$_\odot$, since more massive objects undergo the gravitational collapse which creates a BH. Assuming equilibrium between the gravitational force acting on the infalling material and the radiative pressure, it is possible to calculate the `Eddinghton luminosity', i.e. the maximum luminosity expected from the accreting material on a NS:
\beq
\label{L_Edd}
L_{Edd}\sim \frac{4 \pi G M m_p c}{\sigma_T}
=1.3\times10^{38} \frac{M}{M\odot}\mbox{ erg s}^{-1}
\eeq 
where $M$ is the mass of the accreting star, $m_p$ is the mass of the
proton, $\sigma_T$ is the Thomson cross section.  Thus, the $3.2\times 10^{38}$ erg s$^{-1}$ luminosity is close to the upper limit expected from a 1.4 M$_\odot$ NS. Higher luminosities would destabilize the accretion flow and can be explained only taking into account relativistic effects associated with a BH \citep[ and references therein]{Fab89}.

  \section{PHYSICAL STATUS OF THE HALO}
\label{halo}
The different luminosities observed in bright Early-Type galaxies, i.e. those where the main contribution to the X-ray flux comes from the hot halo, is believed to be due to the capability of the system to `confine' the gas. Large galaxies are usually located at the bottom of the local gravitational well, often also due to their position at the center of clusters or groups of galaxies. 

The X-ray data can be used to derive the physical properties of the hot emitting gas, such as central density, cooling times and total gaseous mass. They also allow to infer the total amount of matter (both luminous and dark) present in such systems and thus to test different galaxy formation scenarios and cosmological models. However we have seen in $\S$ \ref{thermal} that we need to assume the distribution that characterizes the gas density. In this section we will describe some of these models and show how they can be used to define the physical status of the hot halo and the underlying total matter distribution. 

\subsection{Beta isothermal model}
\label{beta}
In general, elastic collision times for ions and electrons in the ISM are much shorter than the time scales for any heating, cooling or dynamical process that occurs in the galactic and cluster halos, so that the gas can be treated as a fluid \citep{Sar88}. Moreover, since the sound crossing time (i.e. the time for a sound wave in the ISM to cross the galaxy) is $t_s\sim 10^8$ yr and thus much smaller than the age of the galaxy of $\sim 10^{10}$ yr, the gas will be hydrostatic and the pressure will be a smoothly varying function of the position, unless very fast heating or cooling mechanisms are present. In this case:
\beq
\label{pressure}
\nabla P=-\rho_g\nabla\phi(r) 
\eeq
where $P=\rho_g k T/\mu m_p$ is the gas pressure\footnote{Due to its high temperature and low density, the physical status of the ISM is very similar to that of an ideal gas.}, $\rho_g$ is the gas mass density and $\phi(r)$ is the gravitational potential of the cluster.
If the system is assumed to be spherically symmetric, equation (\ref{pressure}) reduces to:
\beq
\label{pressure2}
{1\over\rho_g}{\mbox{d}P\over\mbox{d}r}=-{\mbox{d}\phi\over\mbox{d}r}=-{GM(r)\over r^2}
\eeq
where $r$ is the radius from the galaxy center and $M(r)$ is the total cluster mass within $r$.

The simplest distributions of gas temperatures is the isothermal one, i.e. $T_g(r)=\mbox{constant}$. This distribution has been often used since temperature profiles based on low spectral resolution instruments seemed to be almost isothermal \citep[e.g.][]{Kim95,rang95,jones97}. Although recent studies support the presence of multi-phase medium in the galactic halos \citep[e.g.][]{Buo99,BCF99,Buo00}, this remains a simple and useful model to derive the halo physical properties. If an isothermal distribution is assumed equation (\ref{pressure2}) becomes:
\beq
{\mbox{d}\ln\rho_g\over\mbox{d}r}={\mu m_p\over kT_g}{\mbox{d}\phi(r)\over\mbox{d}r}
\eeq
In most models the galaxy potential is assumed to be that of a self-gravitating isothermal sphere \citep{Chandra42,King66}. Since this can not be represented exactly in terms of simple analytic functions, the \cite{King62} approximation is often used:
\beq
\label{beta_star}
\rho(r)=\rho_0\left [1+\left ({r\over r_0}\right )^2\right ]^{-3/2}
\eeq
\beq
\phi(r)=-4\pi\rho_0r_0^2\frac{\ln[(r/r_0)+(1+(r/r_0)^2)^{1/2}]}{r/r_0}
\eeq
where $\rho$ is the total mass density, $\rho_0$ is the central total density and $r_0$ is the core radius of the distribution. These parameters are related to the line-of-sight velocity dispersion of test particles (e.g. the stars) by:
\beq
\sigma_r^2={4\pi G\rho_0r_0^2\over 9}
\eeq
(see \citealp{Sar88} for a detailed derivation of this equation). The resulting gas distribution is given by \citep{Cav76}:
\beq
\label{beta_model}
\rho_g(r)=\rho_{g,0}\left[1+\left({r\over r_0}\right)^2\right]^{-3\beta/2}
\eeq
Here
\beq
\label{betaspec}
\beta={\mu m_p\sigma^2_r\over kT_g}
\eeq
represents the ratio between the energy stored in stars to the one in the gas.
This model is often referred to as {\bf Beta model} in order to distinguish it from a true isothermal distribution \citep[see discussion in Appendix C of ][]{Tri86}.

The X-ray surface brightness at a projected radius $b$, integrated over the whole spectrum, is then (\citealp{Sar77}, \citealp{Gor78}):
\beq
\label{betaproj}
\Sigma(b,T_g)=\Lambda(T_g)\cdot\int n_en_p\mbox{d}l=\Sigma_0(T_g)\left[1+\left({b\over r_0}\right)^2\right]^{-3\beta+1/2}
\eeq
where $\int n_en_p\mbox{d}l$ is the {\it emission measure} of the gas along the line-of-sight $l$ through the galaxy, at projected radius $b$, and
\beq
\Sigma_0(T_g)=\Lambda(T_g)\sqrt{\pi}\left({n_e\over n_p}\right)\rho_{g,0}^2~r_0\frac{\Gamma(3\beta-1/2)}{\Gamma(3\beta)}
\eeq
is the central surface brightness.

To determine the free parameters $\rho_{g,0}$ and $r_0$, that define the gas distribution in the Beta model, the usual approach is to fit the observed X-ray surface brightness profile with expression (\ref{betaproj}). $\beta$, instead, can be derived both by a surface brightness fit or through equation (\ref{betaspec}) if we know the gas temperature and the stellar velocity dispersion. To distinguish the two methods in the following chapters, we will adopt the symbol $\beta$ when we refer to the brightness profile and $\beta_{spec}$ for the spectral determination.
  
The surface brightness fits generally give $\beta\sim 0.4-0.6$ \citep[e.g.][]{Tri86,Fab95,Kim95}, although departures from a simple power law are quite common (see $\S$ \ref{rad_prof}). 
This suggests that the gas is generally hotter than the stars. However
\cite{Matsu01} has recently shown, by means of spectral PSPC data, that this discrepancy is largely reduced if the Early-Type spectrum is modeled with two components: a thermal emission due to hot ISM plus a hard bremsstrahlung which models the contribution of discrete source.  
She obtains $\beta_{spec}\sim 1$ for faint galaxies, which means that the hot halo is probably heated by stellar mass losses. For brighter systems instead $\beta_{spec}\sim 0.5$, thus requiring additional heating. This is believed to be due to gravitational heating caused by the deep potential well which surrounds these galaxies. 

Other heating mechanisms may be: supernova heating \citep{Can87,Ber95}, thermal conduction from the hotter Intra Cluster Medium (ICM) or, when nuclear radio sources are present, heating from relativistic electrons \citep[ and references therein]{Fab89}.

\subsection{Empirical Gas Distributions}
\label{empirical}
The inconsistence of the isothermal Beta model for the brightest galaxies in clusters, discussed in the previous section, and the possible presence of multi-phase gas, has led to develop more empirical methods to derive the gas distribution.

Instead of assuming an a-priori model, the gas distribution can be derived directly from observations of the X-ray surface brightness. The X-ray surface brightness at a projected distance $b$ from the center of a spherical galaxy is:
\beq
\Sigma(b)=2\int^\infty_b{\epsilon(r)\mbox{d}r\over\sqrt{r^2-b^2}}
\eeq
where $\epsilon$ is the X-ray emissivity defined in equation (\ref{emiss3}). This {\it Abell} integral can be inverted to give the emissivity as a function of radius \citep[e.g.][]{Bin87}:
\beq
\label{deproject}
\epsilon(r)=-{1\over\pi}\int^\infty_r{\mbox{d}\Sigma(b)\over\mbox{d}b}{\mbox{d}b\over\sqrt{b^2-r^2}}.
\eeq
Because of the discretized nature of X-ray observations and the sensitivity of integral inversion to noise, the surface brightness data are often smoothed, either by fitting a smooth function to the observations or by applying this equation to the surface brightness averaged in rings about the galaxy center.

The emissivity $\epsilon$ is given by equation (\ref{emiss3}) and depends on the elemental abundances, the gas density and temperature. If the ISM abundances and temperatures are known (e.g. through spectral fits) equation (\ref{deproject}) allows to determine the gas density $n_e(r)$ in the hypothesis that the gas is characterized by a single density and temperature at each radius.  
An iterative procedure which exploits this method has been developed by \cite{kriss83} and will be applied to our data in $\S$ \ref{dens_par}, \ref{dens_par_1404} and \ref{ngc507_dens}.

For the sake of completeness we notice that more general politropic models have also been employed to model the hot gas distribution \citep[e.g.][]{kill88}. We will not discuss them here since they are not employed this work but more informations can be found in \cite{Sar88}.

\subsection{Total Mass Determination}
Once the density and temperature distributions of the gas are known, it is possible to derive the total underlying mass distribution, in the hypothesis that the gas is in hydrostatic equilibrium. We have seen in $\S$ \ref{beta} that this is a reasonable assumption unless the galaxy is undergoing merger events or strong ram pressure stripping from the surrounding medium, and the gas motions remains subsonic.

If the gas is spherically distributed the hydrostatic equilibrium condition is expressed by equation (\ref{pressure2}). Combining it with the ideal gas law we obtain: 
\beq
\label{mass}
M(<r)=-{rkT(r)\over\mu m_p G}\left({\mbox{d}\log\rho_g\over\mbox{d}\log r}+{\mbox{d}\log T\over{\rm d}\log r}\right)
\eeq
where $M(<r)$ is the mass contained within $r$, $T$ is the gas temperature, and ${\rm d}\log\rho_g/{\rm d}\log r$ and ${\rm d}\log T/{\rm d}\log r$ are respectively the logarithmic gradients of the gas density and temperature.

Equation (\ref{mass}) has been used by many authors to discover massive dark halos surrounding Early-Type galaxies with masses of $\sim 10^{13}$ M$_{\odot}$ \citep[e.g]{Fabr80,Kim95,jones97}. This method may be applied also when gas flows due to temperature gradients (see $\S$ \ref{cooling}) develop inside the halo, because dynamical timescales are long compared to the sound crossing time.
However the resulting masses must be considered upper limits since it is possible that other mechanisms, such as the pressure of the ICM, may contribute to lower the mass required to confine the hot gas \citep{Fabr80,Ved88}.

  \section{COOLING FLOWS}
\label{cooling}
If the gaseous halos observed in Early-Type galaxies are in equilibrium with the gravitational potential of the galaxy, the energy released in the X-ray band is compensated by the energy acquired by gravitational infall, keeping the temperature constant and the gas in quasi-hydrostatic conditions. However, we have seen in $\S$ \ref{thermal} that the thermal emissivity is proportional to the square of the density (equation \ref{emiss3}). A reasonable approximation of the gas emissivity is \citep{Sar88}:
\beq
\label{emiss4}
\epsilon\propto T^{a}n_e^2~~~~~\Big\{^{\textstyle a=-1/2~~~~T<3\times 10^7~{\rm K}}_{\textstyle a=1/2~~~~~~T>3\times 10^7~{\rm K}}
\eeq 
where $T$ is the gas temperature and $n_e$ is the electron density. This implies that in the center of the halo the gas cools more rapidly than in the outskirts, and its pressure drops. This process is not compensated by the drop in temperature, since below $3\times 10^7$ K the emissivity is inversely proportional to the temperature and, however, the dependence on $T$ is weaker than the one on density.
The weight of the overlaying gas then causes a slow, subsonic inflow which is usually called ``cooling flow''. 

The presence of cooling flows at the center of gaseous halos has been proposed since the {\it cooling times} $\tau_c$ in the central regions of most galaxies are smaller than the Hubble time (i.e. the epoch of galaxy formation). For an isobaric transformation \citep{Fab81}:
\beq
\label{CT}
\tau_c={5\over 2}{nkT\over n_e n_p\Lambda}
\eeq 
where $n=m_p+n_e$ is the total number density.
Typical cooling times in the center of Early-Type galaxies are $\leq 10^8$ yrs \citep[see for instance][]{Can87}.
  
Even if no direct proof of the presence of cooling flows exist, such as the Doppler shift of X-ray emission lines (the spectral resolution of modern detectors are still too low to measure these slow motions), X-ray observations have allowed to discover both imaging and spectroscopic indirect evidences: a) X-ray images of bright Early-Type galaxies show a strong central emission peak, which is attributed to the emission of the dense cooling component in the galaxy core; b) the gas temperatures inferred from spectral analysis revealed a central drop\footnote{The isothermal models discussed in $\S$ \ref{beta} are still a reasonable approximation of the gas distribution since the temperature drop affects only the central regions of the halo.}, explained with the presence of cooler gas \citep[ and references therein]{Fabian94}. 

If a steady-state spherical flow develops in the halo, we can derive the amount of mass deposited in the galaxy center from the flow equations \citep{FabNul77,Bin81,Can83}:
\beq
\label{Mdot}
\dot{M}=4\pi r^2\rho_g^2\Lambda(T)\left\{ {\mbox{d}\over\mbox{d}r}\left[{5\over 2}{kT\over \mu}+\phi(r)\right ]\right\}^{-1}
\eeq
and compare it with the observed central X-ray peak. This kind of comparisons have shown that the central brightness excess is much smaller than the one expected if all the cooling mass reached the galaxy center.
Moreover observations at other wavelengths have shown no evidence of the presence of the accreted material.
This suggests that the cool material does not remain in the gaseous phase but cools enough to form stars. However this would result in a bluer color of the  Early-Type core, which is not seen. 
To avoid this problem some authors have proposed that the star formation process creates low-mass objects that are more difficult to detect.

Alternative scenarios consider the possibility that the steady-state cooling flow model is wrong, and the gas does not reach the galaxy center but is deposited throughout the galaxy at all radii. This hypothesis is supported by the fact that the mass deposition profiles derived by many X-ray studies are not constant (as would be expected in a steady-state flow) but are proportional to the galactocentric radius: $\dot{M}\propto r$ \citep[e.g][]{Ber95,Kim95,rang95,Matsu01b}.  

If cool material is deposited throughout the halo we expect to see multiphase gas. In fact recent observations suggest that multi-temperature models fit better than single-temperatures ones the observed X-ray spectra of ASCA and ROSAT \citep{Buo99,BCF99,Buo00,Maki01}. However the cooling-flow models are now in serious troubles since recent studies with XMM found no evidence of multi-phase gas and showed that, in several cases, a lower temperature cutoff must be introduced in the cooling flow models to explain the observed spectra \citep{Pet01,Kaa01,Tam01}. The latter point requires the presence of additional heating sources, such as a high supernova rate or an active nucleus, to prevent the gas from cooling.

Finally there are evidences that the hot halo structure is more related to the shape of the gravitational well, than to the presence of a central cooling flow \citep{Maki01,Matsu01,Matsu01b}. In faint Early-Type galaxies the halo temperature seems to be compatible with stellar mass-loss heating, while in brighter galaxies there is an additional extended component that may be produced by a large amount of dark matter in the galaxy outskirts. 

  \section{ENVIRONMENTAL EFFECTS}
\label{environment}
The dependence of galaxy properties on the environmental conditions is a well established fact. Optical studies have shown that galaxy morphology is strongly dependent on the density of galaxies in the sense that while Early-Type galaxies represent the dominant morphological class in groups and clusters of galaxies, Late-Types are more common in low density environments \citep[e.g.][]{Bahcall96}. 
The galaxy luminosity function has a steeper slope in denser environments, thus indicating that different evolutionary processes are at work \cite[e.g.][]{Trent98,Chris00}.
Moreover giant and cD ellipticals are only found in the center of rich systems. Even if it is not clear which are the main parameters (e.g. local density, total density of the cluster, distance from cluster center) that play a fundamental role, it is generally accepted that galaxy encounters, merging and interaction with the ICM are all processes that must affect the properties of galaxies to some extent.

It is thus not surprising that environmental effects may be suspected to contribute in determining also the X-ray properties of galaxies. Although there are several suggested mechanisms by which the environment may affect the X-ray halo of a galaxy, the actual role that these effects play is unclear. 
Ram-pressure stripping is likely to remove gas from galaxies passing through a dense intra-cluster medium \citep{Gunn72}, and turbulent viscous stripping may be as effective in the group environment \citep{Nul82}. It is also likely that the ICM provide reservoirs of gas which can be captured by slow moving or stationary galaxies.

However the observational evidence is conflicting and often difficult to interpret.
For instance \cite{White91} found that, for a sample of Early-Type galaxies observed with {\it Einstein}, galaxies with $\log L_x/L_B<30 $ erg s$^{-1}$ L$^{-1}_B$ had $\sim 50\%$ more neighbors than X-ray bright galaxies. They attribute this to ram pressure stripping, which would be expected to reduce $L_X$ more in higher density environments. An opposite view was presented by \cite{Brown00}, who found that $L_X/L_B$ increased with environmental density. Their explanation is that for the majority of galaxies ram-pressure stripping is a less-important effect than the stifling of galactic winds by a surrounding ICM. In this model the ICM encloses the galaxy, increasing the gas density of its halo and therefore its X-ray luminosity.

\cite{Brown98} claimed an environmental dependence based on a correlation between $L_X/L_B$ and Tully density parameter $\rho$ \citep{Tully88} for their sample. However, \cite{Hels01} show that group dominant galaxies often have X-ray luminosities which are more related to the properties of the group than of the galaxy. Their high luminosities are more likely to be caused by a group cooling flow than by a large galaxy halo. Once these objections are removed from consideration, the correlation between $L_X/L_B$ and $\rho$ becomes very weak.

In a large ROSAT PSPC sample, \cite{O'Sull01} found no evidence of a correlation between $L_X/L_B$ and the environmental density, while the similarity of the $L_X/L_B$ relation in field, groups and clusters suggest that the environment does not change the properties of the whole population. They notice that their result does not mean that there are no effects on the X-ray properties of galaxies but that none of the processes affecting the X-ray halos is dominant or that they counter-balance each other.  

Numerical simulations generally suggest strong environmental influence on the X-ray halo properties. \cite{toni01} showed that interaction with the ICM may produce large fluctuations of the halo luminosity, depending on the galaxy orbit through the cluster. If the galaxy moves at supersonic speeds ram pressure stripping is very effective in removing large quantities of gas from the halo, while slow subsonic motions allow the galaxy to accumulate the gas produced by stellar mass losses. Such variation in halo gas content during the galaxy orbit is large enough to explain the observed scatter in X-ray luminosities.
\cite{D'Ercole00} studied tidal interactions in elliptical galaxies, showing that tidal stripping may produce strong fluctuations in the halo shape and luminosity, as well as removal of hot gas from the galaxy outskirts. \cite{bar00} further showed that such encounters may produce deformations of the X-ray halo and profile.

Finally the location of the larger X-ray halos, associated with giant and cD ellipticals, at the center of groups and clusters of galaxies indicates that the structure of the gravitational potential well must play an important role in retaining the hot gas. As already discussed in the previous section, this scenario is supported by recent works by \cite{Maki01} and \cite{Matsu01} which stress the fact that the properties of most massive system are mainly due to the presence of large potential structures.

  \chapter{NGC 1399}
\label{NGC1399}
\section{INTRODUCTION}
NGC 1399 is the central dominant galaxy of the Fornax cluster. Due to its proximity (19 Mpc for $H_0=75$ km s$^{-1}$ Mpc$^{-1}$), this very regular, almost spherical (E0; \citealt{Ferg89}) cD galaxy has been extensively studied in a wide range of wavelengths, ranging form radio to X-rays. The optical radial profile, 
first studied by \cite{schom86} and later in more detail by \cite{kill88}, reveals a large halo extending up to 250 kpc. The galaxy is surrounded by a large number of globular clusters, 10 times in excess with respect to those of its nearer companion NGC 1404 and of the Fornax galaxy NGC 1380 \citep{Kiss99}. Dynamical studies of the stellar population of NGC 1399 indicate that the central regions of the galaxy are characterized by a high velocity dispersion, decreasing with radius, and slow rotation, as expected in giant ellipticals \citep{Gra98}. At larger radii the velocity dispersion starts increasing again, as shown by studies of the planetary nebulae \citep{Arn94} and globular clusters dynamical properties \citep{Grill94}. This suggests a different dynamical structure, than that of the inner stellar body, for the galaxy envelope and the globular cluster population. 

NGC 1399 hosts a weak nuclear radio source \citep{kbe88} with radio luminosity of $\sim 10^{39}$ ergs s$^{-1}$ between $10^7$ and $10^{10}$ Hz. The radio source has two  jets ending in diffuse lobes, confined in projection within the optical galaxy. \cite{kbe88} suggest that the radio source is intrinsically small and confined by the thermal pressure of the hot ISM. 

X-ray data have shown the presence of an extended hot gaseous halo surrounding NGC 1399. Observations made with the {\it Einstein} IPC by \cite{Kim92} constrained the gas temperature to be $kT>1.1$ keV. The {\it Einstein} IPC data were used also by \cite{kill88} to derive the mass distribution of the galaxy. They found that both models with and without dark matter were compatible with those data, depending on the assumed temperature profile.
\cite{White92} suggested the presence of a cooling flow in the center of the galaxy, depositing 0.8$\pm 0.6$ M$_{\odot}$ y$^{-1}$.
{\it Ginga} observations \citep{ikebe92} led to the detection of extended emission out to a galactocentric radius of $\sim$250 kpc. \cite{ser93} constrained the X-ray temperature to be $1.0<kT<1.5$ keV with the Broad Band X-ray Telescope (BBXT). 
\citet[ hereafter RFFJ]{rang95}, using the ROSAT PSPC, have studied in detail the temperature profile of the hot inter-stellar medium (ISM) out to 220 kpc finding an isothermal profile 
($0.9<kT<1.1$) from 7 to 220 kpc, and a central cooling flow ($0.6<kT<0.85$) of at least 2 M$_{\odot}$ y$^{-1}$.

\cite{ikebe96} were able to identify with ASCA the presence of different components in the X-ray halo, associated respectively with the galaxy and the Fornax cluster potential.
\cite{jones97} used the better resolution of the ROSAT PSPC for studying in detail the cluster X-ray structure and finding a total binding mass of (4.3-8.1)$\times 10^{12}$ M$_{\odot}$ within 100 kpc, with a mass-to-light ratio increasing from 33 M$_{\odot}$/L$_{\odot}$ at 14 kpc to 70 M$_{\odot}$/L$_{\odot}$ at 85 kpc. They also found a metal abundance of 0.6 solar.
\cite{Buo99} analyzed ASCA data finding that either a two temperature spectral model or a cooling-flow model with  solar Fe abundances, are required to fit the thermal X-ray emission, and additional absorption is needed in the galaxy center. \cite{Mat00} found similar near solar metallicities but argued that the metallicity inferred by ASCA spectral fits is dependent on the assumed atomic physics model and is not solved by multi-components spectral models. 

More recently \cite{Sulk01} used ROSAT data to pose an upper limit on the nuclear source brightness. The {\it Chandra} data were used by \cite{Loew01} to further constrain the nuclear source and by \cite{Ang01} to study the point source population hosted by globular clusters.

In this chapter we present the results of a deep observation of the NGC 1399/NGC 1404 field obtained from data collected between 1993 and 1996 with the ROSAT High Resolution Imager (HRI, for a description see \citealp{Dav96}). We take advantage of the $\sim 5"$ resolution of the HRI to study in detail the structure of the galactic halo and relate the results to those obtained at larger scales with poorer resolution instruments. We study the interactions between the nuclear radio source and the galactic halo and discuss the properties of the discrete sources population. A preliminary analysis of {\it Chandra} data supports the ROSAT results.\\
~\\
We adopt $H_0=75$ km s$^{-1}$ Mpc$^{-1}$ and a distance of 19 Mpc (1'=5.5 kpc).

\section{OBSERVATIONS AND DATA ANALYSIS}
The NGC 1399 field, including NGC 1404, was observed at three separate times
with the ROSAT HRI: in February 1993, 
between January and February 1996 and between July and August of the same year.
The total exposure time is 167.6 ks (Table \ref{observations}).
The data were processed with the  SASS7\_8 and SASS7\_9 versions of the ROSAT standard analysis software (SASS).
For our data analysis, we used the IRAF/XRAY and CIAO packages developed
at the Smithsonian Astrophysical Observatory and at the {\it Chandra} X-ray Center (CXC), and other specific software as mentioned in the text.
\begin{table*}[h]
\begin{center}
\footnotesize
\caption{ROSAT HRI observations of NGC 1399.\label{observations}}
\begin{tabular}{ccccrcc}
\\
\tableline
\tableline									
name & \multicolumn{2}{c}{Field center} & sequence id. &
 Exp.time & obs. date & P.I.\tablenotemark{(a)}\\
& R.A. & Dec & & (sec) & &\\
\tableline
NGC 1399  & 03$^{\rm h}$38$^{\rm m}$31$^{\rm s}$ & --35$^\circ$27'00'' & RH600256n00 & 7265 & 1993 Feb 17 & D.-W. Kim\\
''  & '' & '' & RH600831n00 & 72720 & 1996 Jan 04-1996 Feb 23 & G. Fabbiano\\
''  & '' & '' & RH600831a01 & 87582 & 1996 Jul 07-1996 Aug 26 & G. Fabbiano\\
\tableline
\tablenotetext{(a)}{~Principal Investigator}
\end{tabular}
\end{center}
\end{table*}

\subsection{Aspect Correction}
\label{Aspect corrections}
\begin{figure}[t]
\centerline{\psfig{figure=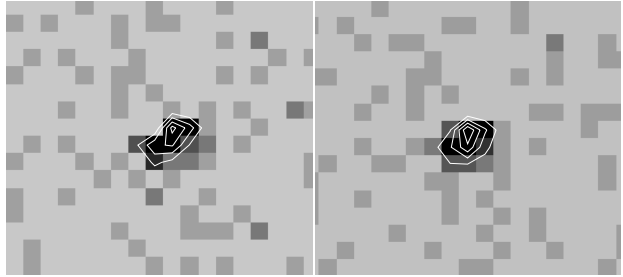,angle=180,width=0.5\textwidth}}
\caption{Comparison between the aspect of the brightest pointlike source in the NGC 1399/1404 field (No.8 in Table \ref{source_tab}) before (left) and after (right) correction. The pixel size is 4". Contour levels are spaced by $5.9\times 10^{-2}$ cnts arcmin$^{-2}$ s$^{-1}$ with the lower one at $6\times 10^{-2}$ cnts arcmin$^{-2}$ s$^{-1}$.}
\label{asp_corr}
\end{figure}

The HRI data processed with the SASS versions prior to SASS7\_B of March 1999 (as it happens for our data) suffer from an error in the aspect time behavior  \citep{Har99}. 
This translates into an error on the position of the incoming photons and thus it results in  a
degradation of the Point Response Function (PRF). Because high resolution analysis is the primary objective of this work we run the correction routine ASPTIME (F. Primini 2000, private communication) on the data, to improve the aspect solution.
Visual inspection of the brightest pointlike sources in our field (Figure \ref{asp_corr}) demonstrates the improvement in the image quality.

Additional problems that may preclude the attainment
of the HRI potential resolution are the imperfect correction of the spacecraft wobble and the wrong tracking of
reference stars due to the variable pixel sensitivity across the detector \citep{Harris98}.
We followed the procedure suggested by Harris and collaborators, of dividing the observations in time bins (OBI) and realigning these segments by using the centroids of a reference pointlike source in each OBI.

Our data can be divided in 4 OBIs in RH600256n00, 43 in RH600831n00 and 33 in RH600831a01.
We checked each OBI individually using the brightest point source in the field 
(No.8 in Table \ref{source_tab}). We found that in the first observation (RH600256n00) the OBI are well aligned and no correction is needed.
The second observation (RH600831n00) has 7 OBIs out of 43 that are 7"-10" displaced; however, the overall PRF is 6" FWHM so that just a slight improvement can be obtained dewobbling the image. The third observation (RH600831a01) has a 8" PRF but the signal to noise ratio of the individual OBIs is worse than in the previous observations so that the dewobbling procedure cannot be applied.
Given these results, we decided not to apply the dewobbling correction, and to
take as good a resulting PRF, in the composite image, of $\simeq$7" FWHM.

\subsection{Composite Observation and Exposure Corrections}
\label{Composite observation}
\begin{figure*}[ph]
\psfig{figure=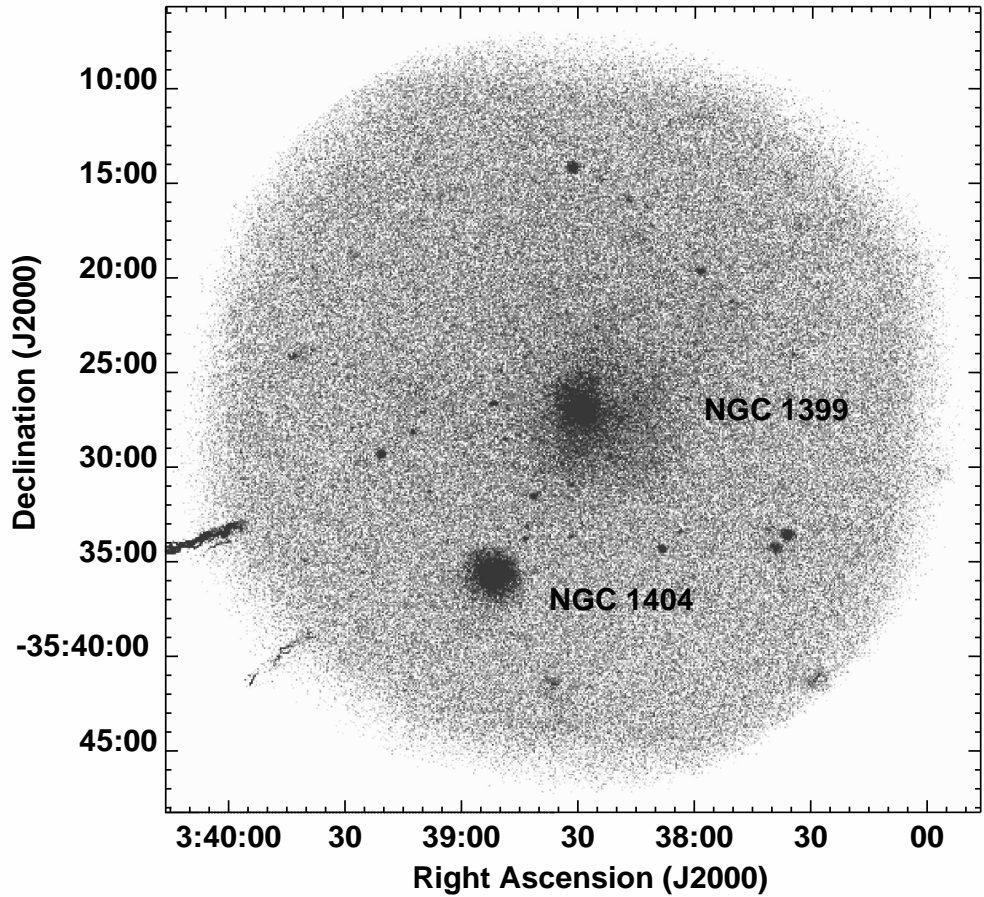,angle=0,width=1.0\textwidth}
\caption{The `raw' NGC 1399/NGC 1404 HRI field. The composite image is here displayed after it was rebinned to 5 arcsec/pixel.
Even without exposure correction the extended emission surrounding NGC 1399 is clearly visible.
The elongated features in the lower left
corner are due to the presence of ``bad pixels'' on the detector. North is up and East is left.}
\label{NGC1399comp}
\end{figure*}

The aspect-corrected observations were co-added to obtain a ``composite
observation''. To make sure that pointing uncertainties did not degrade the image we used the centroids of three bright pointlike sources in the field
(No.8, 19 an 24 in Table \ref{source_tab}) to align the images. 
The applied corrections were all within 2 arcsec. 
The resulting composite image is shown in Figure \ref{NGC1399comp}.
This image was then corrected for vignetting and variations of
exposure time and quantum efficiency across the detector  by
producing an ``exposure map'' for each observation with the software
developed by \citet[ hereafter SMB]{Snow94}. 
\vspace{1cm}

\subsection{Surface Brightness Distribution and X-ray/Optical Comparison}
\label{brightness}
To study the large-scale brightness distribution of the NGC 1399 field we rebinned the exposure corrected data in 5''$\times$5'' pixels. We then adaptively smoothed the image with the CXC CIAO {\em csmooth} algorithm which convolves the data with a gaussian of variable width (depending on the local signal to noise range of the image) so to enhance both small and large scale structures.
\begin{figure*}[p]
\centerline{\hspace{0.5cm}\framebox(200,200)[br]{}\hspace{-1.5cm}\raisebox{1cm}{\parbox[b]{1cm}{\bf(a)}}\hspace{1.cm}}
\vspace{1.5cm}
\centerline{\psfig{figure=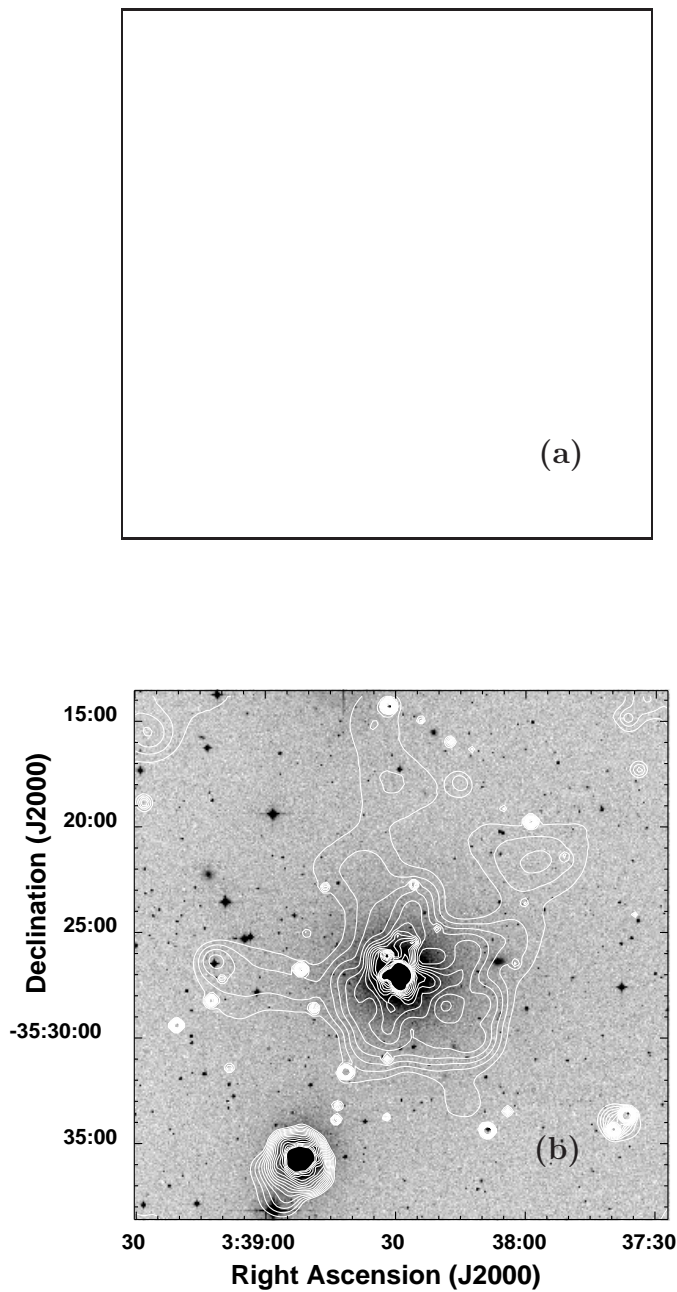,angle=0,width=0.60\textwidth}\hspace{-2.5cm}\raisebox{2cm}{\parbox[b]{1cm}{\bf(b)}}\hspace{2.5cm}}
\caption{{\bf (a)} $5\times 5$ arcsec/pixel adaptively smoothed image of the central part of the NGC 1399/1404 HRI field. Colors from black to yellow represent logarithmic X-ray intensities from $5.6\times 10^{-3}$ to $3.58\times 10^{-1}$ cnts arcmin$^{-2}$ s$^{-1}$. {\bf (b)} X-ray brightness contours overlaid on the 1 arcsec/pixel DSS image (logarithmic grayscale). Contours are spaced by a factor 1.1 with the lowest one at $6.1\times 10^{-3}$ cnts arcmin$^{-2}$ s$^{-1}$. The X-ray emission peak is centered on the optical galaxy for both NGC 1399 and NGC 1404. In the case of NGC 1404 the X-ray isophotes are consistent with the optical distribution while in NGC 1399 the X-ray emission extends further out than the optical one.}
\label{csmooth}
\end{figure*}

The resulting image (Figure \ref{csmooth}a) shows a complex X-ray morphology.
The center of the image is occupied by the extended halo of NGC 1399. 
The galaxy possess a central emission peak and an external extended and asymmetric halo. This halo is not azimuthally symmetric with respect to the X-ray peak, but it extends more on the SW side. The X-ray surface brightness distribution of the halo appears filamentary, with elongated structures and  voids.
As it can be seen from a comparison with the optical Digitized Sky Survey (DSS) image (Figure \ref{csmooth}b), while the X-ray emission peak is centered on the optical galaxy, the X-ray halo extends radially much further than the optical distribution. Moreover most of the features seen in the X-ray image have no direct optical counterpart. The SW clump centered at RA, Dec=$3^{\rm h}38^{\rm m}12^{\rm s}$,--35$^\circ$30'00'' and the one at $3^{\rm h}39^{\rm m}12^{\rm s}$,--35$^\circ$26'12'' were also detected by the wavelets algorithm (sources No.42 and 43 in Table \ref{source_tab} and Figure \ref{det}) presented in $\S$ \ref{sources}.

A number of pointlike source in the field is visible in Figure \ref{csmooth}a. The analysis of these sources is discussed in $\S$ \ref{sources}. The X-ray emission of NGC 1404, instead, will be discussed in detail in Chapter \ref{NGC 1404}.

\begin{figure}[t]
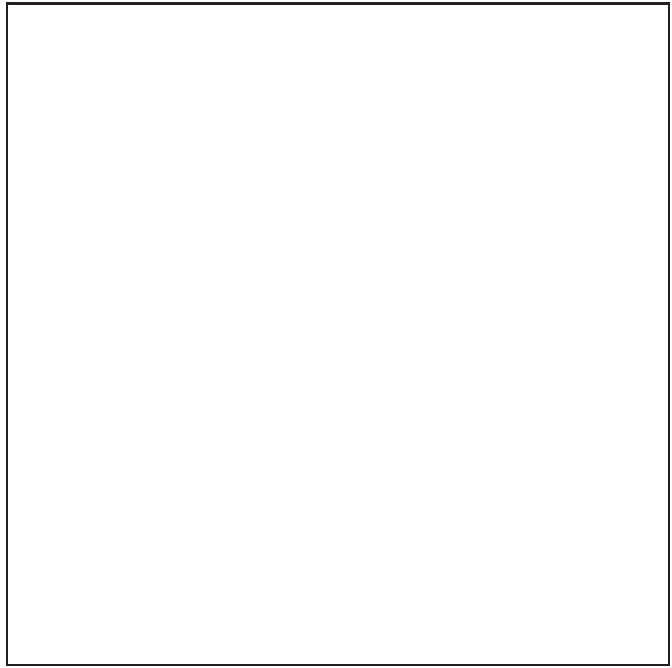

\centerline{\framebox(250,250)[br]{}}
\vspace{0.5cm}
\caption{Adaptively smoothed image of the inner NGC 1399 halo. Colors from black to yellow represent logarithmic X-ray intensities from $6.4\times 10^{-3}$ to $7.7\times 10^{-1}$ cnts arcmin$^{-2}$ s$^{-1}$. }
\label{centbox}
\end{figure}

We examined the inner halo of NGC 1399 in greater detail using a 1"/pixel resolution. The adaptively smoothed image (Figure \ref{centbox}) reveals an elongation of the inner halo structure in the N-S direction plus a large arc protruding to the West side of the halo. Several voids are present in the X-ray distribution, the largest being 1.5 arcmin NW of the emission peak. The central peak appears circular with at most a slight N-S elongation.

These features are all above $3\sigma$ significance, since $3\sigma$ was the minimum significance level requested for an intensity fluctuations to be smoothed on a given scale.  We can rule out that features on scales of 1 arcmin are of statistical nature, because their significance can be as high as $6\sigma$ (see $\S$ \ref{model}). These brightness fluctuations must be due to the local physical conditions of the hot gaseous halo.

\subsection{Radial Brightness Profiles}
\label{rad_prof}
As a first step in the quantitative study of the X-ray emission, we created a radial profile from the HRI data, assuming circular symmetry. Count rates from the exposure-corrected composite image, were extracted in circular annuli centered on the X-ray centroid RA, Dec=$3^{\rm h}38^{\rm m}28^{\rm s}.9$, --35$^\circ$27''02'.1. We took care to remove the contribution of all the detected point-like sources by excluding circles within the $3\sigma$ radius measured by the wavelets algorithm ($\S$ \ref{sources}), from the source centroid\footnote{The counts in each annulus are renormalized to the net observed area by the extraction procedure.}. We also excluded a 150" circle centered on NGC 1404 (see $\S$ \ref{1404_profile}).

\begin{figure}[t]
\centerline{\psfig{figure=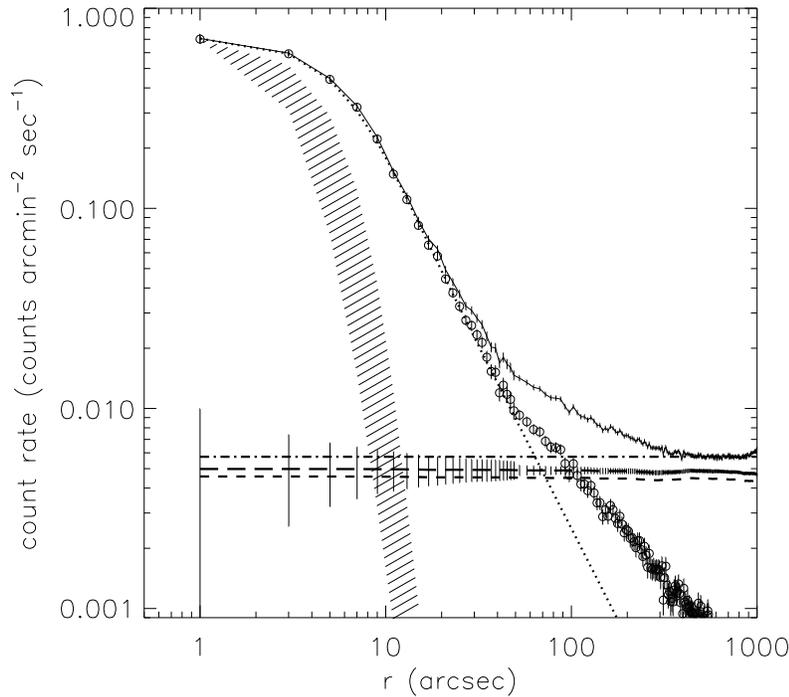,angle=0,width=0.65\textwidth}}
\caption{HRI radial profile of NGC 1399 X-ray surface brightness (continuous line) after excluding all point sources ($\S$ \ref{sources}) and NGC 1404. The dot-dashed line represents the flattening level of the HRI profile measured in the 500''-850'' annulus.
Short and long dashed lines represent respectively the SMB background level before and after rescaling to match PSPC counts.
The rescaled background was used to derive the background-subtracted counts (open circles) and the best-fit Beta model within 50" shown in Figure \ref{fit} (dotted line). 
The HRI PRF range is represented by the shaded region. Radial bins are 2" wide up to r=50", and 5" wide at larger radii.}
\label{profile}
\end{figure}

The X-ray brightness profile is shown as a continuous line in Figure \ref{profile}. The emission extends out to $\sim$ 500'' (46 kpc at the assumed distance of 19 Mpc). The radial profile flattens out past 500" (dash-dotted line in Figure \ref{profile}) suggesting that we have reached the field background level and that any residual halo emission is below our sensitivity limit. Nevertheless we know from previous investigations of the Fornax cluster with {\it Ginga} \citep{ikebe92}, {\it Einstein} \citep{kill88} and ROSAT \citep{jones97}, that extended X-ray emission, bright enough to be detected in our data, is present at galactocentric radii $>500''$. We thus expected to see a smooth gradient in the X-ray profile continuing past 500''.
We checked to see if a very high background was present in some OBIs
or either in some of the HRI spectral channels\footnote{The ROSAT HRI has 11 spectral channels. The spectral accuracy is too low to perform a reliable analysis but can be used to check for instrumental problems.} but we found none that 
could explain our lack of sensitivity. A possible explanation is that we didn't reach the background level around 500'' and the actual background is much lower but, due to uncertainties in the exposure correction near the edge of HRI field of view, we are not able to see the expected decline in the X-ray emission.

\begin{figure}[p]
\centerline{\psfig{figure=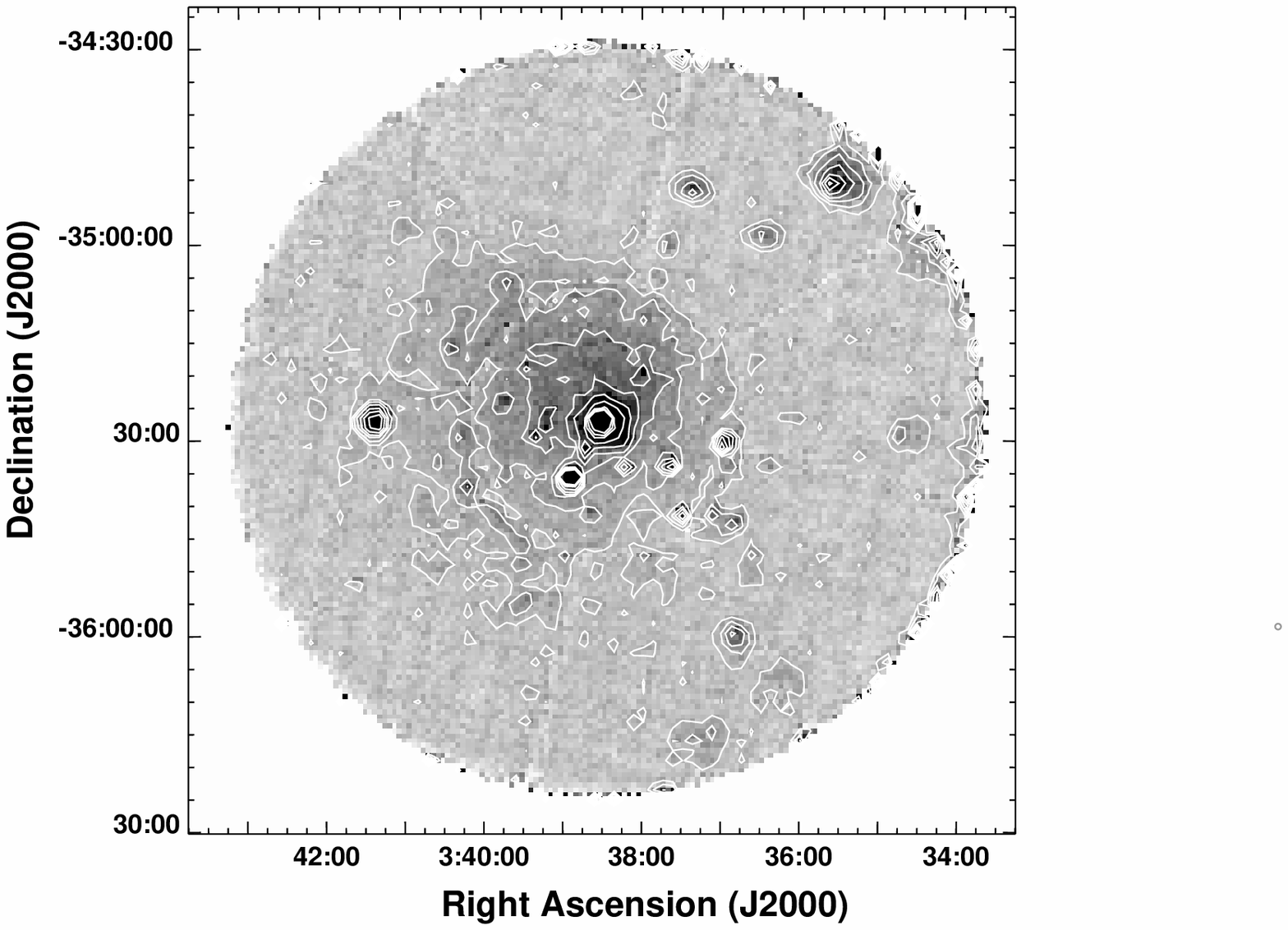,angle=0,width=0.57\textwidth}}
\caption{Exposure corrected PSPC image of the NGC 1399/1404 field. The two brightest central emission peaks represent the two galaxies (NGC 1399 and NGC 1404) visible also in the HRI field. X-ray contours, spaced by a factor of 1.3 with the lowest level at $3.5\times 10^{-3}$ cnts arcmin$^{-2}$ s$^{-1}$, clearly show the presence of an extended halo much larger than the HRI FOV (central $40\times 40$ arcmin). The black line delimits the two sectors used to extract the radial profiles shown in Figures \ref{pies}a, b. Radial ribs are due to the PSPC support structure.}
\label{PSPC}
\vspace{0.5cm}
\centerline{\psfig{figure=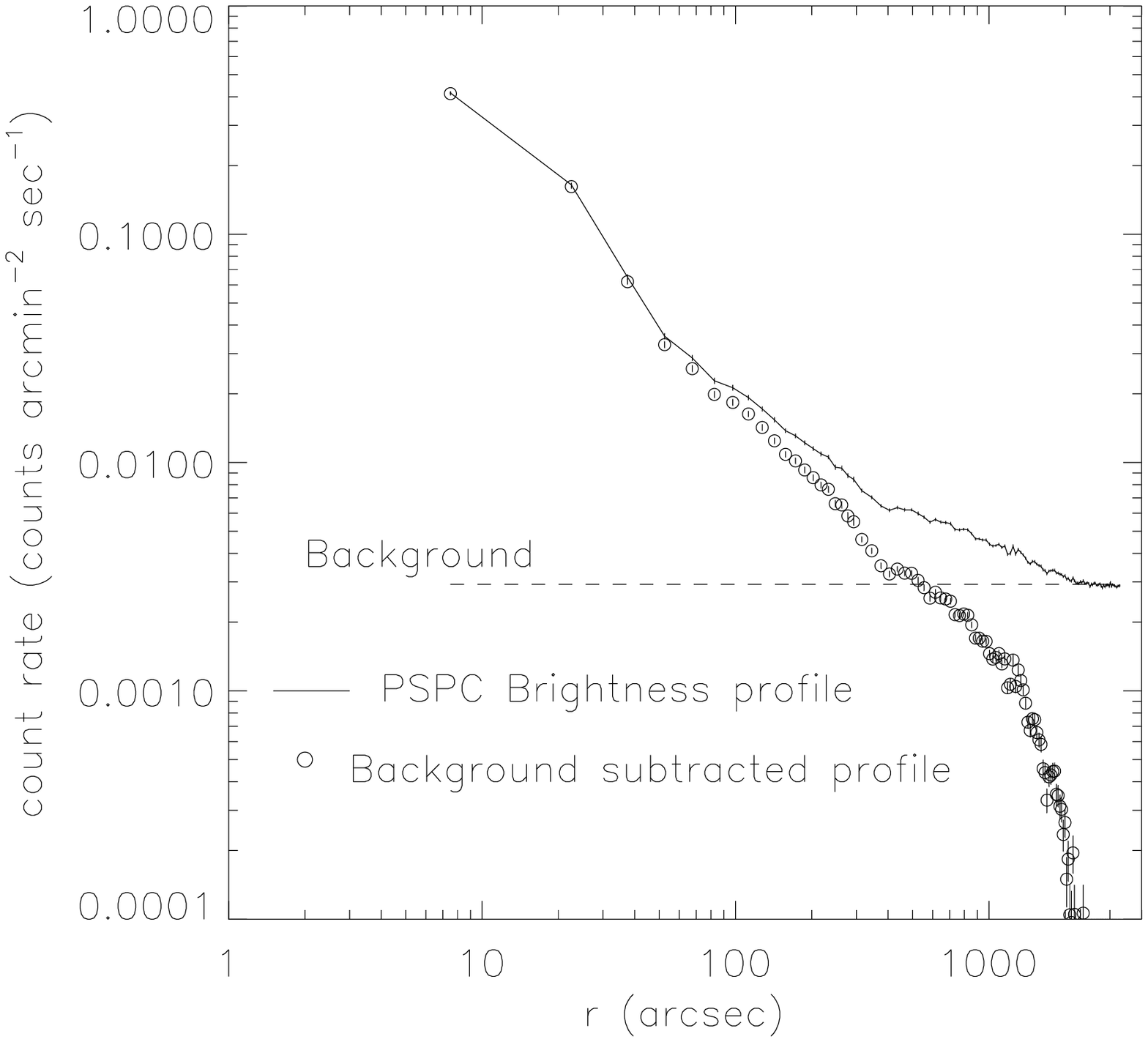,angle=0,width=0.53\textwidth}}
\caption{Exposure corrected PSPC radial profile of NGC 1399 (continuous line). 
The dashed line represents the background level measured in the 2400-3000 arcsec region, used to derive the background-subtracted counts (open circles).
The X-ray halo clearly extends at much larger radii than visible in the HRI FOV but the central region is not well resolved due to the larger PRF.}
\label{PSPC_prof}
\end{figure}

We used the SMB software to calculate a background map for this field. The radial profile derived from this background map is shown as a small dashed line in Figure \ref{profile}.
The SMB model however may underestimate the ``true'' HRI background somewhat because it models only the charged particle contribution. We therefore corrected this value using the 52 ks ROSAT PSPC observation RP600043N00 centered on NGC 1399, taken on the 1991 August 15 (Figure \ref{PSPC}). Even if the PSPC observation is much shorter than our HRI image, the higher sensitivity and lower background of this instrument allows a better study of the large scale X-ray emission.

The exposure corrected PSPC profile is shown in Figure \ref{PSPC_prof}. The background level (dashed line) was estimated from the 2400''-3000'' annulus. 
We compared our radial profile to the one derived by \cite{jones97} and RFFJ, who made use of the same data, finding consistent results. 
We then rescaled the SMB background to find the best agreement between HRI and PSPC counts in the 50-400 arcsec region, after rebinning the HRI data to match the wider PSPC PRF. The ``true'' HRI background level found in this way (long-dashed line in Figure \ref{profile}) lies in between the flattening level of the HRI profile and the SMB background and was used to derive the HRI background-subtracted profile shown as empty circles.
This way of finding the HRI background assumes that the PSPC brightness profile is not affected by systematic errors on the background determination and can thus be used to rescale the HRI counts. However, even if the PSPC background suffered from residual errors (e.g. due to uncertainties in the exposure correction at large radii), they would not significantly affect the central region profile -- used to rescale the SMB HRI background -- because there the X-ray brightness is one order of magnitude larger than the PSPC background level.

\begin{figure}[t]
\centerline{\psfig{figure=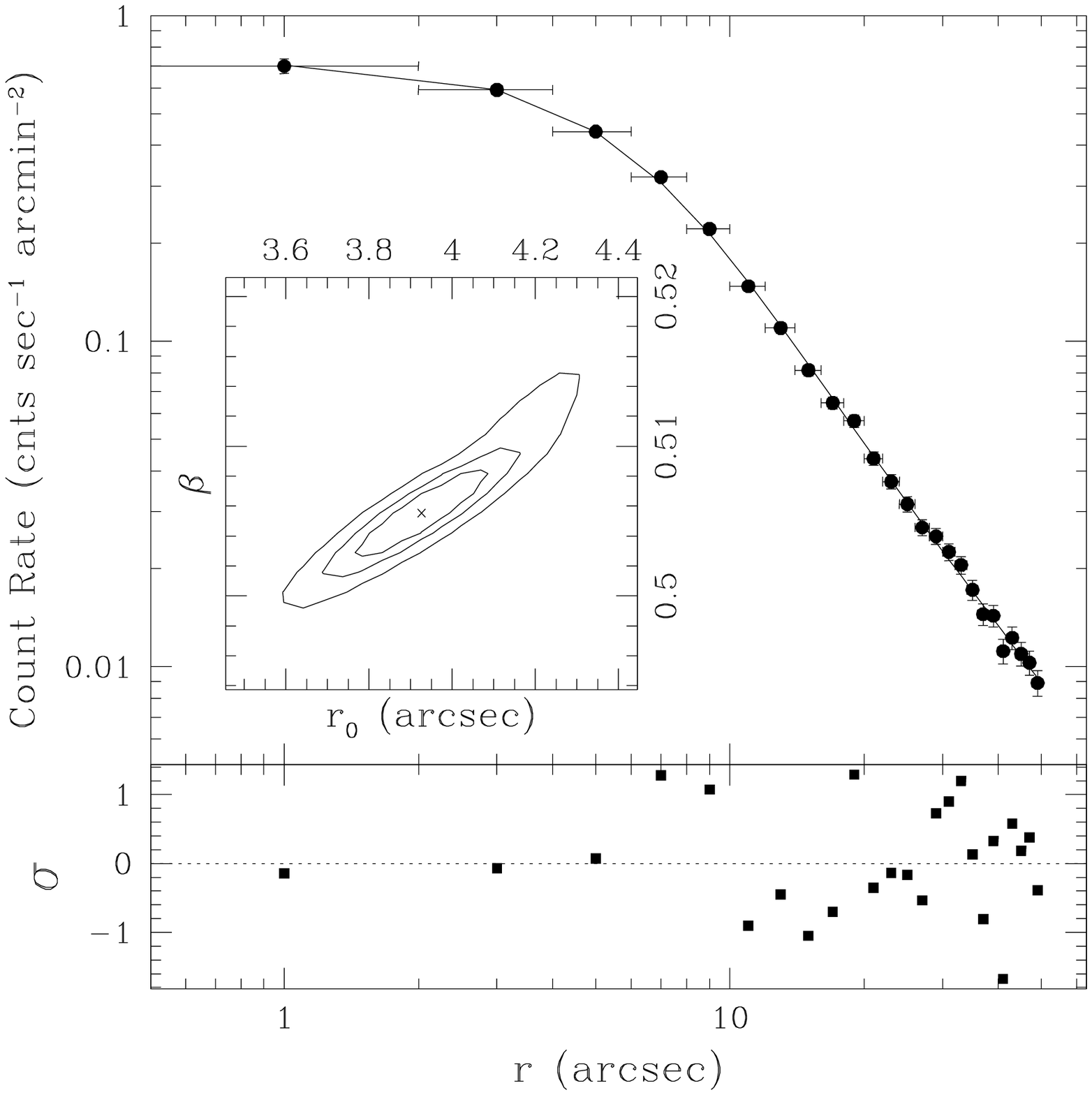,angle=0,width=0.45\textheight}}
\caption{The best-fit model and the residuals of NGC 1399 brightness profile within 50''. The 66\%, 90\% and 99\% contour levels relative to the core radius $r_0$ and slope $\beta$ are shown in the inner panel.}
\label{fit}
\end{figure}

The HRI radial surface brightness profile has a complex structure. In the `central' region, i.e. within 50'' from the emission peak, the X-ray emission is well represented by a simple Beta model (equation \ref{betaproj}) with best fit parameters $r_0=3.93\pm 0.16$ arcsec, $\beta=0.506\pm 0.003$ and $\chi^2=14.5$ for 22 degrees of freedom (Figure \ref{fit}). To determine the best fit parameters the model was convolved with the HRI on-axis PRF \citep{Dav96}. At radii larger than 1 arcmin the surface brightness profile shows a significant excess over this model (see Figure \ref{profile}), indicating the presence of an additional `galactic' component. Restricting the fit to r $<$ 40'' to check if this more extended component contributes around 50'', gives consistent results with $r_0=3.88^{+0.20}_{-0.16}$ arcsec and $\beta=0.504^{+0.004}_{-0.003}$ ($\chi^2=11$ for 17 degrees of freedom). Thus, for $3"<r<50"$ the X-ray emission falls as $r^{-2.04\pm 0.02}$. 
As can be seen by the comparison with the HRI PRF (shaded region in Figure \ref{profile}), whose uncertainty is due to residual errors in the aspect solution \citep{Dav96}, the central component is extended and cannot be due to the presence of a nuclear point source, that in fact is not detected in our data (see $\S$ \ref{Xradio}). 
\begin{figure*}[p]
\centerline{\psfig{figure=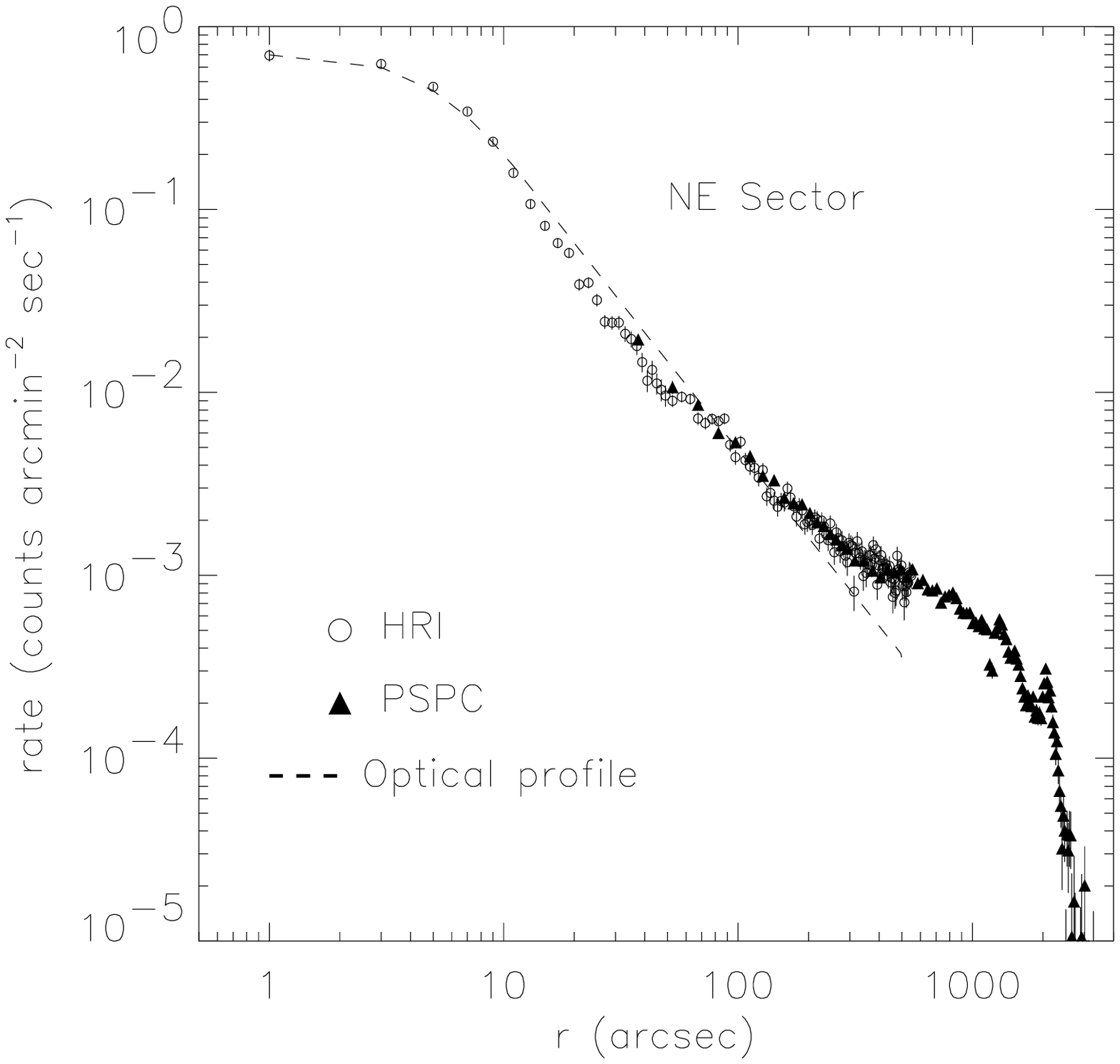,angle=0,width=0.55\textwidth}\hspace{-1.5cm}\raisebox{7cm}{\parbox[b]{1cm}{\bf(a)}}\hspace{1.5cm}}
\centerline{\psfig{figure=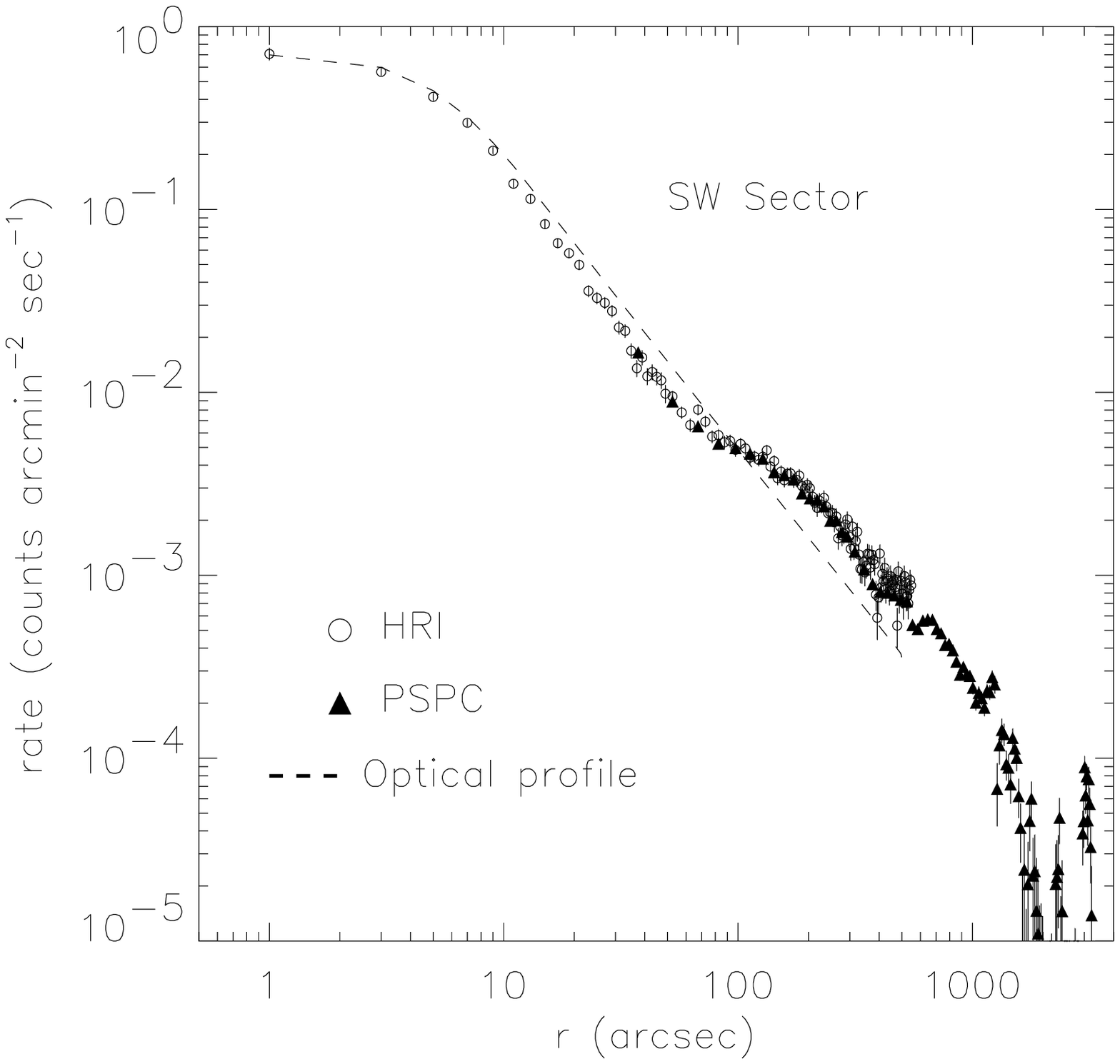,angle=0,width=0.55\textwidth}\hspace{-1.5cm}\raisebox{7cm}{\parbox[b]{1cm}{\bf(b)}}\hspace{1.5cm}}
\caption{Background subtracted radial profile of NGC 1399 in the NE (a) and the SW (b) sectors. Open circles represent HRI data (2" annuli for $r<50$", 5" for $r>50"$), while filled triangles represent PSPC data (15" annuli for $r<300$", 30" for $r>300"$). PSPC counts are reduced by a factor $\sim 0.3$ to match the HRI profile; the inner 30" are not shown because of the lower resolution. The dashed line represents the optical profile (convolved with the HRI PRF) from \cite{kill88}.}
\label{pies}
\end{figure*}

The PSPC profile in Figure \ref{PSPC_prof} shows that X-ray emission extends to much larger radii than visible in the HRI data and suggests the presence of a third component dominant at $r>400"$. Large azimuthal differences are seen in the X-ray profile at radii larger $\sim 1$ arcmin. We derived composite HRI-PSPC X-ray profiles in two sectors dividing the images along the line connecting the X-ray centroids of NGC 1399 and NGC 1404: a NE sector (Position Angle: $331.5^{\circ}<$P.A.$<151.5^{\circ}$) and a SW sector ($151.5^{\circ}<$P.A.$<331.5^{\circ}$). While the emission within 1 arcmin is azimuthally symmetric, the galactic component ($r>1'$) is much more extended in the SW direction and presents a steep decline in the NE sector, as suggested by Figures \ref{csmooth}a and \ref{centbox}. The even more extended component visible in the PSPC data follows the opposite behavior, being 3 times brighter around 1000'' on the NE side than on the SW one. While these asymmetries have been reported by \cite{jones97}, their PSPC data alone do not have enough resolution to clearly separate the central component from the external halo. Jones and collaborators used a single power-law model to fit the whole distribution, thus obtaining a value of $\sim$0.35 for the power law slope.  

\clearpage
\subsection{Bidimensional Halo Models}
\label{model}
Because of the complex halo structure it is not possible to obtain a satisfactory fit of the entire surface brightness distribution of NGC 1399 with a single circularly symmetric model. The {\em Sherpa} modeling and fitting application of the CIAO package allows us to adopt a more realistic approach:  we added together three models to represent the different components seen in the surface brightness profile (Figures \ref{profile} and \ref{PSPC_prof}). Moreover, to take into account the deviations from circular symmetry (Figures \ref{pies}a and \ref{pies}b), each component is represented by a bidimensional extension of the classical Beta model (equation \ref{betaproj}) of the form: 
\begin{equation}
\label{2dmodel}
\Sigma(x,y)=\frac{A}{(1+(r/r_0)^2)^{3\beta-0.5}}
\end{equation}
where
\begin{equation} r(x,y)=\frac{x^2_{new}(1-\epsilon)^2+y^2_{new}}{1-\epsilon}\end{equation}
and
\begin{equation}x_{new}=(x-x_0)\cos(\theta)+(y-y_0)\sin(\theta)\end{equation}
\begin{equation}y_{new}=(y-y_0)\cos(\theta)-(x-x_0)\sin(\theta).\end{equation}
Here $x_{new}$ and $y_{new}$ correspond to the position of the X-ray centroid,  $\epsilon=1-\frac{\rm minor~axis}{\rm major~axis}$ is the ellipticity and $\theta$ the Position Angle. 

The final bidimensional model has to be fitted directly on the X-ray images, rather than on the radial profiles (as done in Figure \ref{profile}). To exploit  both the HRI spatial resolution (to sample the halo core) and the large PSPC FOV (to include the large-scale components) we had to fit the HRI and the PSPC data simultaneously. However, at the time this work was in progress, the CIAO package didn't allow to fit two datasets at the same time. Thus we used the following iterative procedure: {\bf (a)} we rebinned the HRI image in 15''$\times$15'' pixels, in order to have enough counts per pixel to fit the galaxy halo and to smooth out small scale asymmetries; NGC 1404 and all the detected point-like sources ($\S$ \ref{sources}) were masked out before performing the fit.
{\bf (b)} The PSPC data were rebinned in 45''$\times$45'' pixels; the regions containing the support structure where masked, together with all the sources detected in the HRI field. Because the PSPC field of view (FOV) is larger than the HRI one, to mask all the remaining PSPC sources we used the source parameters measured by the wavelets algorithm in the context of the GALPIPE project \citep[ and references therein]{mack96}, which
has produced a list of all sources present in the ROSAT PSPC archive\footnote{An online version of the GALPIPE database and documentation can be found at the D.I.A.N.A. homepage at the Palermo Observatory: http://dbms.astropa.unipa.it/}. 
{\bf (c)} We built a model composed of three bidimensional $\beta$ components represented by expression (\ref{2dmodel}) plus a constant background.
{\bf (d)} The first two components, representing the central and galactic halo emission, were fitted to the 15''$\times$15'' HRI data. Because at this resolution it is not possible to properly determine all the parameters of the central component, we fixed the core radius to the best-fit value obtained from the radial fit shown in Figure \ref{profile}. While the central component shows a slight N-S elongation ($\epsilon\sim 0.05$, $\S$\ref{brightness}) at this resolution our attempts to vary also ellipticity $\epsilon$ and position angle $\theta$ resulted in no improvements of the fit, so that they were fixed to zero.
{\bf (e)} The third component, corresponding to the cluster emission, was fitted to the 45''$\times$45'' PSPC data, fixing the central and galactic halo parameters to the values obtained in step (d). The background value was fixed to the flattening level measured in $\S$\ref{rad_prof}.
{\bf (f)} Steps (d) and (e) were repeated iteratively until the best fit parameters converged within the errors. 
\begin{table}[t]
\caption{Best-fit parameters for the bidimensional multi-component halo model.\label{fit_tab}}
\footnotesize
\begin{center}
\begin{tabular}{lcccccccc}
\tableline								
\tableline									
Component & \multicolumn{2}{c}{Center Position} & $r_0$ &
 $\beta$ & $\epsilon$ & $\theta$ & $\chi^2_\nu$ & $\nu$ \\
& R.A. & Dec & (arcsec) & &  & (rad) & & (d.o.f.)\\
\tableline
\raisebox{-0.1cm}{Central} & \raisebox{-0.1cm}{03$^{\rm h}$38$^{\rm m}$28.9$^{\rm s}$} & \raisebox{-0.1cm}{--35$^\circ$27'01''} & \raisebox{-0.1cm}{3.93\tablenotemark{(a)}} & \raisebox{-0.1cm}{0.54$\pm $0.02} & \raisebox{-0.1cm}{0.0\tablenotemark{(a)}} & \raisebox{-0.1cm}{0.0\tablenotemark{(a)}} & \raisebox{-0.4cm}{\parbox[b]{1cm}{$\Big\}$1.2\tablenotemark{(b)}}} & \raisebox{-0.4cm}{1612\tablenotemark{(b)}}\\
\raisebox{.4cm}{Galactic} & \raisebox{.4cm}{03$^{\rm h}$38$^{\rm m}$25.0$^{\rm s}$} & \raisebox{.4cm}{--35$^\circ$27'40''} & \raisebox{.4cm}{2125$\pm$ 800} & \raisebox{.4cm}{41$\pm$ 25} & \raisebox{.4cm}{0.018$\pm$ 0.018} & \raisebox{.4cm}{1.09$^{+1.97}_{-1.09}$}  & \raisebox{.4cm}{} & \raisebox{.4cm}{}\\
\raisebox{0.1cm}{Cluster} & \raisebox{0.1cm}{03$^{\rm h}$38$^{\rm m}$48.2$^{\rm s}$} & \raisebox{0.1cm}{--35$^\circ$23'05''} & \raisebox{0.1cm}{9000$\pm$ 500} & \raisebox{0.1cm}{24$\pm $ 3} & \raisebox{0.1cm}{0.14$\pm$ 0.02} & \raisebox{0.1cm}{0.0 $\pm$ 0.1}
& \raisebox{0.1cm}{1.5} & \raisebox{0.1cm}{4790}\\
\tableline									
\end{tabular}
\end{center}
\tablenotetext{(a)}{ The Central component parameters $r_0$, $\epsilon$ and $\theta$ have no error because they were held fixed during the fit.}
\tablenotetext{(b)}{ The $\chi^2$ and $\nu$ values are the same for the `Central' and `Galactic' components because they were fitted together on the HRI image.}
\tablenotetext{}{{\bf Note} - uncertainties are $1\sigma$ confidence level for 5 interesting parameters}
\end{table}

\begin{figure*}[!t]
\centerline{\psfig{figure=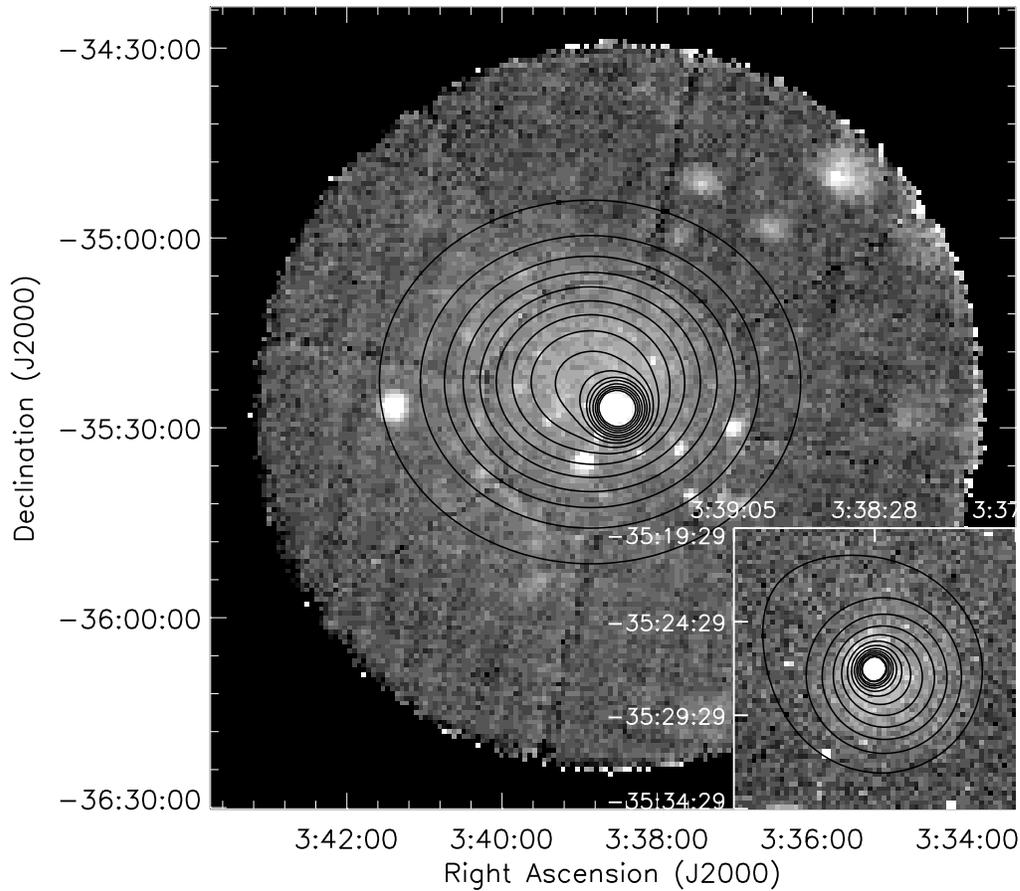,angle=0,width=0.9\textwidth}}
\caption{{\bf Main panel:} contours of the bidimensional halo model superimposed on the 40''/pixel PSPC image. The cluster component is clearly displaced with respect to the galactic halo; the central component is not visible at this resolution. Contours are spaced by a factor of 1.1 with the lowest one at 3.1$\times 10^{-3}$ cnts arcmin$^{-2}$ s$^{-1}$. 
{\bf Bottom-right panel:} bidimensional model contours overlaid on the 15''/pixel HRI image, showing in greater detail the galactic halo region. Contours are spaced by a factor of 1.1 with the lowest one at 6.0$\times 10^{-3}$ cnts arcmin$^{-2}$ s$^{-1}$.}
\label{model_map}
\end{figure*}

The best-fit model parameters are shown in Table \ref{fit_tab}. In Figure \ref{model_map} we show the model contours superimposed on the 45''/pixel PSPC image (main panel); the nuclear region is shown in greater detail in the right-bottom panel, with contours overlaid on the 15''/pixel HRI image. 
The three components have different features: while the central one is centered on the galaxy and is highly peaked, the more extended galactic halo is displaced to the South-West thus accounting of the observed asymmetries shown in Figure \ref{pies}. In contrast with the circular symmetry of the smaller components ($\epsilon_{Central}\sim 0.05, \epsilon_{Galactic}=0.02$) the third component is elongated in the East-West direction and its center is displaced to the North-East of the optical galaxy. These results suggest a different origin of each component, that will be discussed in detail in $\S$ \ref{results}.

The $\chi^2$ values obtained from the fit are quite large. This is not surprising considering the large spatial anisotropy of the X-ray surface brightness (see $\S$ \ref{brightness}). The situation is worse for the PSPC data (see Figure \ref{PSPC}), also because of the poor knowledge of the exposure corrections past $\sim 1100''$ from the center of the FOV (the position of the support ring) and the large and irregular PRF at these radii that makes difficult a correct source masking.
In Table \ref{fit_tab} the `Central' and `Galactic' components have the same $\chi^2$ value because they were fitted together to the HRI data. If we
restrict the fit to the region within 1 arcmin and use higher resolution (2''/pixel) we obtain a $\chi^2_{\nu}\sim 0.6$, in agreement with the results obtained in $\S$ \ref{rad_prof} from the profile fitting.

\begin{figure}[t]
\centerline{\psfig{figure=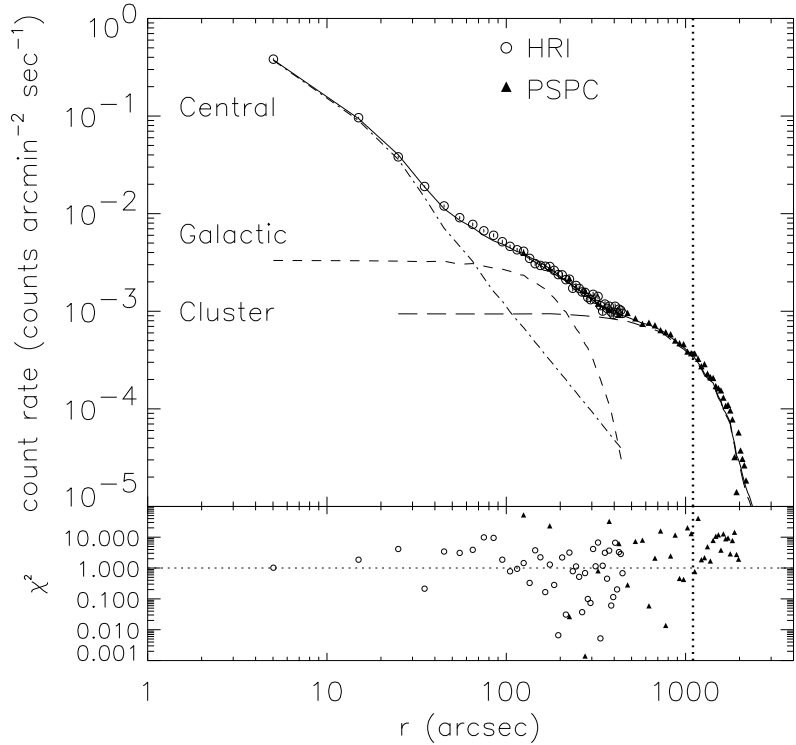,angle=0,width=0.53\textheight}}
\caption{Radial profiles compared to the bidimensional model profiles. Open circles represent HRI counts while triangles represent PSPC counts rescaled to match the HRI. The continuous line represents the best-fit model while dot-dashed, short-dashed and long-dashed lines represent respectively the central, galactic and cluster halo components. The residuals are shown in the lower panel in $\chi^2$ units. The vertical dotted line marks the position of the PSPC support ring; at larger radii the exposure correction is poorly known.}
\label{model_prof}
\end{figure}

The relative contribution of each component to the global profile is shown in Figure \ref{model_prof}. We must notice that here both global and individual components profiles were extracted in annuli centered on the X-ray centroid (rather than centering each one on the respective component centroid) so to allow a comparison of the observed data with the model. The bottom panel shows the radial residuals respect to the best fit model. The high $\chi^2$ values are in agreement with the values obtained from the bidimensional fit; in particular the residuals are systematically higher at radii larger than 1100'' due to the problems explained above.

\begin{figure}[t]
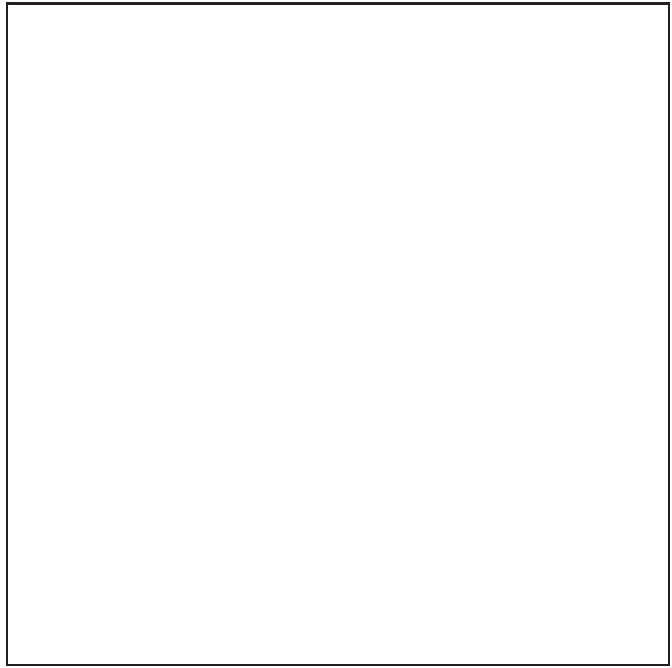

\centerline{\framebox(250,250)[br]{}}
\caption{Residuals of the bidimensional fit on the 15 arcsec/pixel HRI image, after convolving with a gaussian of $\sigma=30''$. Colors from black to white represent residuals from $-2.5 \sigma$ to $22 \sigma$. For comparison we superimposed the contour levels of the adaptively smoothed image (Figure \ref{csmooth}a).}
\label{residuals}
\end{figure}

The residuals of the bidimensional fit to the HRI image are shown in Figure \ref{residuals}. The presence of structures in the gaseous halo is evident and confirms the results obtained with the adaptive smoothing technique. The highest residuals are coincident with the positions of point sources detected in the field, as can be seen from the overlaid contours. Considering the fluctuations on scales of 60'' ($\sim 5$ kpc)
the statistical errors are reduced of a factor 4 with respect to the colorscale levels shown in Figure \ref{residuals}, and the significance of the residual emission is increased of the same factor. This means, for instance, that the deep `hole' on the North-West side of the halo is approximately significant at the $6\sigma$ level above the fluctuations expected from the multicomponent model, while the excess emission regions on the North and South sides of the galactic center are significant at the $3\sigma$ level. Even though the multi-component bidimensional model may not be a proper physical representation of the hot X-ray halo this result shows that the filamentary structures seen in the halo are not due to statistical fluctuation over a smooth surface brightness distribution. This result is further confirmed by the {\it Chandra} data presented in $\S$ \ref{Chandra}. 

\subsection{Density, Cooling Time and Mass Profiles}
\label{dens_par}
To derive the hot gas density of NGC 1399 we used the procedure discussed in $\S$ \ref{empirical}. Since the global gas distribution has a complex and asymmetric structure, we cannot apply this method to the total emission; instead we applied it to the single components separately. For the `central' and `galactic' components we assumed spherical symmetry; this is a reasonable approximation because their ellipticities, derived from the bidimensional models, are small (see previous section). The same approximation is not valid, instead, for the cluster halo so that, in deriving the density profile, we assumed rotational symmetry around the minor axis. We found that assuming symmetry around the major axis the difference is of the order of a few percents (except at very large radii where exposure correction uncertainties are  dominant) and resulted in minor changes to our conclusions. The central, galactic and cluster components were assumed to be isothermal with temperatures of respectively $kT=$0.86, 1.1 and 1.1 keV, in agreement with the PSPC temperature profiles derived by RFFJ. \cite{Buo99} showed that ASCA data suggest the presence of multiphase gas within the central 5' of NGC 1399, better fitted by cooling flows models. However, as he notes, the narrower ROSAT energy range and the lower spectral resolution is unable to clearly discriminate between single and multi-temperature models (even though it reveals the presence of radial temperature gradients) so that our assumption of isothermal components is adequate for the kind of analysis performed here. 

\begin{figure}[t]
\centerline{\psfig{figure=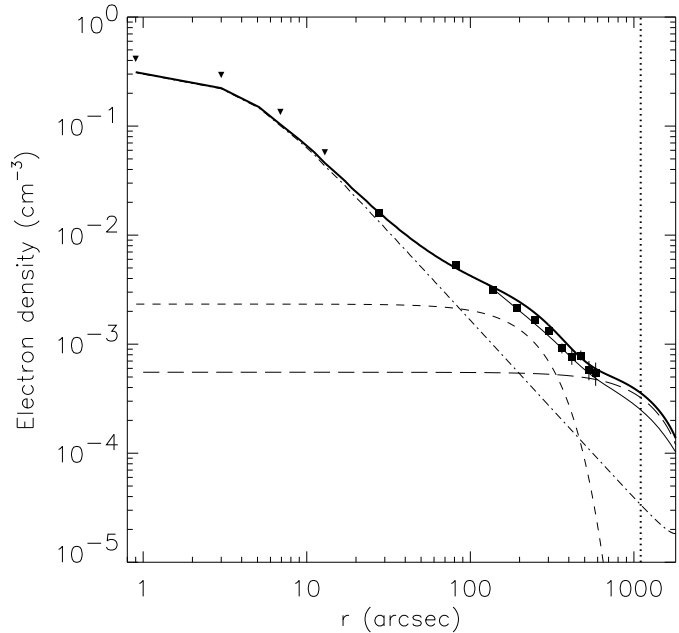,angle=0,width=0.45\textheight}}
\caption{Density profiles derived from deprojection. Dot-dashed, short dashed and long dashed lines represent respectively the central, galactic and cluster halo component. Downward triangles show the upper limit on the central component profile adopting a central absorbing column of $2\times 10^{21}$ cm$^{-2}$. The thick (thin) continuous lines show the total density ignoring (including) the center offset of the different components (see discussion in text). For comparison the values obtained from \cite{rang95} are shown as filled squares.
As in Figure \ref{model_prof} the vertical dotted line marks the position of the PSPC support ring.}
\label{density_prof}
\end{figure}

The electron density profiles of the isothermal components are shown in Figure \ref{density_prof}. The density profiles are centered on the centers of their respective components and thus the total density does not take into account the centers offset. Including the displacements of each component with respect to the galaxy center results in a slightly smoother profile in the outer regions, due to the azimuthal averaging (see Figure \ref{density_prof}). ROSAT and ASCA spectral fits (RFFJ; \citealp{Buo99,Mat00}) revealed that additional absorption, over the galactic value, may be required in the galaxy center. We thus estimated an upper limit for the density of the central component assuming an absorption of $2\times 10^{21}$ cm$^{-2}$ (downward triangles in Figure \ref{density_prof}).

We must notice that adding the single density components, derived from the surface brightness decomposition, to obtain the total density profile is not strictly the correct approach since the surface brightness does not depend linearly on the gas density (see equations \ref{totlum} and \ref{betaproj}). However regions of the image in which more than one component are significant are small and the good agreement of our results with the estimate of RFFJ (who deprojected the total surface brightness) indicates that we are not introducing significant errors in our total density estimate. The same considerations hold for the mass profiles derived below.

\begin{figure}[t!]
\centerline{\psfig{figure=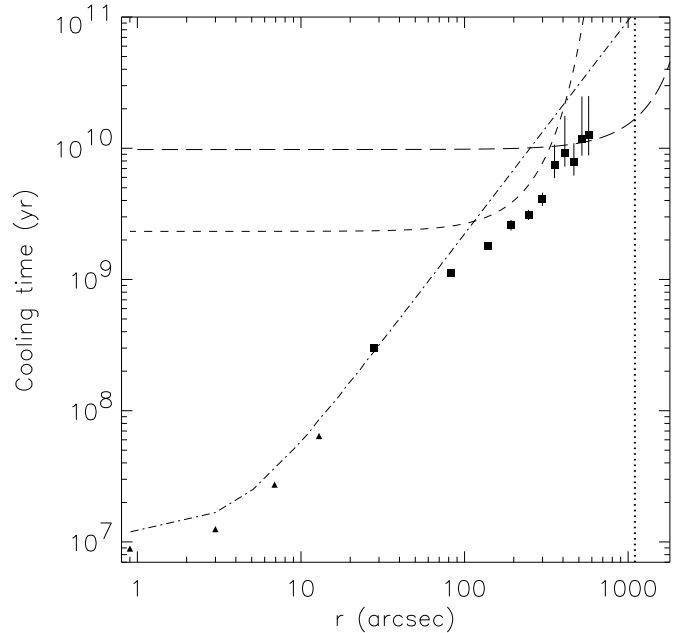,angle=0,width=0.45\textheight}}
\caption{Cooling time profiles derived from deprojection. The symbols have the same meaning as in Figure \ref{density_prof}. Note that the assumption of a high central absorption translates into lower limits on the central cooling time (filled triangles).}
\label{cooling_prof}
\end{figure}

Using the density profiles and the assumed temperatures we estimated the cooling time (equation \ref{CT}) as a function of radius. We adopted the cooling function given by \cite{Sar87}. Figure \ref{cooling_prof} shows that the central and galactic components have cooling times much shorter than the age of the universe, assumed to be $\sim 10^{10}$ years, out to 250'' and 350'' (23 and 32 kpc) respectively.

\begin{figure*}[p]
\centerline{\psfig{figure=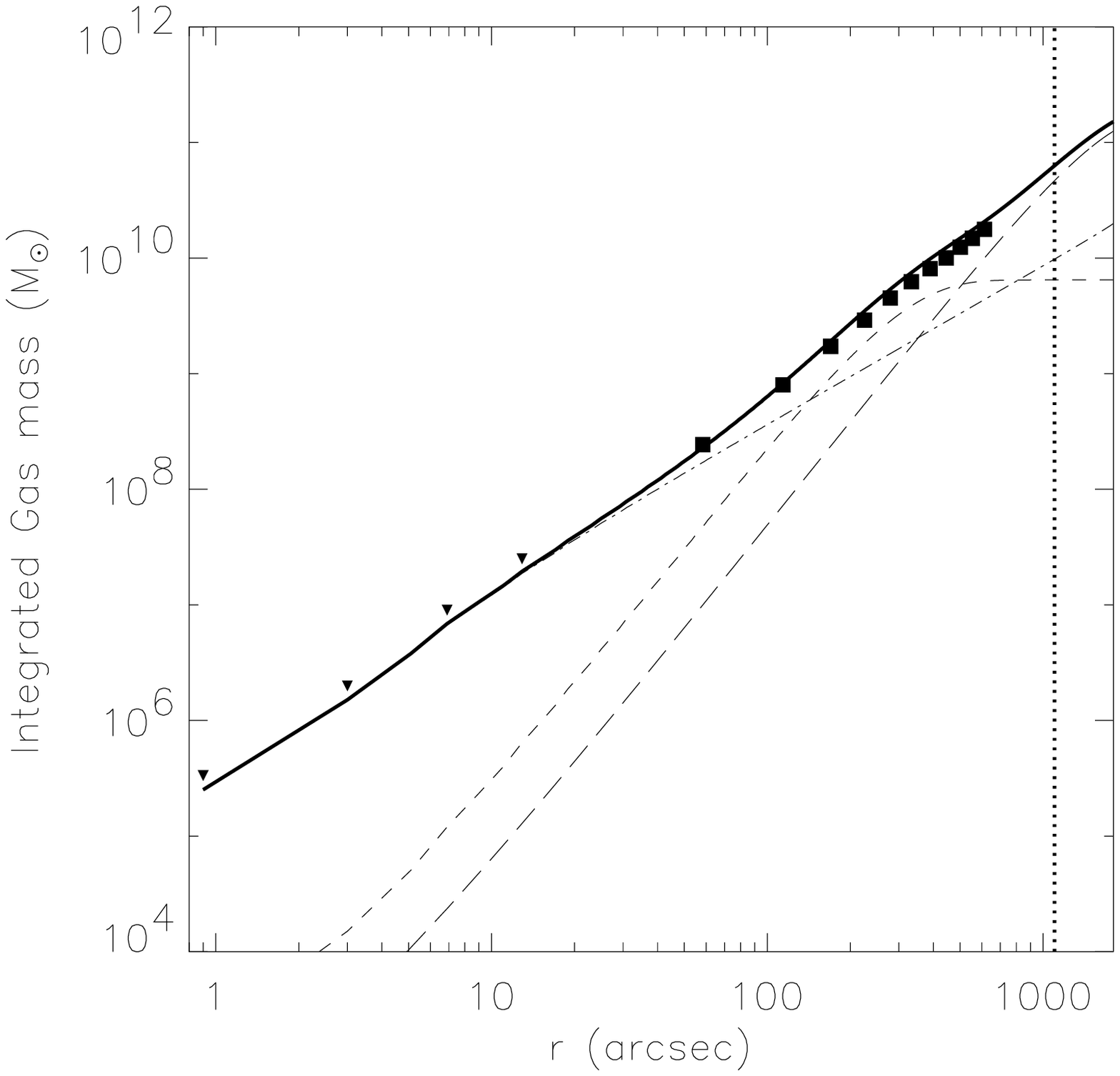,angle=0,width=0.55\textwidth}\hspace{-1.5cm}\raisebox{7.2cm}{\parbox[b]{1cm}{\bf(a)}}\hspace{1.5cm}}
\centerline{\psfig{figure=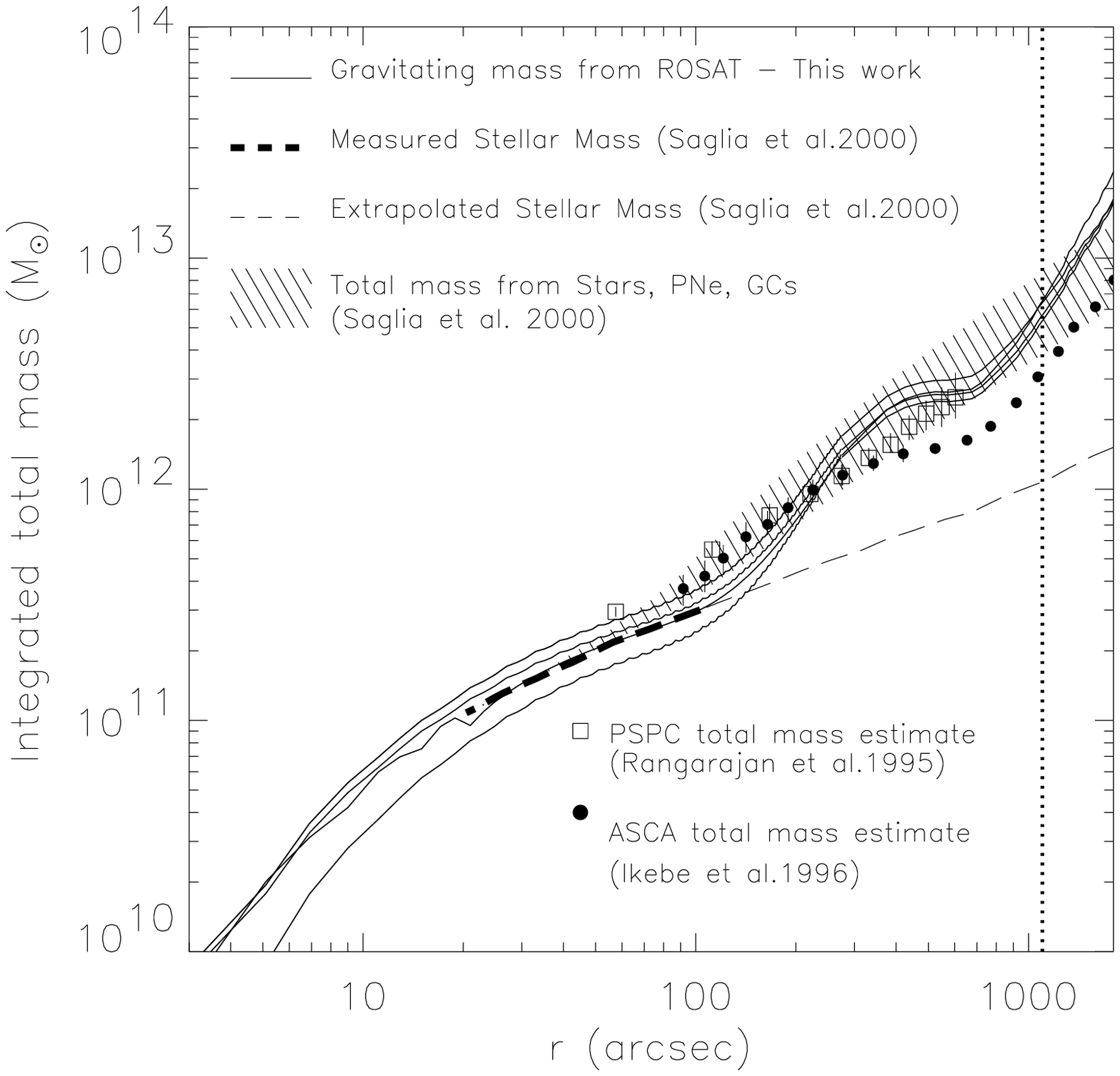,angle=0,width=0.55\textwidth}\hspace{-1.5cm}\raisebox{7.2cm}{\parbox[b]{1cm}{\bf(b)}}\hspace{1.5cm}}
\caption{Integrated mass profiles derived from deprojection. {\bf (a)} Integrated gas mass, with symbols having the same meaning as in Figure \ref{density_prof}. {\bf (b)} Integrated gravitating mass for our different temperature profiles (see discussion in text) are shown as continuous lines. Open squares and filled circles represent the total gravitating mass estimated respectively by \cite{rang95} and \cite{ikebe96}. The gravitating mass range derived by \cite{saglia00} from optical observations of stars, GCs and PNe, is shown as a shaded region. The contribution of the stellar matter measured (extrapolated) by \cite{saglia00} is represented by the thick (thin) dashed line.}
\label{mass_prof}
\end{figure*}

The cumulative gas mass profile is shown in Figure \ref{mass_prof}a as a function of radius. The flat profile of the galactic component over 600'' (55 kpc) reflects the steep slope of the Beta profile. The filled triangles show that the adoption of a high absorbing column in the galaxy center does not affect significantly the gas mass estimates, giving a difference of $\sim $33\%. 
The total gravitating mass was calculated using expression (\ref{mass}).
We tried four different temperature profiles based on published data: we used both the linear and power-law approximations derived by \cite{jones97} based on PSPC spectral fits, a constant temperature of 1.1 keV that approximates the isothermal profile found by RFFJ and a power-law profile $T(r)=T_0 r^\alpha$,
with $T_0=0.6$ keV and $\alpha=0.13$, representing the temperature drop in the inner 150" (14 kpc). The gravitating mass, extracted in concentric annuli centered on the X-ray centroid (i.e. taking into account the offset between the different components), is shown in Figure \ref{mass_prof}b. The similarity of the continuous lines indicates that the gravitating mass is weekly dependent on the assumed temperature profile and is mostly determined by the density profile. The largest deviation is seen in the model taking into account the central temperature drop, which is systematically lower in the inner 100". 
Our profiles are in good agreement with RFFJ and fall within the range predicted by optical observations (shaded region, \citealp{saglia00}). 
We clearly distinguish the transition between the galaxy and cluster halo near 700" (64 kpc), in agreement with the ASCA data (filled circles, \citealp{ikebe96}). Both components exceed the luminous mass, based on an extrapolation of stellar measurements (thin dashed line), indicating the presence of a consistent amount of dark matter.

Thanks to the ROSAT HRI resolution in the inner 100" (9 kpc) we are able to resolve the dynamical behavior of the central component. The gravitating mass follows closely the stellar mass profile, confirming that the central galaxy dynamics are dominated by the luminous matter \citep{saglia00}.

\subsection{Total Fluxes and Luminosities}
\label{fluxes}

To determine the total HRI X-ray flux of NGC 1399 we measured the total net counts within a 500'' radius from the X-ray centroid. The background counts were extracted from the rescaled particle map (see $\S$ \ref{rad_prof}) in the 500''--850'' annulus .
We then corrected the background subtracted counts for the contamination of pointlike sources within the 500'' circle. This was done by subtracting, for each source,  the net counts measured by the wavelets algorithm (see $\S$ \ref{sources}). The final count rate, for the hot halo of NGC 1399, is 0.451$\pm$0.004 counts sec$^{-1}$.

To convert this count rate into a flux we used the best fit Raymond-Smith (RS) model found by RFFJ from the ROSAT PSPC data (a main thermal component with  $kT\simeq$1 keV plus a softer emission with $kT=80.8$ eV contributing for 16\% of the emission in the 0.2--0.3 keV range). Using the HEASARC-XSPEC software, we simulated a RS spectrum fixing the absorbing column to the galactic value ($1.3 \times 10^{20}$ cm$^{-2}$) and the metal abundance to the solar value. With this spectrum, we calculated the conversion factor between counts and fluxes using the PIMMS software, that takes into account the HRI spectral response function. We obtain 1 count sec$^{-1}$=3.158$\times 10^{-11}$ erg s$^{-1}$ cm$^{-2}$ so that $f_X=(1.42\pm 0.01)\times 10^{-11}$ erg s$^{-1}$ cm$^{-2}$ in the 0.1-2.4 keV energy range. Assuming a distance of 19 Mpc, $L_X(0.1-2.4 keV)=(5.50\pm 0.04)\times10^{41}$ erg s$^{-1}$. Taking into account the uncertainties in the exposure correction and background extraction, this result agrees within 10\% with the estimate of RFFJ, converted to our adopted distance.

We also searched for evidence of nuclear variability in our data, finding none.

\subsection{X-ray/Radio Comparison}
\label{Xradio}
\begin{figure*}[!t]
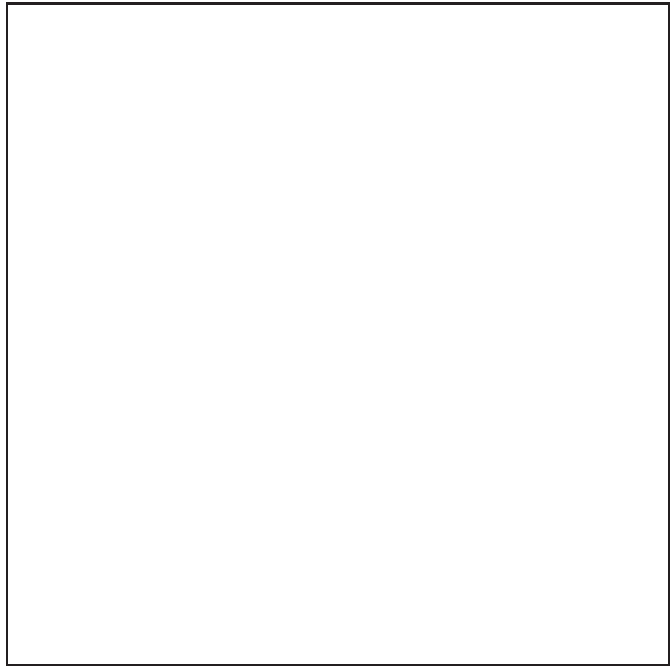

\vspace{2cm}
\centerline{\framebox(250,250)[br]{}}
\vspace{3cm}
\caption{6 cm radio contours (green; from Figure 1 of \citealp{kbe88}) superimposed on the adaptively smoothed 1''/pixel X-ray image.
The logarithmic color scale represents intensities from 8.6$\times 10^{-3}$ (black) to 6.7$\times 10^{-1}$ (white) cnts arcmin$^{-2}$ s$^{-1}$.}
\label{X-radio}
\end{figure*}

NGC 1399 is known to be a radio galaxy with a faint radio core ($S_{core}\sim 10$ mJy at 5 GHz, \citealp{ekers89}). \cite{kbe88} studied in detail the 6 and 20 cm radio emission. The 6 cm radio contours (their Figure 1) are superimposed on the `csmoothed' X-ray image in Figure \ref{X-radio}. The radio core lies in the center of the galaxy, but the nuclear source is not distinguishable from the radio jets at this resolution.

We searched our X-ray data for a central point source that could be the counterpart of the radio core. Trying to model the nuclear emission with a Beta model plus a delta function, convolved with the HRI PRF, resulted in minor variations in the Beta model parameters and no improvement in the fit statistics, thus indicating no need for additional nuclear emission. To estimate an upper limit we varied the central point source brightness and repeated the fit until we reached a $3\sigma$ confidence level. We obtain an upper count rate limit of $2.25\times 10^{-3}$ counts s$^{-1}$. Assuming a power law spectrum with photon index $\alpha_{ph}=1.7$ and galactic absorption $N_H$, we get $f_X^{3\sigma}=1.0\times 10^{-13}$ erg s$^{-1}$ cm$^{-2}$ in the 0.1-2.4 keV band, corresponding to $L_X^{3\sigma}=3.9\times 10^{39}$ erg s$^{-1}$. This value is approximately half of the one derived by \cite{Sulk01} making use of one of our HRI observations (RH600831a01) and 30\% of the limit estimated by RFFJ by means of spectral analysis. Recently \cite{Loew01} were able to further reduce the upper limit on the nuclear source by one order of magnitude making use of {\it Chandra} data.

By comparing the 6 cm radio contours with the X-ray halo (Figure \ref{X-radio}) we identify several interesting features:
i) in the X-rays the nuclear region is elongated in the N-S direction, following the radio jets. This elongation is more pronounced in the Southern direction, where there is an excess of X-ray emission following the Western side of the jet (region A); ii) At the point where the Southern jet ends in the lobe there is a steep gradient in the X-ray emission. The lobe is bent towards the West and aligned with a region of low X-ray emission (region B); iii) The Southern lobe ends in an irregular clump coincident with a region of higher X-ray emission (region C); iv) No features as the ones seen in the Southern lobe are evident in the Northern one. Here the radio jet ends smoothly in a regular lobe with no sign of sharp transition and there is no evidence of cavities (minima) in the X-ray emission. Instead the jet and the lobe are aligned with an X-ray structure that seems to be coincident with the right side of the jet/lobe. At the North-West end of the Northern lobe there are two X-ray clumps (region D): they are detected as a single source by the wavelet algorithm (No.32 in Table \ref{source_tab}). To a more accurate visual inspection we found that the smaller and compact source is coincident with an optical (probably background) counterpart, while there is no counterpart for the diffuse clump ($\S$ \ref{sources}) suggesting that it is a local feature of the hot gaseous halo.

From the residual image of the two-dimensional fit (Figure \ref{residuals}) we can calculate the significance of these features. We find that the depletion corresponding to the Southern lobe (region B) and the Northern diffuse clump (region D) are significant respectively at the $2.5\sigma$ and $\sim 4\sigma$ level. Region C corresponds to a $3\sigma$ brightness enhancement ($\S$ \ref{model}).

\begin{figure}[t]
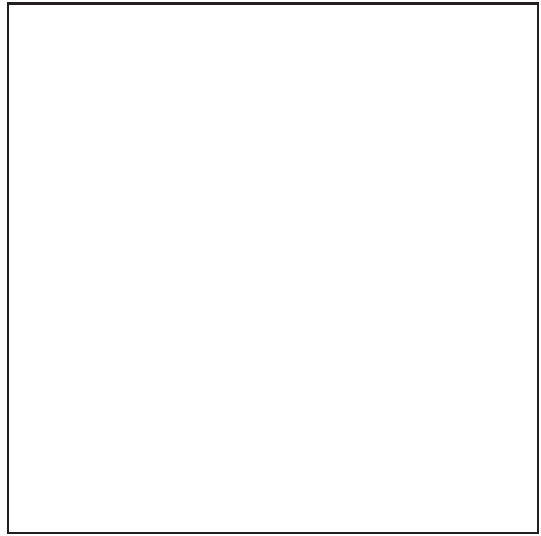

\vspace{1cm}
\centerline{\framebox(200,200)[br]{}}
\vspace{1cm}
\caption{6 cm radio contours superimposed on the 2''/pixel residuals X-ray image of the nuclear region. The image has been smoothed with a gaussian of $\sigma= 4"$. In the logarithmic color scale black represents negative residuals while colors from blue to white represent increasing positive residuals. The residuals range from $-2.2 \sigma$ to $+4 \sigma$. 
The white circles on the left show the positions of X-ray sources detected by the wavelets algorithm (No.2 and 11 in Table \ref{source_tab}). The radius of the central source (corresponding to the central X-ray peak) is $\sim 22"$.}
\label{X-radio_nuclear}
\end{figure}

To examine in greater detail the nuclear region we subtracted the halo model developed in $\S$ \ref{model} from the X-ray image to produce a residual map of the nuclear region with a 2"/pixel resolution (Figure \ref{X-radio_nuclear}).
As discussed at the beginning of this section, there is no evidence of a nuclear point source aligned with the radio centroid. Instead residual emission is visible on both sides of the radio jets a few arcseconds ($\leq 1$ kpc) from the nucleus. There is no evidence that any of these excesses may be due to the nuclear source because they are displaced with respect to the radio emission. Moreover the radio emission seems to avoid the X-ray excess on the eastern side of the nucleus, suggesting that these features may be correlated. 

In coincidence with the direction of the radio jets there are two regions of negative residuals ($\sim -2\sigma$), extending up to $\sim 20"$, where
additional excess emission, on both the Northern and Southern sides of the X-ray core and aligned with the radio jets, is clearly visible. The Southern excess corresponds to region A of Figure \ref{X-radio}.
Residual emission is also present on the Western side of the nuclear region. 
Visual inspection of the {\it Chandra} image (Figure \ref{Chandra_csmooth}) shows that this excess is probably due to the presence of point sources unresolved in the HRI image.

\begin{figure}[t]
\centerline{\psfig{figure=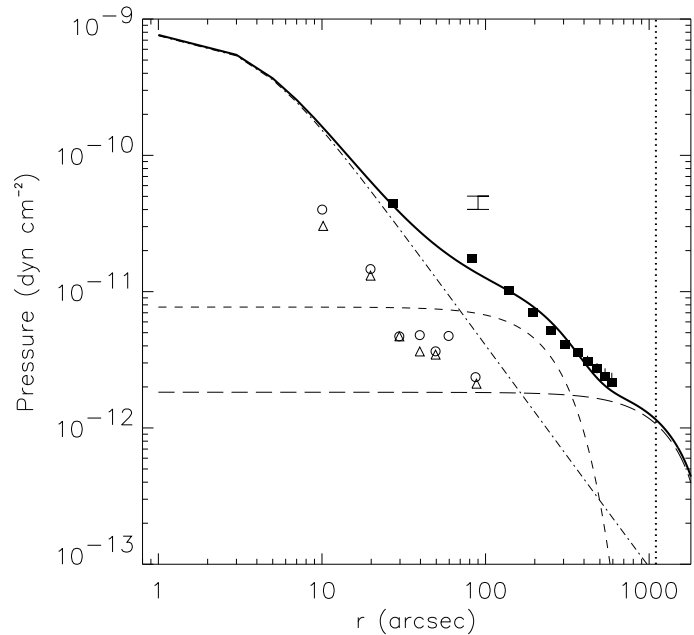,angle=0,width=0.45\textheight}}
\caption{Gas pressure profiles derived from deprojection. The symbols have the same meaning of Figure \ref{density_prof}. Open circles and triangles represent respectively the radio pressure of the Northern and Southern lobes measured by \cite{kbe88}. The vertical error bar in the Figure indicates the pressure range of the D region (see discussion in text).}
\label{press_prof}
\end{figure}

From the density profiles obtained in $\S$ \ref{dens_par} we were able to obtain the pressure profiles of the three halo components shown in Figure \ref{press_prof}. The pressure profiles are compared to the minimum radio pressure measured by \cite{kbe88} in the Northern (circles) and Southern (triangles) radio jets/lobes. The Figure confirms their result that the radio source can be confined by the ISM whose pressure is higher than the radio one. We must notice that they found a difference between radio and thermal pressure of more than one order of magnitude, while we find $P_{Thermal}/P_{Radio}= 3 \div 4$. This is mostly due to the poor ISM temperature constraint obtained from the {\it Einstein} IPC data and to our better spatial resolution. 

The vertical error bar in the Figure indicates the pressure estimate for the diffuse clump observed in the D region. To calculate this value we assumed that the brightness enhancement is due to a homogeneous sphere of $\sim 9"$ (800 pc) radius at the same temperature of the galactic halo (1.1 keV). Depending on whether the background is extracted locally (an annulus 9" wide) or from the Beta model, the `bubble' pressure can vary up to 25\%, so that the allowed range is shown as an error bar.
This estimate is dependent on the assumed geometry of the emitting region and on projection effects, so that it must be considered as an upper limit on the actual pressure.

\subsection{Discrete Sources}
\label{sources}
We used the algorithm developed at the Palermo Observatory by F. Damiani and collaborators \citep{Dam97a} to detect the discrete sources present in the HRI field. The algorithm measures the relevant parameters in 
a wavelet transformed space, using different wavelet scales to optimize the detection of both compact and extended sources.  
The wavelet algorithm found 43 sources in our field, using a signal to noise threshold of 4.65. The threshold was chosen to assure a contamination of one/two spurious detections per field (F. Damiani private communication, see also \citealp{Dam97b}). Three more sources were found with the algorithm, but being clearly associated to the ribbon-like features due to the presence of bad pixels in the detector (Figure \ref{NGC1399comp}), were excluded from the subsequent analysis. 
After visual inspection of the adaptively smoothed image (Figure \ref{csmooth}) three more detections (No.44, 45, 46) with a lower S/N ratio but showing clear features of pointlike sources, were added to the list.

\begin{figure*}[p]
\centerline{\psfig{figure=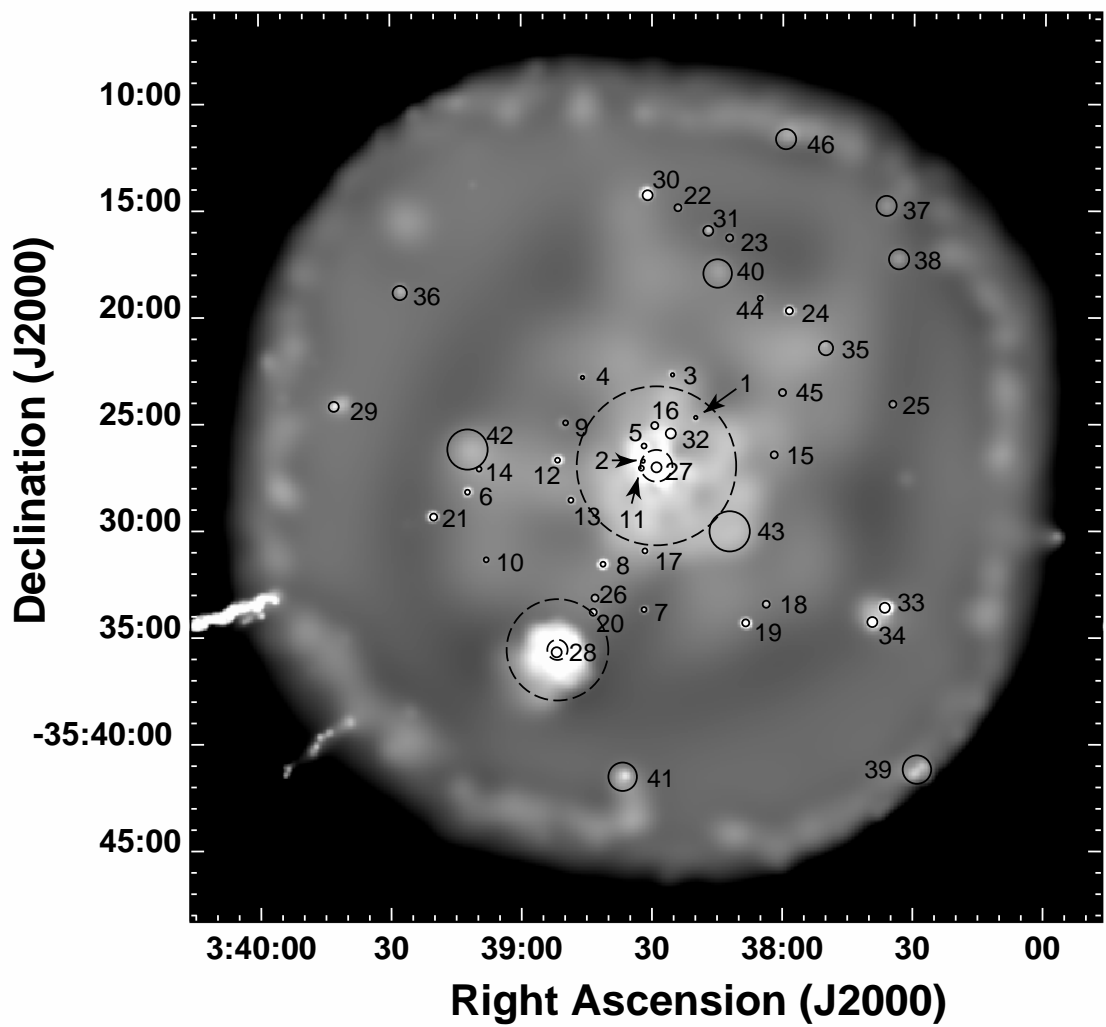,angle=0,width=1.1\textwidth}}
\caption{The sources detected with the wavelets algorithm of \cite{Dam97a,Dam97b} superimposed on the 5''/pixel adaptively smoothed HRI image. Solid circles show the region of maximum S/N ratio for each source. Dashed circles represent $r_{eff}$ and $5r_{eff}$ for NGC 1399 and NGC 1404 \citep{gou94}.}
\label{det}
\end{figure*}

The source positions in the field are shown in Figure \ref{det}. Sources No.42 and 43 are clearly extended and seem to be associated to structures in the NGC 1399 halo revealed by the adaptive smoothing algorithm ($\S$ \ref{brightness}).

Table \ref{source_tab} lists the parameters measured with the wavelets algorithm for each source: position, 3$\sigma$ radius, number of counts with statistical error, maximum signal to noise ratio (for sources detected at different spatial scales), count rate and flux in the 0.5-2.0 keV band for both HRI and PSPC data. We computed the flux of each source assuming a power law spectrum with photon index $\alpha_{ph}=1.96\pm0.11$ ($\mathit{f}^{pow.law}$ in Table \ref{source_tab}), as found by Hasinger and collaborators for faint sources. This gives a conversion factor of 1 cnt s$^{-1}$=$2.087\times 10^{-11}$ ergs s$^{-1}$ cm$^{-2}$ in the 0.5-2.0 keV band (the same used by Hasinger). We also tried a Raymond-Smith spectrum with $kT=0.52$ KeV and abundance $Z\sim 0.2~Z_\odot$, following the results of \cite{Kim95} in their study of pointlike sources near NGC 507, and obtaining an almost identical result. In Table \ref{source_tab} $\mathit{f}^{max~abs.}$ represent the flux estimated with the power law model fixing the absorption to the upper limit of $4.5\times 10^{20}$ cm$^{-2}$, allowed from the spectral analysis performed by RFFJ. 

The the last three columns of Table \ref{source_tab} report the source short-term variability obtained with a Kolmogorov-Smirnov test of the HRI light curves, long-term variability from comparison with PSPC fluxes (as explained below),
and possible optical counterparts (within the 3$\sigma$ radius measured by the wavelets algorithm) found by both visual inspection of the DSS photographic plates and cross-checking with known sources present in the NASA Extragalactic Database (NED). The names of the counterparts correspond to those reported by \cite{Hilk99b} or \cite{Smith96} in their optical and NUV studies of the Fornax field.
Sources No.27 and 28, representing respectively NGC 1399 and NGC 1404, are not included in the table.

\begin{landscape}
\thispagestyle{empty}
\topmargin=1.in
\begin{table}[p]
\caption{Discrete Sources Properties\label{source_tab}}
\scriptsize
\begin{tabular}{cccrrrccccccc}
\\
\tableline
\tableline
Source & R.A.         & Dec.         & 3$\sigma$ Radius & Counts & Max S/N & Count Rate                   & $\mathit{f}_{HRI}^{pow.law}$\tablenotemark{(a)} & $\mathit{f}_{HRI}^{max~abs.}$\tablenotemark{(a)} & $\mathit{f}_{PSPC}^{pow.law}$\tablenotemark{(a)} & Time Var.\tablenotemark{(b)} & Opt.id. & NED obj.\tablenotemark{(c)}\\
 No.  & \multicolumn{2}{c}{(J2000)} & (arcsec)         &        &         & ($10^{-4}$ Cnts sec$^{-1}$) &   &  &   & short-term / long-term &         & \\
\tableline
1 & 3:38:20.0 	  & -35:24:42    & 8.7      & 35$\pm$12   & 4.7  & 2.1$\pm$0.7  & 4.3$\pm 0.2$ 	 &   5.6 & 28$\pm 7$ 	 & n/y & - & - \\
	
2 & 3:38:32.0 	  & -35:26:45    & 8.7      & 54$\pm$17   & 6.1  & 3.2$\pm$1.0  & 6.7$\pm 0.3$ 	 &   8.6 & - 	 & n/$\cdots$ & - & - \\
	
3 & 3:38:25.3 	  & -35:22:42    & 8.7      & 49$\pm$14   & 6.7  & 2.9$\pm$0.8  & 6.1$\pm 0.3$ 	 &   7.8 & - 	 & n/$\cdots$ & - & - \\
	
4 & 3:38:45.9 	  & -35:22:50    & 8.7      & 42$\pm$12   & 6.1  & 2.6$\pm$0.7  & 5.3$\pm 0.3$ 	 &   6.9 & 9$\pm 3$ & n/n & y & CGF 0202\\
	
5 & 3:38:31.8 	  & -35:26:02    & 11.9     & 165$\pm$23  & 13.0 & 10$\pm$2  & 20$\pm 1$	 &  26.0 & -	 & n/$\cdots$ & - & CGF 0233\\
	
6 & 3:39:12.3 	  & -35:28:12    & 13.2     & 172$\pm$21  & 14.9 & 10$\pm$1 & 21$\pm 1$	 &  27.5 & 25$\pm 4$	 & n/n & y & - \\
	
7 & 3:38:31.8 	  & -35:33:42    & 12.2     & 56$\pm$16   & 6.0  & 3.3$\pm$0.9  & 6.9$\pm 0.3$	 &   8.9 & 8$\pm 3$	 & n/n & - & - \\
	
8 & 3:38:41.2 	  & -35:31:35    & 10.8     & 1000$\pm$35 & 60.6 & 59$\pm$2 & 123$\pm 6$     & 159.2 & 170$\pm 8$ & 99\%/y & y & - \\
	
9 & 3:38:49.8 	  & -35:24:57    & 12.2     & 47$\pm$14   & 5.0  & 2.8$\pm$0.9  & 5.8$\pm 0.3$ 	 &   7.5 & - 	 & n/$\cdots$ & - & - \\
	
10 & 3:39:08.1	  & -35:31:22    & 12.2     & 56$\pm$16   & 6.2  & 3.3$\pm$0.9  & 6.9$\pm 0.4$ 	 &   8.9 & 22$\pm 6$ 	 & n/n & - & - \\
	
11 & 3:38:32.4	  & -35:27:05    & 13.7     & 171$\pm$30  & 11.5 & 10$\pm$2 & 21$\pm 1$	 &  27.2 & -	 & n/$\cdots$ & - & - \\
	
12 & 3:38:51.7	  & -35:26:42    & 11.6     & 295$\pm$22  & 24.7 & 17$\pm$1 & 36$\pm 2$	 &  47.0 & 58$\pm 6$	 & 95\%/y & y & CGF 0102\\
	
13 & 3:38:48.6	  & -35:28:35    & 12.2     & 77$\pm$20   & 7.7  & 4.5$\pm$1.1  & 9.3$\pm 0.5$ 	 &  12.0 & - 	 & n/$\cdots$ & y & - \\
	
14 & 3:39:09.7	  & -35:27:07    & 12.2     & 46$\pm$14   & 5.0  & 2.8$\pm$0.8  & 5.7$\pm 0.3$ 	 &   7.4 & - 	 & n/$\cdots$ & y & - \\

15 & 3:38:01.9    & -35:26:27    & 17.3     & 62$\pm$19   & 4.7  & 3.6$\pm$1.1  & 7.5$\pm 0.4$ 	 &   9.7 & - 	 & n/$\cdots$ & - & - \\
	
16 & 3:38:29.4    & -35:25:05    & 17.3     & 75$\pm$23   & 4.8  & 4.4$\pm$1.3  & 9.1$\pm 0.5$ 	 &  11.7 & - 	 & n/$\cdots$ & - & - \\
	
17 & 3:38:31.6    & -35:30:58    & 12.2     & 70$\pm$18   & 6.8  & 4.1$\pm$1.1  & 8.6$\pm 0.5$ 	 &  11.1 & - 	 & n/$\cdots$ & - & - \\
	
18 & 3:38:03.7    & -35:33:27    & 17.3     & 65$\pm$19   & 5.4  & 4.1$\pm$1.2  & 8.5$\pm 0.4$ 	 &  10.9 & - 	 & n/$\cdots$ & y & CGF 0354\\
	
19 & 3:38:08.5    & -35:34:20    & 15.0     & 908$\pm$34  & 52.3 & 57$\pm$2	 & 119$\pm 6$    & 152.7 & 127$\pm 7$ & 99\%/n & y & - \\

20 & 3:38:43.5    & -35:33:50    & 15.2     & 105$\pm$18  & 8.7  & 6.3$\pm$1.1  & 13.1$\pm 0.7$	 &  16.8 & -	 & n/$\cdots$ & y & - \\

21 & 3:39:20.1    & -35:29:22    & 16.5     & 527$\pm$30  & 33.0 & 32$\pm$2	 & 68$\pm 4$  &  86.9 & 83$\pm 7$  & 99\%/n & y & - \\
 
22 & 3:38:24.1    & -35:14:52    & 17.3     & 60$\pm$18   & 4.9  & 3.8$\pm$1.2  & 8.0$\pm 0.4$ 	 &  10.3 & 12$\pm 4$ 	 & n/n & y & - \\

23 & 3:38:12.2    & -35:16:17    & 17.3     & 59$\pm$18   & 4.8  & 3.7$\pm$1.1  & 7.7$\pm 0.4$ 	 &   9.9 & - 	 & n/$\cdots$ & y & - \\
\tableline
\end{tabular}
\tablenotetext{(a)}{ Flux in the 0.5-2.0 KeV band in units of $10^{-15}$ erg s$^{-1}$ cm$^{-2}$.}
\tablenotetext{(b)}{\small ~Short-term = probability of the Kolmogorov-Smirnov test: `n' means no variability detected at the 95\% level.\\
		   ~Long-term = difference between PSPC and HRI counts: `y' means variable source ($>3\sigma$), `n' means no variability detected, `$\cdots$' means no available data.}
\tablenotetext{(c)}{\small ~CGF=\cite{Hilk99b}; FCCB=\cite{Smith96}}
\end{table}
\end{landscape}
\begin{landscape}
\thispagestyle{empty}
\topmargin=1.in
\begin{table}[p]
\addtocounter{table}{-1}
\caption{ - Continued}
\scriptsize
\begin{tabular}{cccrrrcccccccc}
\\
\tableline
\tableline
Source & R.A.         & Dec.         & 3$\sigma$ Radius & Counts & Max S/N & Count Rate                   & $\mathit{f}_{HRI}^{pow.law}$\tablenotemark{(a)} & $\mathit{f}_{HRI}^{max~abs.}$\tablenotemark{(a)} & $\mathit{f}_{PSPC}^{pow.law}$\tablenotemark{(a)} & Time Var.\tablenotemark{(b)} & Opt.id. & NED obj.\tablenotemark{(c)}\\
 No.  & \multicolumn{2}{c}{(J2000)} & (arcsec)         &        &         & ($10^{-4}$ Cnts sec$^{-1}$) &   &  &   & short-term / long-term &         & \\
\tableline
24 & 3:37:58.5   & -35:19:42    & 16.7    & 379$\pm$27   & 24.9  & 24$\pm$2      & 48$\pm 2$              &  61.6 & 75$\pm 7$  & n/y & y & - \\
 
25 & 3:37:34.8   & -35:24:04    & 17.3    & 64$\pm$19    & 5.3   & 4.0$\pm$1.2   &  8.3$\pm 0.4$	   &  10.6 &  -	        & n/$\cdots$ & - & - \\
	
26 & 3:38:43.1   & -35:33:10    & 17.3    & 73$\pm$21    & 5.6   & 4.3$\pm$1.2   &  9.0$\pm 0.5$	   &  11.6 &  -	        & n/$\cdots$ & y & - \\
	
29 & 3:39:43.0   & -35:24:11    & 24.3    & 143$\pm$27   & 8.6   & 10$\pm$2      & 21$\pm 1$	           &  26.7 & 62$\pm 8$	& n/y & y & - \\
	
30 & 3:38:31.0   & -35:14:17    & 26.4    & 586$\pm$38   & 26.5  & 38$\pm$2      & 79$\pm 4$              & 102.1 & -          & 99\%/$\cdots$ & y & FCCB 1263\\
	
31 & 3:38:17.1   & -35:15:57    & 24.5    & 124$\pm$31   & 7.2   & 8$\pm$2       & 16.4$\pm 0.8$	   & 21.1  & -	        & n/$\cdots$ & y & - \\
	
32 & 3:38:25.7   & -35:25:27    & 24.5    & 145$\pm$37   & 6.2   & 8.5$\pm$2.2   & 17.7$\pm 0.9$	   & 22.8  & 27$\pm 7$	& n/n & y & - \\
	
33 & 3:37:36.5   & -35:33:37    & 24.2    & 921$\pm$40   & 41.4  & 60$\pm$3      & 125$\pm 6$             & 161.5 & -          & n/$\cdots$ & y & - \\
	
34 & 3:37:39.4   & -35:34:17    & 21.9    & 426$\pm$31   & 24.2  & 28$\pm$2      & 58$\pm 3$            &  74.5 & -          & n/$\cdots$ & y & - \\
	
35 & 3:37:50.1   & -35:21:27    & 34.6    & 139$\pm$37   & 5.8   & 8.4$\pm$2.3   & 17.6$\pm 0.9$	   &  22.6 & -	        & n/$\cdots$ & y & - \\
	
36 & 3:39:27.8   & -35:18:52    & 34.6    & 129$\pm$35   & 5.6   & 8.6$\pm$2.3   & 18.0$\pm 0.9$	   &  23.2 & 46$\pm 8$	& n/y & y & - \\
	
37 & 3:37:36.3   & -35:14:47    & 49.0    & 153$\pm$44   & 5.1   & 11$\pm$3      & 24$\pm 1$	           &  30.7 & 25$\pm 7$  & n/n & - & - \\
	
38 & 3:37:33.4   & -35:17:17    & 49.0    & 190$\pm$49   & 6.0   & 13$\pm$3      & 27$\pm 2$	           &  35.2 & -	        & n/$\cdots$ & - & - \\
	
39 & 3:37:29.0   & -35:41:11    & 69.3    & 580$\pm$119  & 14.1  & 67$\pm$14     & 140$\pm 7$            & 180.1 & 232$\pm 12$ & n/$\cdots$ & y & - \\
	
40 & 3:38:14.7   & -35:17:57    & 69.3    & 244$\pm$67   & 5.2   & 15$\pm$4      & 31$\pm 2$	           &  40.3 & -	        & n/$\cdots$ & - & - \\
	
41 & 3:38:36.7   & -35:41:32    & 62.9    & 366$\pm$61   & 8.9   & 24$\pm$4      & 51$\pm 3$	           &  65.8 & 57$\pm 7$  & n/n & - & - \\
	
42 & 3:39:12.3   & -35:26:12    & 98.0    & 312$\pm$91   & 4.7   & 19$\pm$5      & 39$\pm 2$	           &  50.7 & -	        & n/$\cdots$ & - & - \\
	
43 & 3:38:12.2   & -35:30:02    & 98.0    & 365$\pm$101  & 5.0   & 22$\pm$6      & 46$\pm 2$              &  59.1 & 60$\pm 14$ & n/n & - & - \\

44 & 3:38:05.1   & -35:19:05    & 12.2    & 42$\pm$13    & 4.6   & 2.5$\pm$0.8   & 5.3$\pm 0.3$	   &   6.8 & -	        & n/$\cdots$ & - & - \\

45 & 3:37:59.9   & -35:23:30    & 17.3    & 56$\pm$18    & 4.4   & 3.3$\pm$1.0   & 6.8$\pm 0.4$	   &   8.8 & -	        & n/$\cdots$ & - & - \\

46 & 03:37:59.2   & -35:11:37    & 49.0    & 134$\pm$42   & 4.4   & 9.9$\pm$3.1   & 20$\pm 1$	           &  26.6 & 28$\pm 8$	& n/n & - & - \\
\tableline
\end{tabular}
\tablenotetext{(a)}{ Flux in the 0.5-2.0 KeV band in units of $10^{-15}$ erg s$^{-1}$ cm$^{-2}$.}
\tablenotetext{(b)}{\small ~Short-term = probability of the Kolmogorov-Smirnov test: `n' means no variability detected at the 95\% level.\\
		   ~Long-term = difference between PSPC and HRI counts: `y' means variable source ($>3\sigma$), `n' means no variability detected, `$\cdots$' means no available data.}
\tablenotetext{(c)}{\small ~CGF=\cite{Hilk99b}; FCCB=\cite{Smith96}}
\end{table}
\end{landscape}
\topmargin=0in

We tried to determine which X-ray sources belong to the Fornax cluster both by visual inspection of the photographic DSS plates and by comparing their positions with those of published catalogs of the Fornax region.
Sources No.2 and 11 are likely to belong to NGC 1399 because of their position within the optical effective radius ($r_{eff}=44.7"$, \citealp{gou94}) while sources No.16 and 32 appear to be coincident with the position of two globular clusters \citep{Kiss99}. 
Two sources are clearly associated with background objects: source No.4 corresponds to a background galaxy located at $z=0.1126$ and represents the X-ray counterpart of the strongest component of the radio source PKS 0336-35 \citep{car83}; source No.30 is coincident with two optical background galaxies classified as interacting by \cite{Ferg89} and clearly visible on the DSS plates.
For the remaining sources there is no clear indication of whether they belong to the Fornax cluster or not.  

Sources No.8, 12, 19, 21, 30 are variable at the 95\% confidence level, according to our KS tests. Source No.30
is associated with a short X-ray burst occurred in the background galaxy between January and August 1996. In fact the source is not detected in both the August 1991 PSPC (RP600043n00) and the February 1993 HRI observations (RH600256n00). It is present, instead, in the last two HRI observations (RH600831n00 and RH600831a01) with a decrease in the observed count rate of a factor 8 between the two. 

We must notice that our HRI background level was also variable so that the presence of short-term temporal variability was  clearly detect only for the strongest sources ($\mathit{f}_{HRI}^{pow.law}\geq 3\times 10^{-14}$ erg s$^{-1}$ cm$^{-2}$). However, since the background subtraction was performed locally (i.e. extracting the background in a region surrounding the source, see \citealp{Dam97a}), this variability does not affect the source counts reported in Table \ref{source_tab}. To further check the reliability of the wavelets algorithm, we extracted the counts manually for each source using a local annulus to measure the background, finding consistent results (within 1$\sigma$) with those derived by the wavelets technique.

We cross-checked our source list with the list of sources detected on the PSPC image by the GALPIPE project (see $\S$\ref{model}). Several sources detected in the HRI field were bright enough and sufficiently isolated to allow a comparison of the PSPC flux with the HRI results and to check for long-term variability. We converted the PSPC count rates to fluxes in the 0.5-2.0 keV band (1 cps=6.397$\times 10^{-12}$ ergs cm$^{-2}$ s$^{-1}$), 
assuming the same power law spectral model  used for converting the HRI counts (see below). The PSPC fluxes and long-term variability (i.e. when PSPC and HRI flux estimates differ more than 3$\sigma$) are included in Table \ref{source_tab}.

This comparison also confirmed that sources No.39 and 46, both near the very edge of the HRI FOV, are not an artifact of the poor S/N ratio or the steep emission gradient because both are present in the PSPC data as well and detected with high significance.

\begin{figure*}[p]
\centerline{\psfig{figure=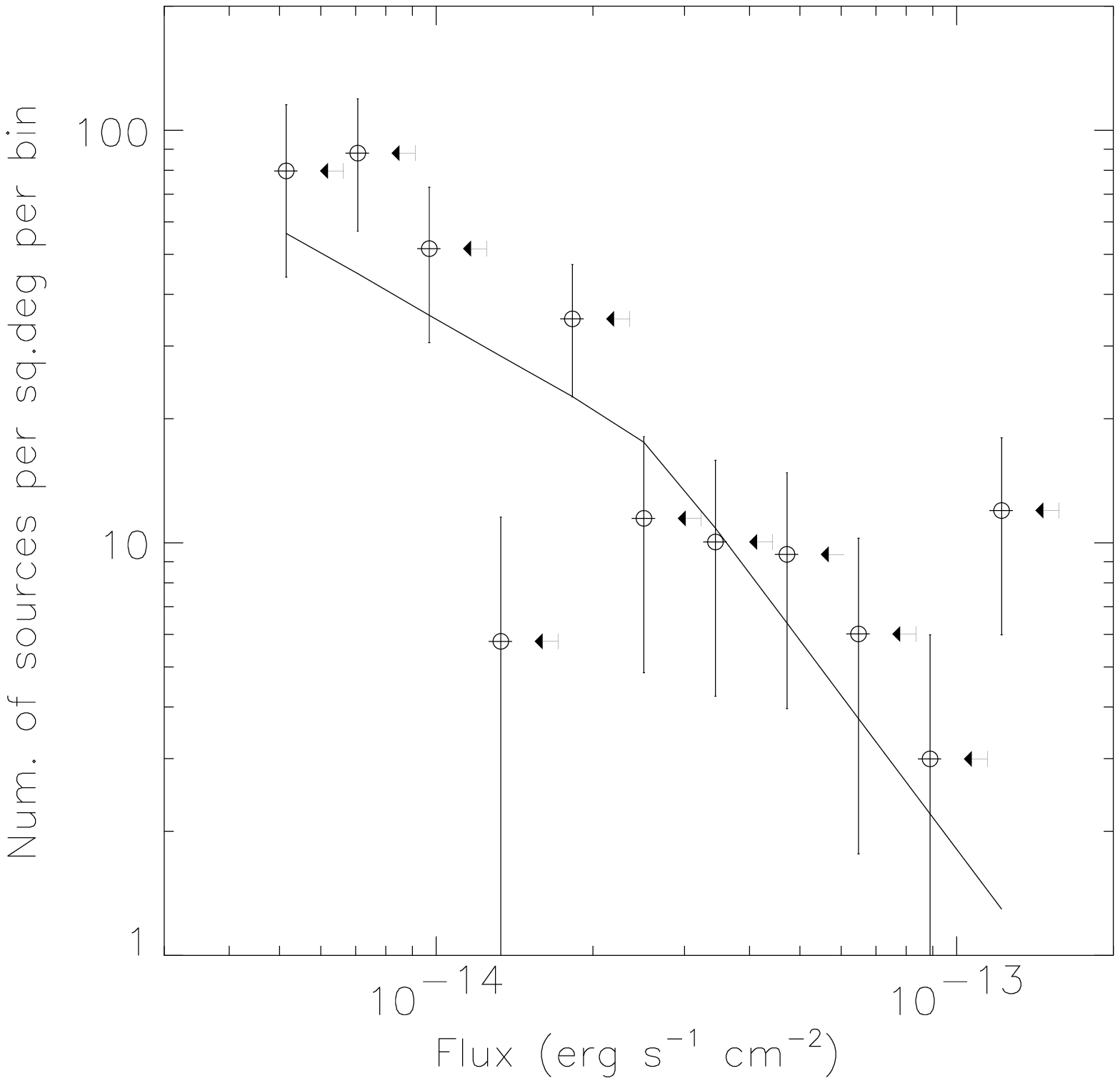,angle=0,width=0.54\textwidth}\hspace{-1.5cm}\raisebox{7.2cm}{\parbox[b]{1cm}{\bf(a)}}\hspace{1.7cm}}
\centerline{\psfig{figure=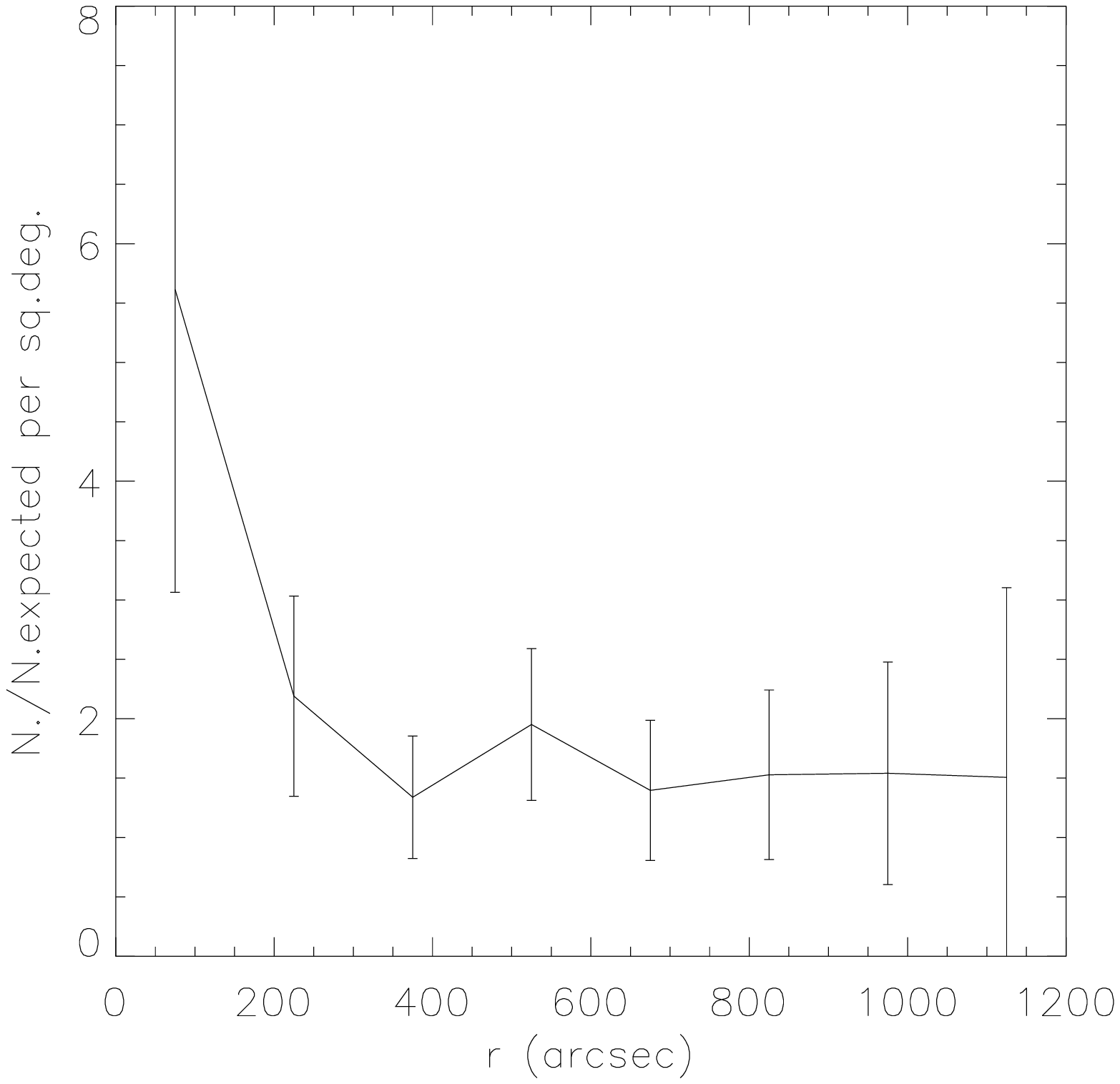,angle=0,width=0.52\textwidth}\hspace{-1.5cm}\raisebox{7.2cm}{\parbox[b]{1cm}{\bf(b)}}\hspace{1.5cm}}
\caption{{\bf (a)} Measured vs expected source counts. Measured counts, corrected for sensitivity variations across the detector, are represented by empty circles, while the continuous line is the expected distribution from \cite{Has93}. Horizontal error bars take into account uncertainties in the power law slope. Arrows represent upper limits obtained fixing the absorption to $4.5\times 10^{20}$ cm$^{-2}$ \citep{rang95}.{\bf~(b)} Radial excess of sources over the expected number from \cite{Has93}. The excess is concentrated around NGC 1399, thus suggesting to be due to sources associated with the galaxy.}
\label{src_flux}
\end{figure*}

We compared our source flux distribution to the expected background counts obtained from a deep survey of the Lockmann hole \citep{Has93}. We used the sensitivity map generated by the wavelet algorithm to weight each source for the effective detector area above the corresponding limiting flux. In this way we are able to normalize the measured counts at each flux to the total surveyed area.
The flux distribution is shown in figure \ref{src_flux} where our differential counts are compared to the Hasinger ones. The horizontal error bars take into account uncertainties in the the power law slope. Our counts are higher than the Hasinger estimate, giving a total of $312\pm 47$ sources vs $230\pm 15$ expected counts, but still compatible within 2$\sigma$. To take into account absorption uncertainties we estimated a lower limit on the number of expected sources adopting the higher absorption value found by RFFJ in the NGC 1399 field (see above). In this case the number of expected counts decreases to $184\pm 14$ leading to an upper limit of the observed excess of $\sim 2.6\sigma$. 

The spatial distribution of detected sources is not uniform. In Figure \ref{src_flux} we show the source excess in circular annuli centered on NGC 1399. We find that the excess is peaked on the dominant galaxy, suggesting that the inner sources are likely to be associated with NGC 1399 rather than being background objects. In fact, excluding the central 300'' (28 kpc), the measured counts drop to $244\pm37$, in agreement within less than $1\sigma$ from the expected value. 
The X-ray luminosity of the sources within the central 300" range from $1.7\times 10^{38}$ erg s$^{-1}$ to $1.4\times 10^{39}$ erg s$^{-1}$, if they are in NGC 1399. Thus all of them have luminosities in excess of $1.3\times 10^{38}$ erg s$^{-1}$, i.e. the Eddinghton luminosity for a 1 $M_\odot$ neutron star ($\S$ \ref{discrete}), suggesting that they may have massive Black Hole companions if they are accretion binaries.

\subsection{The {\it Chandra} data}
\label{Chandra}
While this work was in progress the {\it Chandra} data on NGC 1399 became public. 
A full data reduction of the {\it Chandra} data, which is however already in progress \citep{Ang01,Loew01}, is beyond the aim of the present work.
However we decided to compare the ROSAT data with the {\it Chandra} image to see if our conclusions are supported by the improved {\it Chandra} resolution and sensitivity.

\begin{figure}[t!]
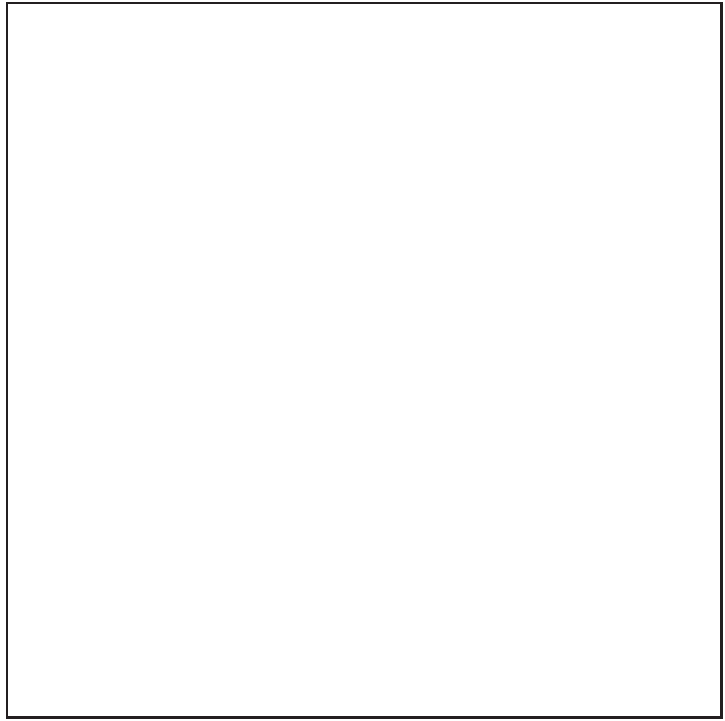

\vspace{1.5cm}
\centerline{\framebox(270,270)[br]{}}
\vspace{2cm}
\caption{Adaptively smoothed {\it Chandra} ACIS-S image of NGC 1399 in the energy range 0.3-10 keV. Colors from black to white represent count rates from 0.021 cnts s$^{-1}$ arcmin$^{-2}$ to 7.7 cnts s$^{-1}$ arcmin$^{-2}$. Circles represent the position of sources detected by the wavelets algorithm in the HRI image.}
\label{Chandra_csmooth}
\end{figure}

We used the longest {\it Chandra} observation of 60 ks (Obs. ID 319, 55.942 ks live time), performed on 2000 January 18 with the ACIS-S detector.
We produced a 0.5 arcsec/pixel image in the energy band 0.3-10 keV, from the ACIS-S3 chip, which covers the central $8\times 8$ arcmin of NGC 1399.
The image was then adaptively smoothed with the {\it csmooth} algorithm included in the CIAO package. The final image is shown in Figure \ref{Chandra_csmooth}.
No exposure correction was applied because the {\it csmooth} algorithm works better on the uncorrected data. However the exposure map is quite flat and we further checked that any feature present in the exposure map does not affect our conclusions. 

Figure \ref{Chandra_csmooth} shows a large number of pointlike sources present in the NGC 1399 field, most of which were unresolved in the HRI image. We marked with circles the sources detected in the HRI field (see $\S$ \ref{sources}).
The {\it Chandra} data show that sources 2, 11 and 5, identified as single sources in the HRI image, are instead multiple sources. 

The filamentary structures found in the HRI image (Figure \ref{centbox}) are present in the {\it Chandra} image as well. The `arm' protruding on the Western side and the `voids' in the X-ray emission are clearly seen. The hypothesis that the filamentary structures may be due to the presence of a large number of pointlike sources unresolved in the HRI image, is ruled out by {\it Chandra} data. In fact the {\it csmooth} algorithm clearly separates point sources from the extended structures (Figure \ref{Chandra_csmooth}) and shows that the two components are not spatially correlated.

\begin{figure}[t!]
\centerline{\psfig{figure=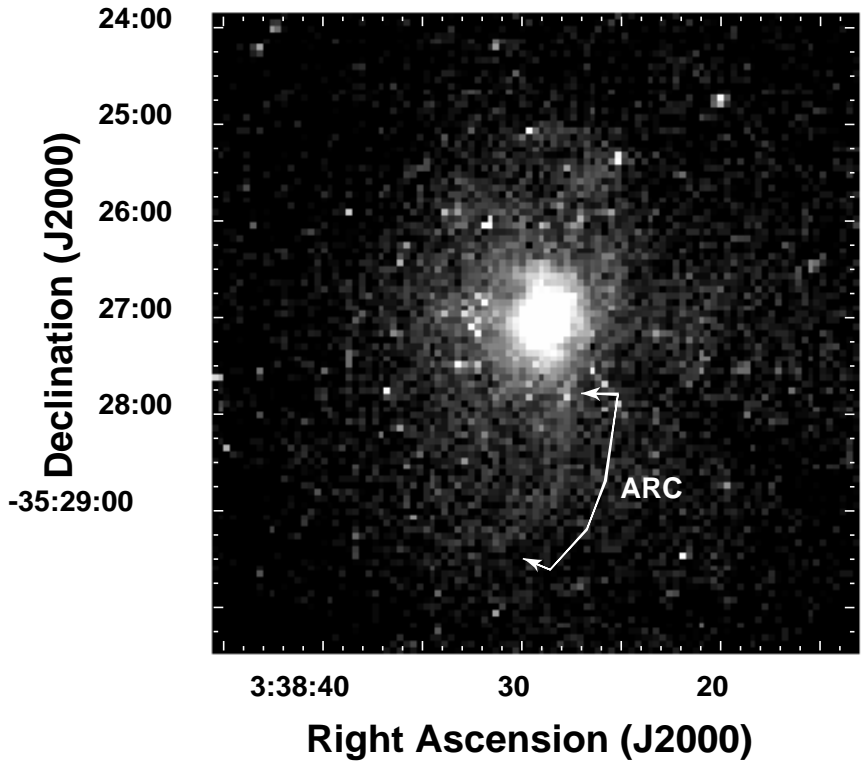,angle=0,width=0.75\textheight}}
\caption{Raw 4 arcsec/pixel {\it Chandra} image of NGC 1399. The long arc following the Western edge of the radio lobe is visible South of NGC 1399.}
\label{chandra_arc}
\end{figure}

The correspondence between the halo features and the radio jets/lobes, discussed in $\S$ \ref{Xradio}, are confirmed. Moreover the {\it Chandra} data revealed the presence of a long arc extending South that follows closely the Western edge of the radio lobe. The arc is better visible on the raw {\it Chandra} image shown in Figure \ref{chandra_arc}. This feature has a low statistical significance ($2\sigma$) but the correlation with the radio lobe argues in favor of a real structure.
\vspace{1cm}

\section{DISCUSSION}
\label{results}

The study of galactic and cluster gaseous halos has often been based on the assumption of homogeneity and spherical symmetry. The deep HRI image of the NGC 1399 halo has shown instead significant departures from these simplified assumptions. Here we briefly summarize the result of our analysis, and then discuss their physical implications.

We model the X-ray brightness distribution of NGC 1399 with three distinct components: (1) a central component (dominating for $r<50"$), coincident with the optical galaxy; (2) a `galactic' component ($50"<r\leq 400"$) displaced South-West of NGC 1399 and (3) an ellipsoidal `cluster' component ($r>400"$) centered North-East of the optical galaxy. The cooling times of the central and galactic components are smaller than the Hubble time, suggesting the presence of a central cooling flow. The binding mass estimates, obtained assuming hydrostatic equilibrium, revealed that the dynamics of the central component are dominated by stellar matter while, at radii larger than 2 arcmin there is a large dark matter halo with two distinct components dominating respectively on galactic and cluster scales.  
 
We also found significant filamentary structures and cavities in the X-ray emission. Some of these structures are correlated with the morphology and the position of the radio jets and lobes of NGC 1399, suggesting interactions between the radio emitting plasma and the hot gas.
 
Finally we used a wavelets based algorithm to detect 46 sources in the HRI field, 8 of which show significant time variability. Their distribution is consistent with the background source distribution at the $3\sigma$ level.  However there is excess over background counts centered on NGC 1399. The presence of a large number of pointlike sources near the galaxy center is further confirmed by {\it Chandra} data.

\subsection{Origin of the Central X-ray Peak}
\label{inner}

The center of the gaseous halo is occupied by a strong emission peak (see Figure \ref{csmooth}a) centered on the optical galaxy. Such emission is often found in cooling flow galaxies (\citealp{Can83,Kim95}; see also \citealp{fab91}) because of the higher density of cooler hot gas in the galactic center. RFFJ revealed the presence of a cooler region within the central 1.5 arcmin (14 kpc); this is compatible with the extent of the region dominated by the X-ray peak, as shown by the radial profile in Figure \ref{model_prof}. Moreover both the central and galactic components have cooling times below the Hubble time (assumed to be $\sim 10^{10}$ years) within respectively 250'' (23 kpc) and 350'' (32 kpc), so that significant cooling must have taken place.
The cooling time of the cluster component, instead, is compatible with the Hubble time, suggesting that the cooling flow is mainly of galactic origin.

The alternative hypothesis that the central peak can be due to the presence of a nuclear source \citep{jones97} is clearly ruled out by our data, because the observed profile (Figure \ref{profile}) is wider than the HRI PRF even taking into account the PRF uncertainties. The difference between the optical (with slope $\alpha\sim -1.6$, \citealp{Hilk99a}) and X-ray profiles ($\alpha\sim -2$ for the central component, see $\S$ \ref{rad_prof}) seen in Figure \ref{pies} also suggests that is unlikely that the central peak could be due to stellar sources. \cite{allen00} analyzing ASCA data, found that the addition of a power-law component to their NGC 1399 spectral model significantly improves the quality of the fit. They notice that the luminosity of the power-law component is compatible with the contribution expected from a population of unresolved hard X-ray binaries. However, the HRI count rate that would be produced by this power-law component falls more than an order of magnitude below the inner component
count rate. We can thus exclude that this hard discrete population dominates the central component emission.

The gravitational mass estimate derived from the hydrostatic equilibrium (equation \ref{mass}) shows that the hot gas within 100" follows closely the stellar mass profile. This result is independent on the assumed temperature profile (Figure \ref{mass_prof}) being mostly determined by the hot gas density profile. These X-ray data are thus in excellent agreement with the optical mass determinations of \cite{saglia00} who find no evidence of dark matter in the galaxy core.

The origin of the central density enhancement may be due to hot gas inflow within the inner 1 arcmin. The deprojection technique described in $\S$ \ref{dens_par} allowed to derive the cumulative gas mass profile of the central component (Figure \ref{mass_prof}). According to the  mass deposition rate measured by RFFJ, which is in good agreement with our data, we expect to find $\sim 2.5\times 10^{9} {\rm ~M}_\odot$ of deposited gas within the central 60'' ($\sim 5$ kpc) over $10^{10}$ yr, while we only observe $\sim 2\times 10^8 {\rm ~M}_\odot$ in the hot phase. Taking into account a possible larger than line-of-sight central absorption, the observed mass would increase of $\sim 30\%$, but the deposition rate would also increase (see RFFJ), so that the result remains almost unchanged. 
This difference suggests that $\sim 90\%$ of the hot gaseous mass that reached the inner regions of NGC 1399 has cooled enough to become invisible in X-rays. However RFFJ argue that the missing mass is not seen in any other phase.
The situation gets even worse considering that we are ignoring stellar mass losses which must be considerable since the metallicity of the ISM in the galactic center is consistent with stellar values \citep{Buo99,Mat00}. If the age of the cooling flow is lower than $10^{10}$ yr, this disagreement would be reduced. Equating the observed hot gas mass to the amount expected from the cooling flow, we estimate a lower limit of $10^9$ years to the  age of the cooling flow. 

We must notice that the latest studies of cooling flows in clusters and galaxies with Newton-XMM are revealing that the classical cooling flow model is not able to reproduce the observed temperature distributions \citep{Pet01,Kaa01,Tam01}.
\cite{Matsu01} has shown that taking into account only the central region ($r< 1.5 r_e$) of NGC 1399, $\beta_{spec}\sim 1$ indicating that the gas temperature is in good agreement with the stellar velocity dispersion ($\S$ \ref{beta}). This result suggest a correlation of the central component with the stellar population. Moreover \cite{Maki01} proposed that the cooler halo component found in the center of galaxy clusters is due to the structure of the gravitational well around cD galaxies, rather than to the presence of a cooling flow. 
An alternative to the cooling-flow scenario is thus emerging, which may provide a good explanation for the presence of the central emission peak. We will discuss this possibility in detail in section \ref{centralpeak_discuss}.

\subsection{Dynamical Status of the Halo}
\label{NGC1399_dynamical}
\subsubsection{Gravitational Effects}
The X-ray brightness profiles (Figure \ref{pies}) shows that past 1 arcmin the hot gas distribution is more extended than the stellar distribution. If the gas is in hydrostatic equilibrium, additional matter is required to explain this broader distribution. The change in slope of the X-ray brightness profile for $r>400"$ further requires the presence of two components that we identified with the `galactic' and `cluster' halo. This complex distribution was previously found by \cite{ikebe96} using the ASCA data, and explained in terms of a different concentration of dark matter on galactic and cluster scales \citep{ikebe96,Maki01}. Our data further show that this galactic component is not correlated with stellar matter, which is even more concentrated (Figure \ref{mass_prof}b) and dominates the dynamics of the galactic core.

The kinematics inferred from the GCs and PNe (see Figure 6 of \citealp{Kiss99})  indicate that for $1'<r<16'$ (5 kpc$<r<$90 kpc) from the galaxy center, the velocity dispersion increases with radius. This result is consistent with our measurements and suggests either the presence of a significant amount of dark matter associated with the outer radii of NGC 1399 or equivalently that the outer galactic dynamics is dominated by the cluster potential.

We must notice however that while on galactic scales the mass profiles may be explained by the presence of dark matter, the deviations from symmetry, i.e. the offset between the stellar component (consistent with the central component) and the galactic dark halo, are hardly explained in the galaxy formation framework. For this reason in next sections we take into account possible departures from equilibrium.

\subsubsection{Galaxy-Galaxy Interactions}
\label{interactions}
An alternative scenario has recently been advanced by Napolitano, Arnaboldi \& Capaccioli (2001, in preparation), that could explain our results by means of galaxy interactions.
 Their calculations show that the energy injection due to galaxy-galaxy encounters can account for the outer galaxy dynamics as seen in PNe and GCs, and is in agreement with the observed X-ray temperature profile. In this case there would be no need for additional dark matter in the `galactic' halo. 
Support for the tidal interaction picture comes from \cite{Kiss99} who showed that the GC system of NGC 1399 is 10 times more abundant than those of neighbors galaxies, while their properties are very similar. In the hypothesis that all galaxies initially had the same number of GCs per unit luminosity, the excess around NGC 1399 may be compatible with tidal stripping of GCs from nearby Fornax galaxies.  

The effects of tidal interactions on the gaseous halos of elliptical galaxies have been explored by \citet[ hereafter DRC]{D'Ercole00}. They found that the interaction strongly affects the gas flow, altering the ISM density and thus increasing the cooling speed and luminosity. This mechanism is enhanced when energy injection from SNe Ia is included and the gas remains turbulent independently from the epoch of the encounter. In this process the surface brightness is scarcely affected and the global isophotes remain circular, although locally distorted. 
The DRC model predicts that the gas inflow produces steeper profiles in the central regions of the galaxy in the moment of maximum luminosity, indicating that the tidal perturbations may contribute to the formation of central brightness peaks such as the one seen in the NGC 1399 core. The steep slope of the galactic component (Table \ref{fit_tab}) too may be due to tidal gas stripping because the galaxy undergoes significant mass losses (up to 50\%) in the outer regions.
However, we must notice that in the case of NGC 1399, the effect of tidal stripping may be smaller than what expected from the DCR models because PSPC data \citep{jones97} suggest a lower SN rate than assumed by D'Ercole and collaborators.

For these models to be consistent with the X-ray observations, tidal interaction must strongly affect the halo density distribution: the increase in the M/L ratio past 2 arcmin (Figure \ref{mass_prof}) is mainly determined by the shallower slope of the gas density profile in the outer galactic regions, since the isothermal model still exhibits the multi-component behavior ($\S$ \ref{dens_par}). Thus a simple temperature increase would not produce the observed agreement with the optical data. Such strong profile alterations are not seen in the DRC simulations. Conversely \cite{bar00} found that encounters between galaxies of different masses can produce both radial profiles with a central peak and a `shoulder' similar to our multi-component structure and azimuthal asymmetries as those seen in the NGC 1399 galactic halo. However his simulations also show that the stellar profile follows the gas profile, which is not seen in our data. 

The X-ray surface brightness analysis ($\S$ \ref{brightness}) revealed the presence of complex structures in the galactic halo. We have shown in $\S$ \ref{model} that these brightness fluctuations are not due to artifacts of the smoothing algorithm, because they are present in the residuals of the bidimensional model, and have high statistical significance. Due to the weak dependence of the HRI spectral response on the temperature around 1 keV (see \citealp{clar97}), they must be due to local density fluctuations of the hot gas. While some of these features can be due to the interaction between the radio jets and the surrounding ISM ($\S$ \ref{Xradio}), most of them are not related to any evident radio or optical feature and are spread all around the galaxy, suggesting a `disturbed' nature of the whole halo.

It is possible that these structures are the signature of interaction between the NGC 1399 halo and other Fornax galaxies. In fact DRC predict significant density fluctuations in the hot gas and the formation of filamentary structures near the galaxy center as a consequence of galaxy-galaxy encounters. It's not easy to determine if these perturbations are due to the nearby NGC 1404 or to more distant members because, according to DRC, the perturbing effects lasts several Gyrs. We searched for possible signatures of interactions between NGC 1399 and NGC 1404 looking for excess emission in the region between the two galaxies. The slight excess we found is only significant at the 2$\sigma$ level, so that no significant conclusion can be drawn from our data.

\subsubsection{Ram Pressure}
A different scenario that could explain the cluster and galactic halo displacement is one in which NGC 1399 is moving in the Fornax cluster potential, thus experiencing ram pressure from the cluster hot gas. The radial profiles in Figure \ref{pies} show a high degree of asymmetry with respect to the optical galaxy. The 2D models indicate that the galactic component is displaced $\sim 1'$ (5 kpc) South-West with respect to the X-ray peak (and thus optical galaxy), while the cluster component seems located $\sim 5.6$ arcmin (31 kpc) North-East of NGC 1399. \cite{jones97} argue that this displacement is compatible with the hypothesis that the galaxy is slowly moving (40 km s$^{-1}$) South-East in a larger potential. This would explain the displacement of the galaxy with respect to the cluster component but not the one of the galactic halo. In the ram pressure scenario, the displacement and large scale asymmetry of the `galactic' halo may suggest that NGC 1399 is not sitting at the very center of the cluster potential and is slowly moving North-East toward the `cluster' component.

The effect of ram stripping has already been observed in nearby galaxies. For instance \cite{irw96} observed the presence of a very irregular halo in the Virgo galaxy NGC 4472. They found a steeper gradient in the X-ray emission in the direction of the cluster center associated with a temperature enhancement. They calculate that ram stripping is able to explain the observed halo features if the galaxy is moving at $\sim 1300$ km s$^{-1}$ in the cluster potential. Their ROSAT HRI images also revealed filamentary structures and holes in the galactic halo, very similar to the features that we observed in the adaptively smoothed image of the inner halo (Figure \ref{centbox}). 
Additional evidence that galaxies moving in the cluster environment can produce filaments of cooler gas behind them were found by \cite{Dav94} or \cite{fabian01}. 

The accurate simulation of dynamic stripping in cluster and group elliptical galaxies performed by \cite{toni01}, show that the effect of dynamic stripping depends on the galaxy orbit and the duration of the subsonic and supersonic phases which determine the balance within gas losses and accretion onto the galaxy. In general their simulations reveal the formation of complex and elongated tails behind the galaxy (similar to those seen in NGC 4472) and strong decentering of X-ray isophotes due to supersonic stripping.
Such irregular structures are not seen in the NGC 1399 galactic halo which seems instead rather circular ($\epsilon_{Galactic}=0.02$, see Table \ref{fit_tab}).

A very rough estimate of the velocity of NGC 1399 can be obtained supposing that the displacement of the optical galaxy, coincident with the X-ray centroid, with respect to the galactic halo is due to the deceleration of the halo caused by ram pressure from the cluster component. In the hypothesis that the motion is highly subsonic, we can treat the different components as homogeneous spheres and calculate the initial velocity required to produce the observed displacement after a time $t=10^9$ yr due to the ram pressure of the cluster halo $P\sim \rho_{cluster} v^2$. Using the parameters derived form the bidimensional model ($\S$ \ref{model}): $\rho_{galaxy}/\rho_{cluster}\sim 4$, $R_{galaxy}\sim 4'$ and a displacement $\Delta S\sim \frac{1}{2} R$, we obtain an initial velocity of $\sim 10$ km s$^{-1}$, consistent with the initial assumption of subsonic motion.
Even considering a more recent interaction where $t=10^8$ the galaxy motion remains highly subsonic ($\sim 100$ km s$^{-1}$). 
This result is in agreement with the observations, showing that the halo morphology is only weakly distorted. 
In conclusion, even though extreme projection effects may hide stronger  stripping features, ram pressure may be causing the galaxy halo displacement but seems unlikely to be responsible for the observed inner halo structure of NGC 1399.

\subsubsection{Cluster Merging effects}
For what concerns the elongated cluster halo component, in many cases irregular features in the cluster halos seems associated with merger events (see for instance \citealp{sun00,mazz01}). Simulations of cluster merging \citep{roett96} show that elongation and twisting of the X-ray isophotes are a general result of such interactions. In the case of the Fornax cluster \cite{drink00} found evidence of a subcluster 3 degrees South-West of NGC 1399 centered on NGC 1316 (Fornax A). They claim that the two components are at the early stage of merging; in this case merging is unlikely to be the cause of the elongation of the cluster X-ray isophotes. To further study the dynamical status of the cluster halo we would need temperature maps of the whole cluster, as can be obtained with {\it Chandra} or XMM.

It is worth noticing that the X-ray centroid of the cluster component is displaced with respect to the optical distribution of Fornax cluster galaxies, which tend to concentrate South-West of NGC 1399 (see Figure 1 from \citealp{Hilk99b} and Figure 3 from \citealp{drink00}). If the merging picture is confirmed, this may imply that the galaxies in the two subclusters are infalling in the global potential well faster than the associated gaseous halos.

\cite{Kim95} compared three X-ray dominant galaxies in poor groups: NGC 1399, NGC 507 and NGC 5044. They show that even if these galaxies have similar properties (temperatures, cooling core) their X-ray brightness profiles are very different. This suggests that the shape of the potential and the mass deposition mechanism may vary from galaxy to galaxy. If environmental effects (tidal interactions, ram pressure stripping) are important factors in determining the NGC 1399 halo structure, they must be considered as well in order to explain the observed differences.

\subsection{X-ray/Radio Interactions}
\label{Xradio_discuss}

Signs of interaction between radio jet/lobes and the surrounding gaseous medium have been reported in powerful radio galaxies (see for instance \citealp{car94} for Cygnus A; \citealp{bor93,McNam00,Fab00} for evidence of such interaction in clusters hosting powerful radio sources).
NGC 1399 is a faint radio galaxy whose radio lobes are entirely contained within the optical galaxy. Nevertheless the structures seen in Figure \ref{X-radio} suggest that the same mechanisms at work in more powerful sources may be responsible for some of the galactic halo structures. 
The residual image in Figure \ref{X-radio_nuclear} shows excess emission on both sides of the nucleus and a lack of X-ray emission along the radio jets in the inner 15" ($\sim 1.4$ kpc), as if the radio jets are displacing the ambient gas. For $r\sim 25"$ ($\sim 2.3$ kpc) more X-ray emission is seen in the North and South part of the galaxy, aligned with the direction of the radio jets. 

\cite{clar97} have studied, by means of hydrodynamical simulations, how these features are produced during the propagation of the radio jet trough the ICM. They show that, for a $\sim 1$ keV gaseous halo and a detector response similar to the ROSAT HRI (see their Figure 8), we expect to see excess emission located on the side of the jets, due to the shocked gas near the tail of the radio lobes, as the one seen in  Figure \ref{X-radio} (region {\bf A}) and in Figure \ref{X-radio_nuclear}. 

In coincidence with the radio lobes, X-ray enhancements or deficits are expected depending on the thickness of the shocked gas $\Delta r$ with respect to the lobe radius $r$. Near the edge of the radio lobe, where $r\geq 3.4\Delta r$, the shocked gas density is high enough to produce a brightness enhancement. On the contrary, in the tail of the lobes the compressed gas had time to expand, so that $r<3.4\Delta r$ and the X-ray brightness is diminished, producing a `cavity'. This picture is very similar to what we see in our data.
Features {\bf B} and {\bf C} in Figure \ref{X-radio} may suggest respectively the presence of a cavity produced by the Southern lobe in the ICM and the excess emission associated with the edge of the lobe. The alternative hypothesis, that the excess is due to the presence of a bow shock at the leading end of the lobe, seems unlikely both because Carilli and collaborators showed that the bow shock is hardly seen at energies $< 4$ keV and because the signatures of bow shocks are usually sharper than the one seen in our data, showing up as compact bright spots or arcs.  The {\it Chandra} image also revealed the presence of a long arc South of NGC 1399. The steeper gradient of the radio emission on the Western side of the Southern lobe, with respect of the Eastern side, suggests that this feature may represent a layer of hot gas compressed by the radio emitting plasma.

Signatures of interaction with the ICM do not show up so clearly in the Northern lobe, but this could be due to projection effects (see for instance \citealp{bru01,fino01}). The brightness enhancement near the West side of the lobe could be due to the thin layer of compressed
gas surrounding the lobe while the steeper radio gradient near region {\bf D} suggests that the lobe is avoiding a region of higher gas density. Indeed the crude estimate  indicates that the pressure of the bubble ($\S$ \ref{Xradio}) can be up to 4 times higher than that of the surrounding ISM. An upper limit of $k\simeq 420$ on the ratio of the electron to heavy particle energy (or alternatively a lower limit of 0.005 on the filling factor) can be derived by matching the radio pressure to the thermal pressure of the bubble (Figure \ref{press_prof}), in agreement with \cite{kbe88}.

\begin{table}[t]
\caption{Southern lobe magnetic field strenght.\label{IC_tab}}
\begin{center}
\begin{tabular}{cccccccc}
\\
\hline
\hline
$\eta$ & $\phi$ & $\nu_1$ & $k$ & $S_X$ & $B_{ME}$ & $B_{IC}$ & $B_{ME}/B_{IC}$\\
& (rad) & (GHz) & & (ergs s$^{-1}$cm$^{-2}$Hz$^{-1}$) & ($\mu$G) & ($\mu$G) & \\
\hline
1   & $\pi /2$ & $1.00\times 10^{-2}$ & 1 & $2.90\times 10^{-32}$ & 4.90 & 0.37 & $13.3$\\
1   & $\pi /6$ & $1.00\times 10^{-2}$ & 1 & $2.90\times 10^{-32}$ & 6.59 & 0.54 & 12.2\\
1   & $\pi /2$ & $1.00\times 10^{-2}$ & 100 & $2.90\times 10^{-32}$ & 15.0 & 0.37 & 40.9\\
1   & $\pi /2$ & $1.00\times 10^{-4}$ & 1 & $2.90\times 10^{-32}$ & 4.90 & 0.37 & $13.4$\\
0.1 & $\pi /2$ & $1.00\times 10^{-2}$ & 1 & $2.90\times 10^{-32}$ & 9.46 & 0.37 & 25.7\\
0.01& $\pi /2$ & $1.00\times 10^{-2}$ & 1 & $2.90\times 10^{-32}$ & 18.3 & 0.37 & 49.7\\
1   & $\pi /2$ & $1.00\times 10^{-2}$ & 1 & $3.64\times 10^{-33}$ & 4.90 & 1.16 & $4.21$\\
1   & $\pi /6$ & $1.00\times 10^{-2}$ & 1 & $3.64\times 10^{-33}$ & 6.59 & 1.7 &  3.85\\
\hline
\end{tabular}
\end{center}
\tablenotetext{}{{\bf Note} - The calculation uses equations (\ref{B_IC}) and (\ref{B_ME}) assuming the following parameters: radio spectral index $\alpha_r=-0.8$, radio flux density $S_r=0.0825$ Jy at $\nu_r=1.465$ GHz, lobe diameter $\theta=28"$ and path lenght through the source $s=25.7$ kpc. The X-ray flux $S_X$ density is calculated at 1 keV assuming a galactic absorption of $1.3\times 10^{20}$ cm$^{-2}$; the upper syncrotron cutoff is taken to be $\nu_2=100$ GHz. }
\end{table}

As discussed in $\S$ \ref{IC}, Inverse Compton (IC) scattering of Cosmic Background photons by the radio emitting particles is often considered as an alternative process to explain the X-ray brightness enhancements near radio lobes \citep{Har79,Kan95}. We have evaluated this possibility for the X-ray excess near the edge of the Southern radio lobe of NGC 1399, following \cite{Fei95}, and comparing the minimum energy field strength $B_{ME}$ (equation \ref{B_ME}) with the magnetic field $B_{IC}$ (equation \ref{B_IC}) required to produce a measured ratio of radio to X-ray flux. We estimated the ratio $B_{ME}/B_{IC}$ for a grid of values of the involved parameters:  the filling factor $\eta$ of the syncrotron emitting region, the angle $\phi$ between the magnetic field (assumed to be uniform) and the line of sight, the lower cutoff frequency $\nu_1$ of the radio syncrotron spectrum,  the ratio $k$ of energy in heavvy particles with respect to electrons and  the X-ray flux density $S_X$. We used the radio flux for the Southern lobe measured by \cite{kbe88}, corrected for the fact that just half of the lobe contributes to the X-ray excess. The results are shown in Table \ref{IC_tab}. The X-ray flux was measured  from the residuals shown in Figure \ref{residuals}, assuming that all the excess over the 2D model is due to IC scattering; this value may vary of $~25\%$ depending on the region used for the extraction, so that we included both the upper ($3.64\times 10^{-33}$ ergs s$^{-1}$cm$^{-2}$Hz$^{-1}$) and lower ($2.90\times 10^{-32}$ ergs s$^{-1}$cm$^{-2}$Hz$^{-1}$) flux limits in Table \ref{IC_tab}. 
We find that it is unlikely that more than a few percent of the excess near the edge of the Southern radio lobe may be due to IC scattering, since the ratio $B_{ME}/B_{IC}$ is always $>1$ for any choise of the parameters involved in the calculation.

Pressure confinement of the radio lobes \citep{kbe88} is supported by our data (Figure \ref{press_prof}). If the inclination of the radio source is $< 45^\circ$, the radio jets remain collimated up to the radius within which the high density central component dominates ($\sim 50"$, Figure \ref{press_prof}) and diffuse in lobes as soon as they enter in the lower density environment. A larger inclination is unlikely because the thermal pressure of the cluster component (long-dashed line in Figure \ref{press_prof}) is of the same order of magnitude as the radio lobes pressure, and would probably not be able to confine the radio emitting plasma, if this was extending at greater distances.

Finally, we analyzed the possibility that the power-law component found in the spectral model of \cite{allen00} for NGC 1399 (see $\S$ \ref{inner}) may be due to an active nucleus. Their estimated flux (converted to our adopted distance) is twice the value expected from our upper limit on a possible nuclear source, adopting their best fit parameters. Considering that \cite{Loew01} further reduce this upper limit by an order of magnitude it appears that the bulk of the  power-law emission comes from the discrete source population revealed by the {\it Chandra} data (see Figure \ref{Chandra_csmooth}).

\subsection{Discrete sources}
The presence of a population of individual X-ray sources in E and S0 galaxies was first suggested by \cite{Tri85}. Additional evidence was then provided by ROSAT and ASCA observations (e.g. \citealp{Fab94,Kim96,Matsumo97}). These X-ray source populations are now been detected down to much lower luminosities with {\it Chandra} (e.g. \citealp{Sar00,Sar01}). The HRI data demonstrate the presence of such population, possibly of accretion binaries, also in NGC~1399. This result is confirmed by the {\it Chandra} data which reveal a large number of pointlike sources surrounding the galaxy, many of which associated with globular clusters \citep{Ang01}. This population of accretion binaries is probably responsible for the hard spectral component \citep{Buo99,Mat00,allen00} required to fit the high energy spectrum of NGC 1399.  
As demonstrated by comparing the 5" resolution ROSAT HRI image with the 10 times sharper {\it Chandra} ACIS image, subarcsecond resolution is essential for the study of X-ray source populations in galaxies.

  \chapter{NGC 1404}
\label{NGC 1404}
\section{INTRODUCTION}
NGC 1404 is a bright early-type located $\sim 10$ arcmin South-East of NGC 1399, within the Fornax cluster. The galaxy is classified as an elliptical in the the RC3 catalogue \citep[E1;][]{deVau91} and in the RSA catalogue \citep[E2;][]{RSA}. However kinematical studies have shown that NGC 1404 is likely to be a rotationally supported system and a thus misclassified S0 galaxy \citep{D'Onofrio95,Gra98}.

Due to her proximity to the dominant Fornax galaxy, in the X-ray band NGC 1404 has been usually studied in association with its brighter companion. The first detection of X-ray emission was obtained by the HEAO-1 satellite \citep{Nugent83} but the EXOSAT satellite was the first able to separate the NGC 1404 emission from NGC 1399 \citep{Mason85}.
The {\it Einstein} satellite confirmed the presence of an extended halo with $L_X\sim 3\times 10^{41}$ erg s$^{-1}$ in the 0.2-4.0 keV band \citep{Thomas86,Fab92}, and a temperature $> 1$ keV at the 90\% confidence level \citep{Kim92}. 

\cite{Loew94} analyzed ASCA data finding evidence of a very subsolar metal abundance ($Z<0.2 Z_\odot$ at the 90\% level) in the NGC 1404 halo and a temperature of $0.75$ keV, suggesting a strong radial metallicity gradient. These findings were confirmed by \cite{jones97} using ROSAT PSPC data ($kT=0.65$ keV and $Z=0.16 Z_\odot$). They also found an elongated feature protruding on the SE side of the galaxy interpreted as a signature of ram pressure stripping of the NGC 1404 corona infalling toward the denser NGC 1399 halo.
\cite{Arimoto97} noticed that the ASCA data are compatible with the presence of a hard spectral component due to unresolved point sources. \cite{Buo98} found that a higher metallicity is allowed if the spectrum is fitted with a two temperature model. The low metallicity is invoked by \cite{Irw98b} to explain the soft ROSAT PSPC spectrum. 

In their study of a sample of early-type galaxies \cite{Brown00} and \cite{Matsu01} found that the NGC 1404 corona has a higher temperature than other galaxies with similar velocity dispersion and luminosity; they explain this result as a strong environmental influence on the physical status of the halo. 
\cite{toni01} simulated the effect of gas stripping on galaxies similar to NGC 1404. Comparing their surface brightness profile to the one measured by \cite{Fab92} they found that their models tend to predict a steeper profile than it is observed.     
Finally, the need for additional heating, over the one provided by gravity alone, is found by \cite{Bregman01} using the {\it Far-Ultraviolet Spectroscopic Explorer} (FUSE) satellite, since the measured cooling rate is lower than the one expected by X-ray measurements.

~\\
In this chapter we use the ROSAT HRI resolution to explore the structure of the NGC 1404 corona and the interactions with the surrounding intracluster medium.
As for NGC 1399, we adopt $H_0=75$ km s$^{-1}$ Mpc$^{-1}$ and a distance of 19 Mpc (1'=5.5 kpc).

\section{OBSERVATIONS AND DATA ANALYSIS}

NGC 1404 was observed at the same time of NGC 1399 since, being located near the dominant Fornax galaxy, it falls within the HRI field of view. For this reason the basic data reduction (aspect correction, exposure correction and composition of the three observations) is the same described in c$\S$ \ref{Aspect corrections} and \ref{Composite observation}. The position of NGC 1404 inside the HRI field is shown in Figure \ref{NGC1399comp}.

\subsection{Brightness Distribution and X-ray/Optical Comparison}
From Figure \ref{NGC1399comp} it is evident that NGC 1404 is missing the large extended halo which surrounds its nearby companion. This is was confirmed by the adaptively smoothed image shown in Figure \ref{csmooth}a: the X-ray emission of NGC 1404 is almost symmetric and centered on the optical galaxy (Figure \ref{csmooth}b).

\begin{figure*}[p]
\vspace{0.1cm}
\centerline{\hspace{1.5cm}\framebox(210,210)[br]{}\hspace{-1.5cm}\raisebox{1cm}{\parbox[b]{1cm}{\bf(a)}}\hspace{1.cm}}
\vspace{1.5cm}
\centerline{\psfig{figure=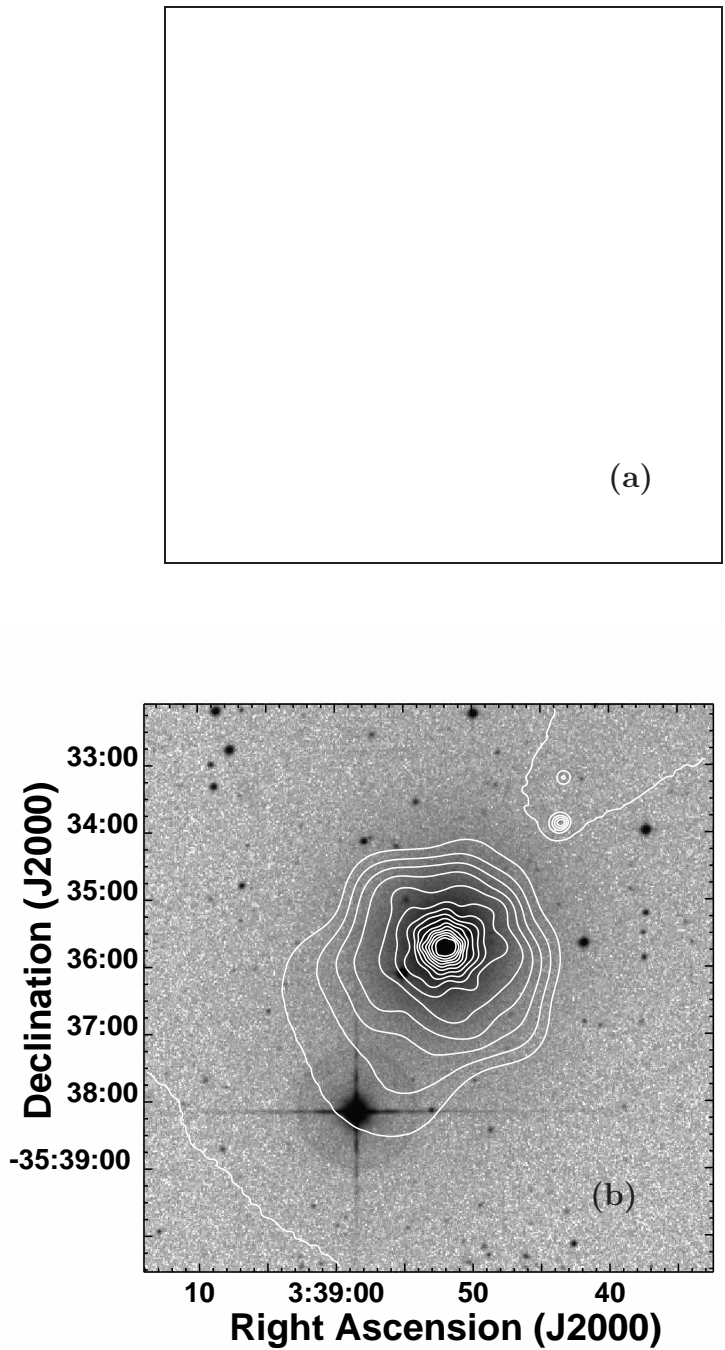,angle=0,width=0.6\textwidth}\hspace{-2.cm}\raisebox{2cm}{\parbox[b]{1cm}{\bf(b)}}\hspace{2.cm}}
\caption{{\bf (a)} $2\times 2$ arcsec/pixel adaptively smoothed image of NGC 1404 HRI field. Colors from black to yellow represent logarithmic X-ray intensities from $1.1\times 10^{-3}$ to $1.3\times 10^{-1}$ cnts arcmin$^{-2}$ s$^{-1}$. {\bf (b)} X-ray brightness contours overlaid on the 1 arcsec/pixel 
DSS image (logarithmic grayscale). Contours are spaced by a factor 1.3 with the lowest one at $1.4\times 10^{-3}$ cnts arcmin$^{-2}$ s$^{-1}$.}
\label{csmooth_ngc1404}
\end{figure*}

We examined the NGC 1404 halo in greater detail creating an adaptively smoothed image of the galaxy with a 2 arcsec/pixel resolution. Figure \ref{csmooth_ngc1404}a shows that the hot halo is circular in the center and becomes slightly elongated in the NE-SW direction past 40" from the center, with an axial ratio of $\sim 0.8$ at $r=80"$. 
At larger radii the surface brightness asymmetries are more pronounced: there is a steep gradient on the NW side of the halo and a `tail' protruding on the SE side of the galaxy. The tail corresponds to the elongated feature reported by \cite{jones97} in the PSPC data.  

The comparison with the DSS image (Figure \ref{csmooth_ngc1404}b) shows that the extension of the X-ray halo is consistent with the optical distribution. A more accurate analysis also reveals that the stellar distribution is flattened in the NW-SE direction, i.e. perpendicular to the elongation of the X-ray isophotes. This indicates that the ellipticity of the X-ray isophotes does not reflect the underlying stellar distribution, but is probably due to ram pressure effects.	

We notice that the presence of the bright star visible near the edge of the NGC 1404 tail (Figure \ref{csmooth_ngc1404}b), does not affect our analysis since we didn't detect any stellar X-ray emission in the raw (unsmoothed) HRI data.

\subsection{Radial Brightness Profiles}
\label{1404_profile}
\begin{figure*}[p]
\centerline{\psfig{figure=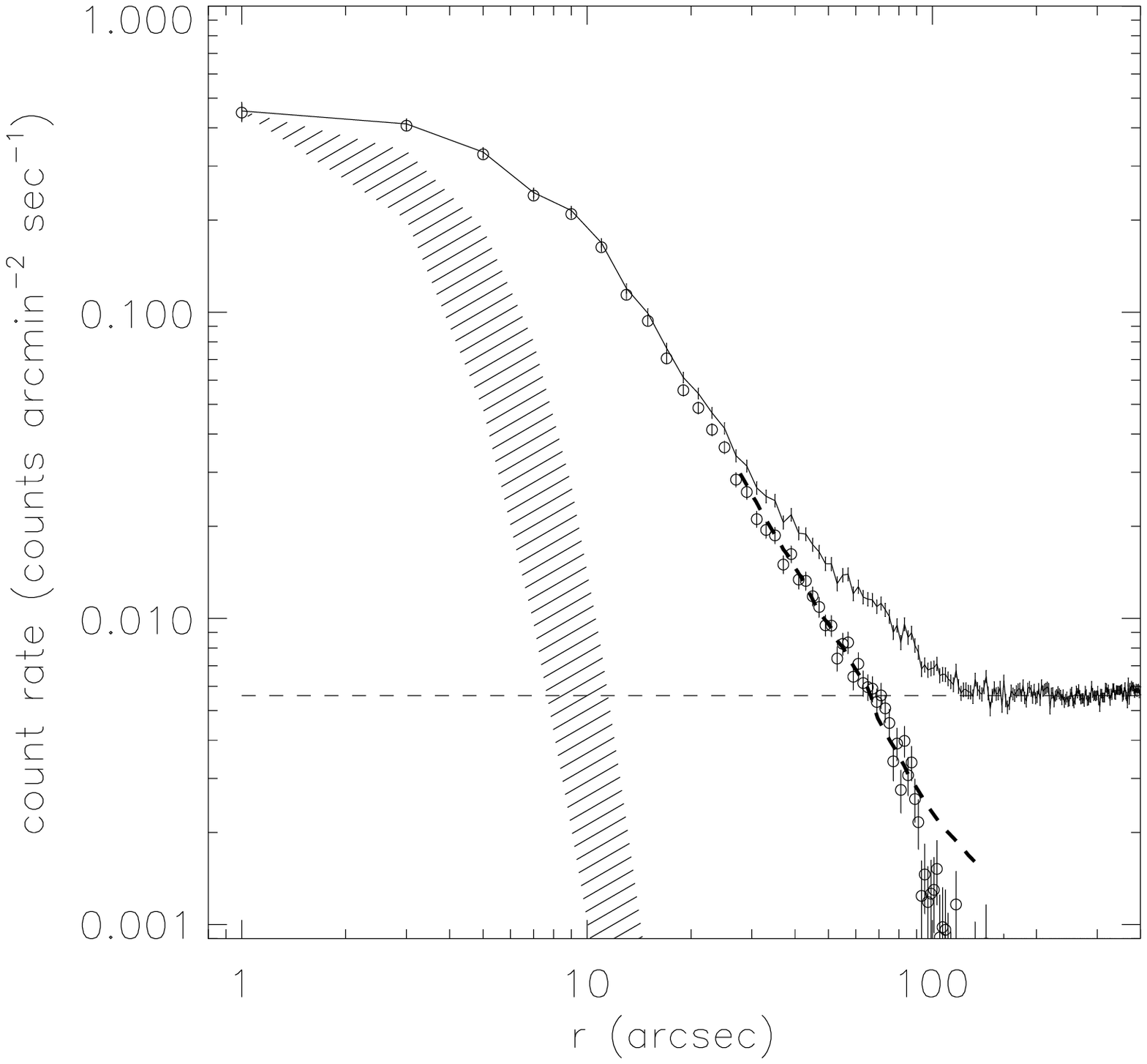,angle=0,width=0.6\textwidth}\hspace{-1.5cm}\raisebox{7cm}{\parbox[b]{1cm}{\bf(a)}}\hspace{1.5cm}}
\centerline{\psfig{figure=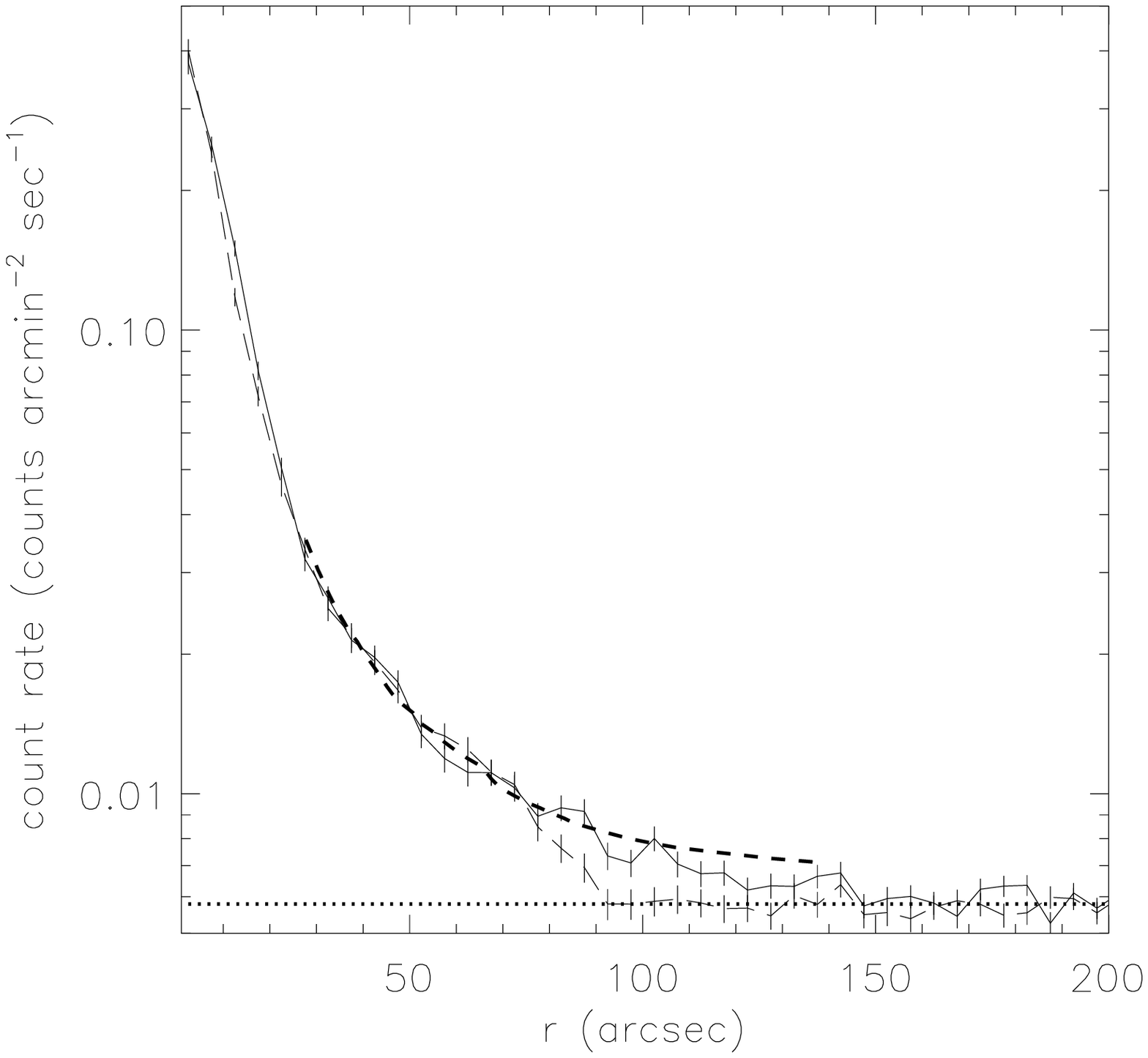,angle=0,width=0.6\textwidth}\hspace{-1.5cm}\raisebox{7cm}{\parbox[b]{1cm}{\bf(b)}}\hspace{1.5cm}}
\caption{{\bf (a)} Radial profile of NGC 1404 extracted in 2'' annuli (continuous line). Circles represent the background subtracted profile using the count rate measured in the 200''-300'' interval (dashed line). The optical profile from \cite{Forb98} is shown as a dashed line. The HRI on-axis PRF range is represented by the shaded region.{\bf~(b)} Radial profiles extracted in 5'' annuli in the North-West ($270^\circ<$P.A.$<360^\circ$, thin dashed line) and the South-East ($90^\circ<$P.A.$<180^\circ$, continuous line) sectors. The optical profiles and background levels are represented respectively by the thick dashed line and the dotted line.}
\label{1404profile}
\end{figure*}
We derived the brightness profile of NGC 1404 extracting count rates in 2'' annuli centered on the galaxy centroid. The background determination for this galaxy is even more difficult than for NGC 1399 (see $\S$ \ref{rad_prof}), due to the fact that it is embedded in the halo of its bright companion. Thus we had no other choice than to estimate the background level from the region between 200'' and 300'' from the galaxy (see Figure \ref{1404profile}a), obtaining $5.60\times 10^{-3}$ counts arcmin$^{-2}$ sec$^{-1}$. The measured count rates are shown in Figure \ref{1404profile}a as a continuous line while the background subtracted profile is represented by open circles.
The NGC 1404 emission extends out to 110'' ($\sim$10 kpc) from the galaxy center and shows no sign of a multi component structure as the one detected in NGC 1399. The surface brightness is almost symmetric (Figure \ref{csmooth_ngc1404}a) and declines with a power law profile out to 90''. 
Past this radius the effect of ram pressure stripping becomes evident as a sharp cutoff in the radial profile. In Figure \ref{1404profile}b we can see that in the North-West sector (thin dashed line) the galactic halo suddenly disappears in the X-ray background for $r>90"$ ($r>8$ kpc) while in the South-East one (continuous line) the emission from the tail, seen in the adaptively smoothed images (Figure \ref{csmooth}), extends up to $\sim 150"$ (14 kpc). 

\begin{figure}[t]
\centerline{\psfig{figure=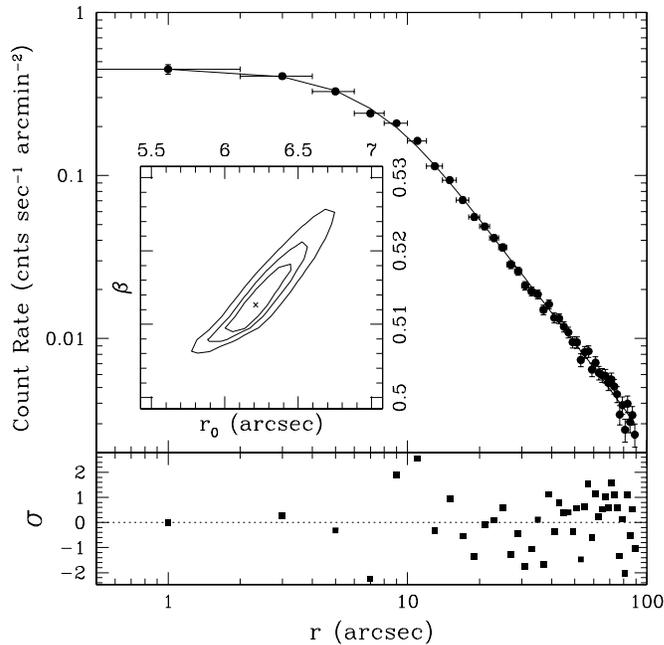,angle=0,width=0.45\textheight}}
\caption{The best-fit model and residuals of NGC 1404 brightness profile within 90''. The 66\%, 90\% and 99\% contour levels relative to the core radius $r_0$ and slope $\beta$ are shown in the inner panel.}
\label{1404fit}
\end{figure}

We fitted the observed radial brightness distribution, within a 90'' radius, with a Beta model (equation \ref{betaproj}). In this case the on-axis PRF we used for NGC 1399 is no longer valid because NGC 1404 is located 10 arcmin South-West of NGC 1399. Although an analytical correction of the on-axis PRF is possible \citep{Dav96}, it doesn't take into account the PRF azimuthal asymmetry.
We thus convolved the Beta model with the on-axis PRF and considered the resulting core radius as an upper limit to the correct value, obtaining
$r_0=6.21^{+0.25}_{-0.2}$ arcsec and $\beta=0.513^{+0.005}_{-0.004}$ with $\chi^2=51.5$ for 42 d.o.f. The best fit model is shown in Figure \ref{1404fit}.

\subsection{Density, Cooling Time and Mass Profiles}
\label{dens_par_1404} 
\begin{figure*}[p]
\centerline{\psfig{figure=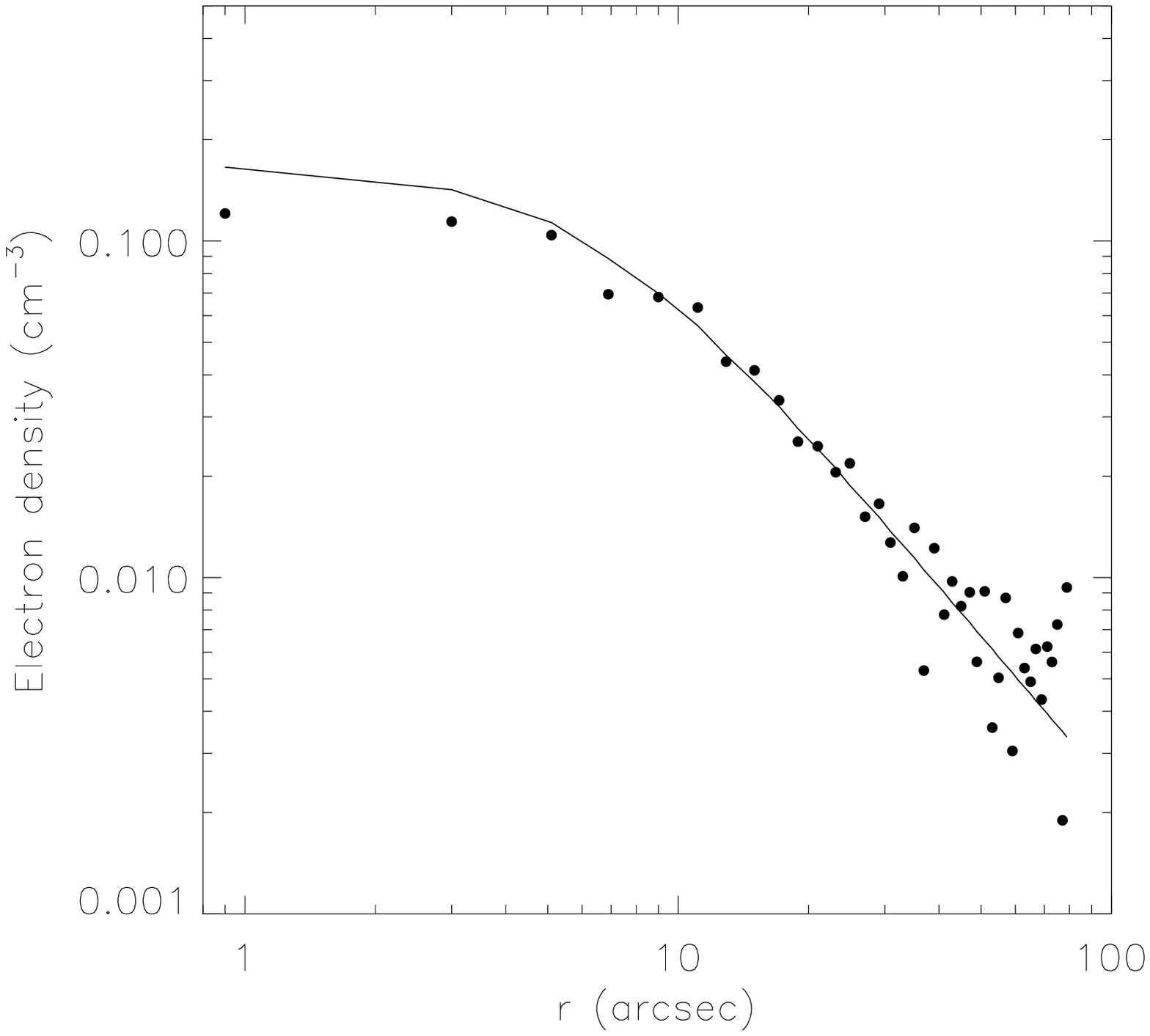,angle=0,width=0.6\textwidth}\hspace{-1.5cm}\raisebox{8cm}{\parbox[b]{1cm}{\bf(a)}}\hspace{1.5cm}}
\centerline{\psfig{figure=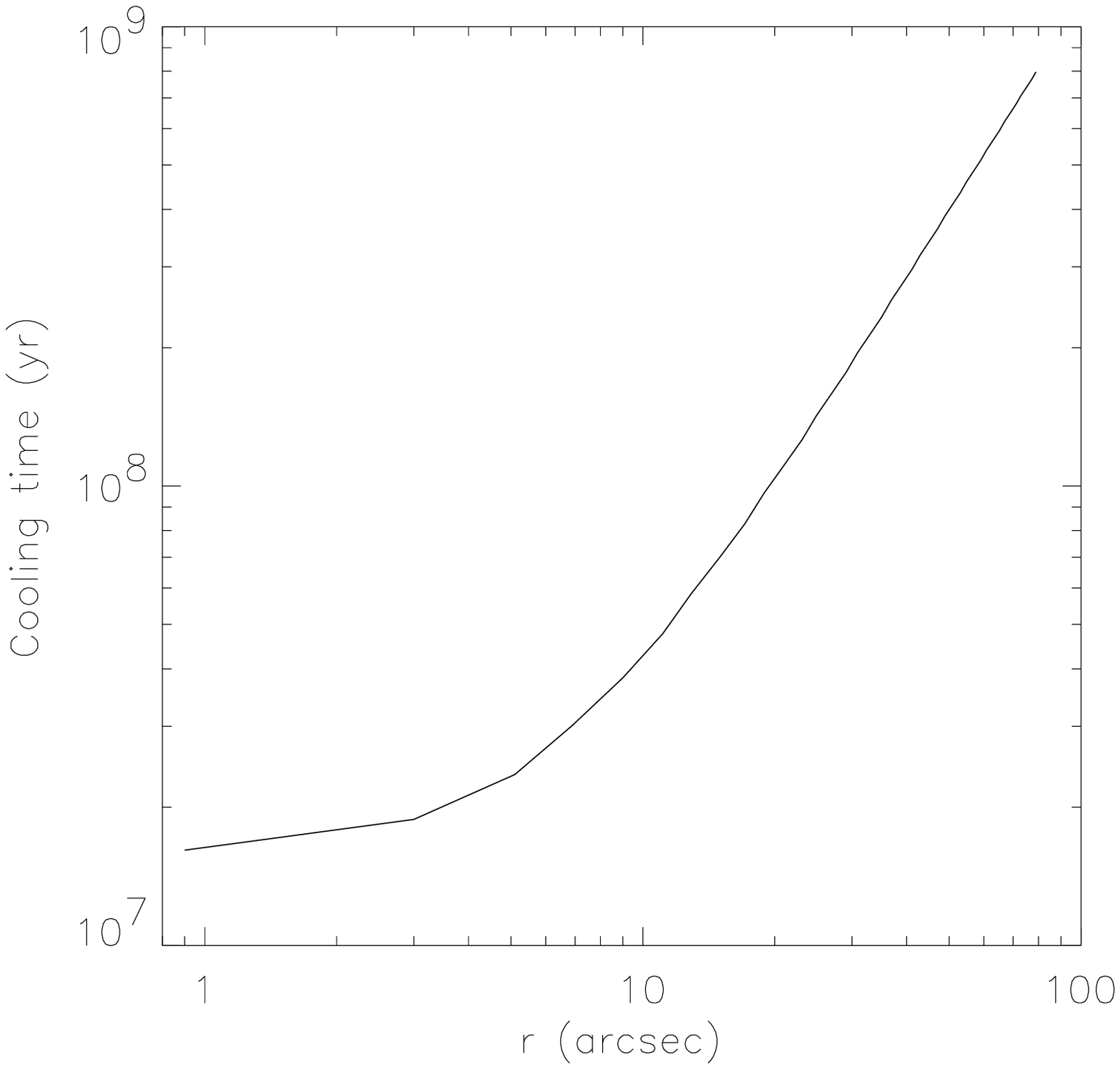,angle=0,width=0.6\textwidth}\hspace{-1.5cm}\raisebox{2cm}{\parbox[b]{1cm}{\bf(b)}}\hspace{1.5cm}}
\caption{{\bf (a)} Deprojected electron density profile of NGC 1404. The filled circles represent the profile obtained from direct surface brightness deprojection while the continuous line shows the smoother profile obtained by deprojecting the best-fit Beta model. {\bf (b)} Cooling time profile derived from the best fit Beta model.}
\label{density_prof_1404}
\end{figure*}

Following the method outlined in $\S$ \ref{dens_par} we derived the electron density, cooling time and mass profiles for NGC 1404. Since the deprojection 
algorithm tends to enhance the fluctuations present in the brightness profile,
we used the best fit model to derive the deprojected electron density. This is possible because the NGC 1404 radial brightness profile was well fitted by a single symmetric Beta model out to 90" ($\S$ \ref{1404profile}). 
Figure \ref{density_prof_1404}a shows the good agreement between the deprojected density derived from the Beta model (continuous line) and the one from the surface brightness profile (filled circles). The discrepancy in the inner 5" is due to the fact that the Beta model is deconvolved for the instrumental PRF. 
We stopped the deprojection at 80" to avoid uncertainties related to the background subtraction and to the asymmetries due to ram pressure stripping ($\S$ \ref{1404profile}), both of which tend to increase the fluctuations at large radii.

The cooling time profile derived from the Beta model is shown in Figure \ref{density_prof_1404}b and suggests that significant cooling must be present in the NGC 1404 halo because the cooling time is smaller than $10^9$ yr within the inner 80" ($\sim$7 kpc). We adopted the cooling function of \citep{Sar87}.

\begin{figure}[t!]
\centerline{\psfig{figure=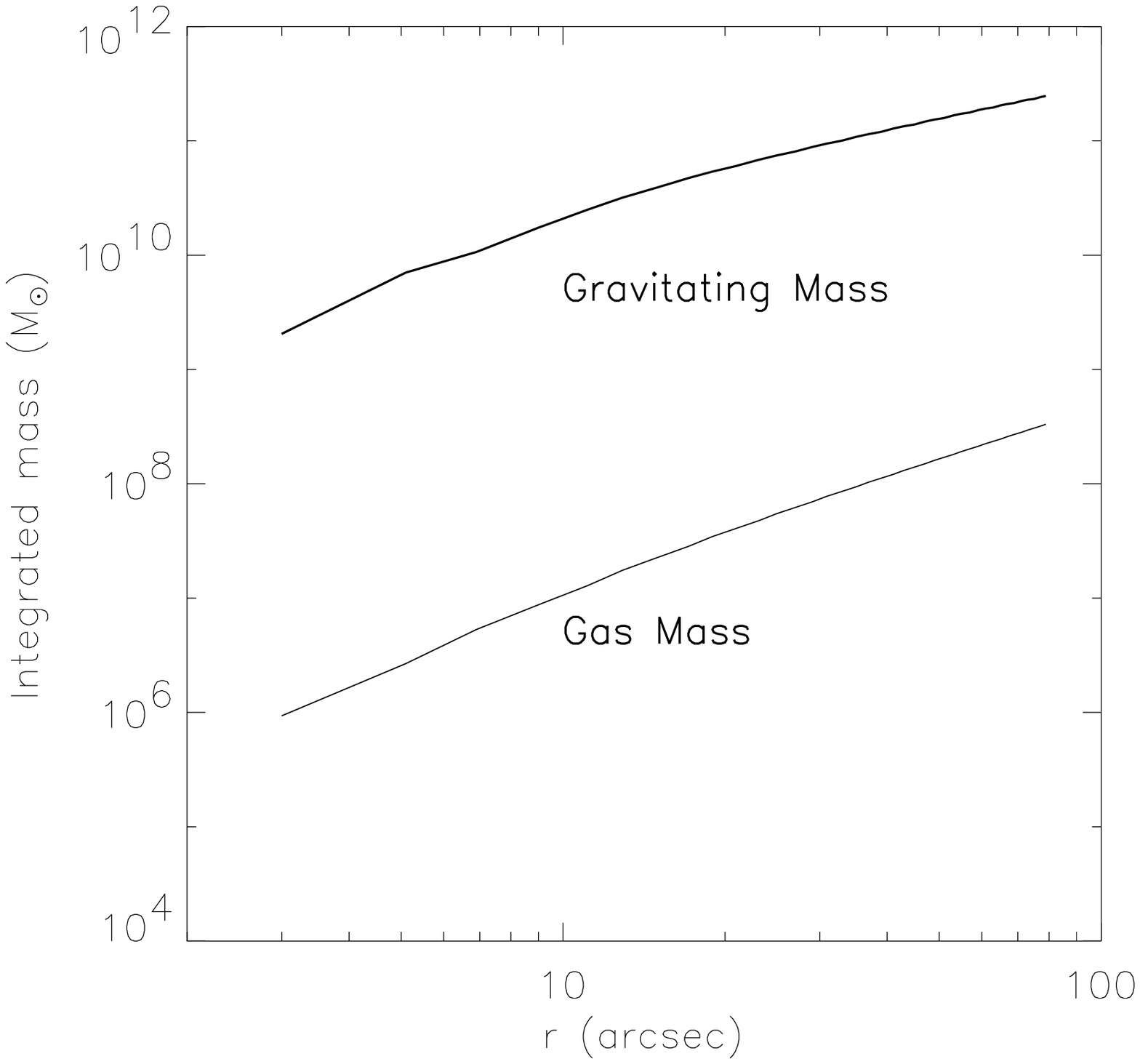,angle=0,width=0.5\textheight}}
\caption{Integrated mass profiles of NGC 1404 derived from deprojection. The total gas mass is represented by the lower line while the total mass derived using equation (\ref{mass}) is shown by the upper line.}
\label{mass_prof_1404}
\end{figure}

The integrated gas mass and the total gravitating mass derived from equation (\ref{mass}) are shown in Figure \ref{mass_prof_1404}. We assumed NGC 1404 to have an isothermal profile with $kT=0.6$ keV \citep{jones97}.

\subsection{Total Fluxes and Luminosities}
In the case of NGC 1404 we used a single temperature Raymond-Smith spectrum with $kT=0.6$ keV and metal abundance 0.2 solar, following \cite{jones97}. We measured a count rate of $0.117\pm0.001$ counts sec$^{-1}$ within a 150'' radius, extracting the background from the 200''-300'' annulus. The corresponding (0.1-2.4 keV) flux and luminosity are $f_X=(4.06\pm 0.03)\times 10^{-12}$ erg s$^{-1}$ cm$^{-2}$ and $L_X=(1.57\pm 0.01)\times10^{41}$ erg s$^{-1}$.

\section{DISCUSSION}
\label{discussion1404}
In contrast with NGC 1399, NGC 1404 possesses a very regular and symmetric halo, well represented by a Beta profile within 90'' (8 kpc). 
The X-ray brightness profile falls as $\sim r^{-2.08\pm 0.03}$, in good agreement with the optical brightness profile $\Sigma\propto r^{1.9\pm 0.1}$ \citep{Forb98}. Since the X-ray emission measure is a function of $n_e^2$ (equations \ref{emiss3} and \ref{totlum}) it is easy to see that $\rho^2_g\sim\rho_{stars}$. If the gas and the stars are two isothermal spheres experiencing the same gravitational potential, equations (\ref{beta_star}) and (\ref{beta_model}) imply that $T_g\sim 2T_{stars}$ \citep{Fab89}.
This is further supported by the ROSAT PSPC spectral data, which yield $\beta_{spec}\sim 0.6$ \citep{Brown98,Matsu01}, suggesting that the gas distribution is produced by the same gravitational potential that is binding the stellar population, out to $r\sim 80"$ ($\sim 7$ kpc; Figure \ref{1404profile}b).

This scenario is in disagreement with the finding that luminosities and temperatures of X-ray compact galaxies are well explained by kinematical heating of gas supplied by stellar mass losses \citep{Matsu01}, which should give $\beta_{spec}\sim 1$. However \cite{Matsu01} notice that NGC 1404 is a peculiar galaxy, in the sense that it is significantly brighter than other compact ellipticals with the same optical luminosity. In fact, without being a central cluster galaxy, its luminosity is of the same order of magnitude of NGC 1399 ($10^{41}$ erg s$^{-1}$). She explains this result in term of interaction with the ISM.

Indeed at radii larger than 80'' the influence of dynamic stripping is evident from  the presence of an elongated tail on the South-East side of the galaxy and the steep surface brightness gradient on the opposite side (Figure \ref{csmooth} and \ref{1404profile}b). These features are aligned with the position of NGC 1399, suggesting that they may be due to the infall of the galaxy toward the cluster center. Ram pressure may be also responsible for the flattening of the X-ray isophotes in direction perpendicular to this motion.
In fact, if the hot gas is just experiencing the same potential of the stellar component, we would expect an elongation in the same direction of the optical isophotes.

However the effect of ram stripping from the ICM is to reduce the X-ray luminosity, removing hot gas from the galactic halo \citep{toni01}. 
The regularity of the X-ray brightness profile and its similarities to the optical profile within 90" ($\sim 8$ kpc) implies that either the stripping efficiency is low or that it affects significantly only the outer galactic regions. This is further supported by the fact that our brightness profile is shallower than expected from simulations of dynamically stripped galaxies performed by \cite{toni01}, even though they notice that their models may be overestimating the importance of the cooling processes.
An alternative and more promising explanation is represented by the stifling of galactic winds ($\S$ \ref{environment}). In fact the presence of the dense NGC 1399 halo may both increase the X-ray luminosity and raise the halo temperature (see for instance \citealp{Brown00}).

The gas, cooling time and total mass profiles derived in $\S$ \ref{dens_par_1404} are in the same range of those of the central component of the NGC 1399 halo, within the inner 80''. This seems in conflict with the different dynamical status of the halo core of NGC 1399, suggested by the $\beta_{spec}\sim 1$ value found by \cite{Matsu01}. The explanation may lie in the presence of the additional and extended halo components (i.e. galactic and cluster components) generally found in cD galaxies and related to the central position inside the cluster potential well (e.g. \citealp{Maki01}). We will discuss further this possibility in $\S$ \ref{ngc1404_final_discuss}.

  \chapter{NGC 507}
\label{NGC507_chapter}
\section{INTRODUCTION}
NGC 507 is the brightest member of a group of galaxies belonging to a poor system: the Pisces cluster. The latter in turn is part of the Perseus-Pisces supercluster \citep[e.g.][]{Wegner93,Sakai94}, located at $\sim 66$ Mpc\footnote{Assuming H$_0=75$ km s$^{-1}$ Mpc$^{-1}$} from our galaxy \citep{Huchra99}.  
NGC 507 is classified as E/S0 galaxy \citep{deVau91}. \cite{Gonzalez00} showed that, while the surface brightness profile follows a {\it de Vaucouleur} law (equation \ref{deVau_eq}) out to $\sim 15"$ from the galaxy center, at larger radii there is significant excess emission over the $r^{1/4}$ profile.
\cite{Barton98} studied the galaxy distribution in the surrounding of NGC 507 finding that it is part of a group including $\sim 70$ members.

The first studies in the X-ray band based on {\it Einstein} data \citep{Kim92,KFT92} measured a luminosity comparable to those of poor clusters ($L_X\simeq 10^{43}$ ergs s$^{-1}$) with a temperature $kT>1.5$ keV at the 90\% level, and suggested the presence of a cooler core. These findings where  confirmed by Position Sensitive Proportional Counter (PSPC) mounted on the ROSAT satellite which revealed extended X-ray emission out to a radius of at least 16 arcmin. \cite{Kim95} were able to resolve the cooler region within the central 150", thus supporting the presence of a massive cooling flow, and suggested that cooling clumps may be distributed at large radii.
They estimated a hot gas temperature $\sim 1.1$ keV with no sign of excess absorption over the galactic value. The data constrained the metallicity to be not higher than the solar one.

\cite{Matsumo97} exploited the higher spectral resolution and larger energy range of the ASCA satellite to separate two distinct components in the NGC 507 X-ray spectrum: a soft thermal emission with $kT\simeq 1$ keV and a harder component with $kT\sim 10$ keV probably due to unresolved discrete sources. The ASCA data seem to favor a subsolar metallicity of $\sim 0.3~Z_\odot$.
The different metal abundance obtained by ROSAT and ASCA is explained in part by the new analysis performed by \cite{Buo00b} on the ROSAT PSPC data, which found a steep gradient in the metallicity profile of the hot halo.

NGC 507 is a radio galaxy with a steep radio spectrum and weak core \citep{Colla75,deRuiter86,parma86}. It may be a source with particularly weak jets, or possibly a remnant of a radio galaxy whose nuclear engine is almost inactive and whose luminosity has decreased due to synchrotron or adiabatic losses \citep{Fanti87}. 
\cite{Canosa99} used the ROSAT HRI data to study the correlation between the radio and X-ray emission of a possible nuclear source, obtaining an upper limit for the nuclear X-ray flux which exceeds the expectation based on the radio data by more than two orders of magnitude.

~\\
In this work we use the data collected by the ROSAT HRI to study the structure of the NGC 507 hot halo. In particular we concentrate on the properties of the nuclear region, where the HRI resolution allows to resolve small scale structures. We then relate the results to those obtained by the ROSAT PSPC at larger scales. 

\section{OBSERVATIONS AND DATA ANALYSIS}
	\subsection{The Data}
The NGC 507 field, including NGC 499, was observed on January 1995 by the ROSAT HRI \citep{Dav96} for a total time of $\sim 28$ ks (Table \ref{ngc507observations}).
\begin{table*}[h]
\begin{center}
\footnotesize
\caption{The ROSAT HRI observation of the NGC 507/NGC499 field.\label{ngc507observations}}
\begin{tabular}{ll}
\\
\tableline
\tableline									
name & NGC 507\\
Field center & \\ 
(R.A., Dec) & 01$^{\rm h}$23$^{\rm m}$39$^{\rm s}$ 33$^\circ$15'36''\\ 
sequence id. & RH600680n00\\
Exp. time & 28310 s\\
obs. date & 1995 Jan 14\\
P.I. & D.-W. Kim\\ 
No. OBI & 2\\
Distance\tablenotemark{~(a)} & 65.8 Mpc\\
$N_H$\tablenotemark{(b)} & $5.3\times 10^{20}$ cm$^{-2}$\\
\tableline
\end{tabular}
\tablenotetext{(a)}{Estimated assuming $H_0=75$ km s$^{-1}$ Mpc$^{-1}$ and a velocity of 4934 km s$^{-1}$ \citep{Huchra99}}
\tablenotetext{(b)}{Galactic line-of-sight column density \citep{Stark92}}
\end{center}
\end{table*}
\begin{figure}[b!]
\centerline{\psfig{figure=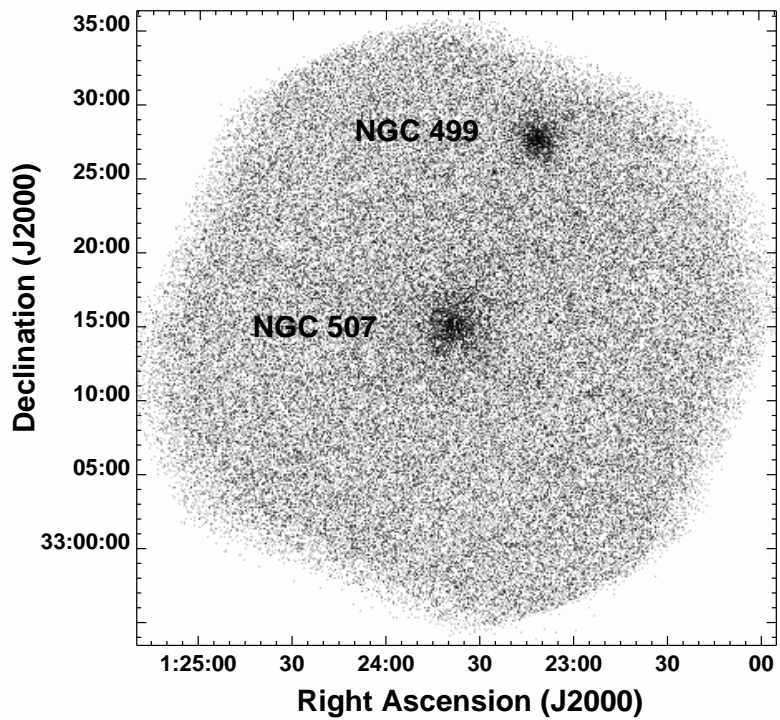,width=0.75\textwidth}}
\caption{Raw HRI image of the NGC 507/NGC 499 field. The data are rebinned in $5\times 5$ arcsec pixels. North is up and East is left.}
\label{ngc507}
\end{figure}
The data, consisting of 2 Observing Bin Intervals (OBI), were reduced with the  SASS7\_5 versions of the ROSAT standard analysis software (SASS). To correct for the errors in the aspect time behavior \citep{Har99}, discussed in $\S$ \ref{Aspect corrections}, we run the ASPTIME routine (F. Primini 2000, private communication). 

We further tried to correct for problems related to the spacecraft wobble  \citep[ also see $\S$ \ref{Aspect corrections}]{Harris98}; however the pointlike sources detected in our field (Table \ref{ngc507_source_tab}) are not bright enough to allow an accurate determination of the X-ray centroid (crucial for the correction), when divided in time bins. Since the global PRF is $\sim 7"$ FWHM we decided to use the uncorrected data. The HRI field of view (FOV) is shown in Figure \ref{ngc507}.

We have run the software developed by \citet[ hereafter SMB]{Snow94}, to produce an exposure map and a background map for our observation. We then corrected for exposure time and quantum efficiency variations across the detector by dividing the raw data by this exposure map. 

\subsection{Surface Brightness Distribution}
\label{ngc507_brightness}
\begin{figure}[!t]
\vspace{1cm}
\centerline{\psfig{figure=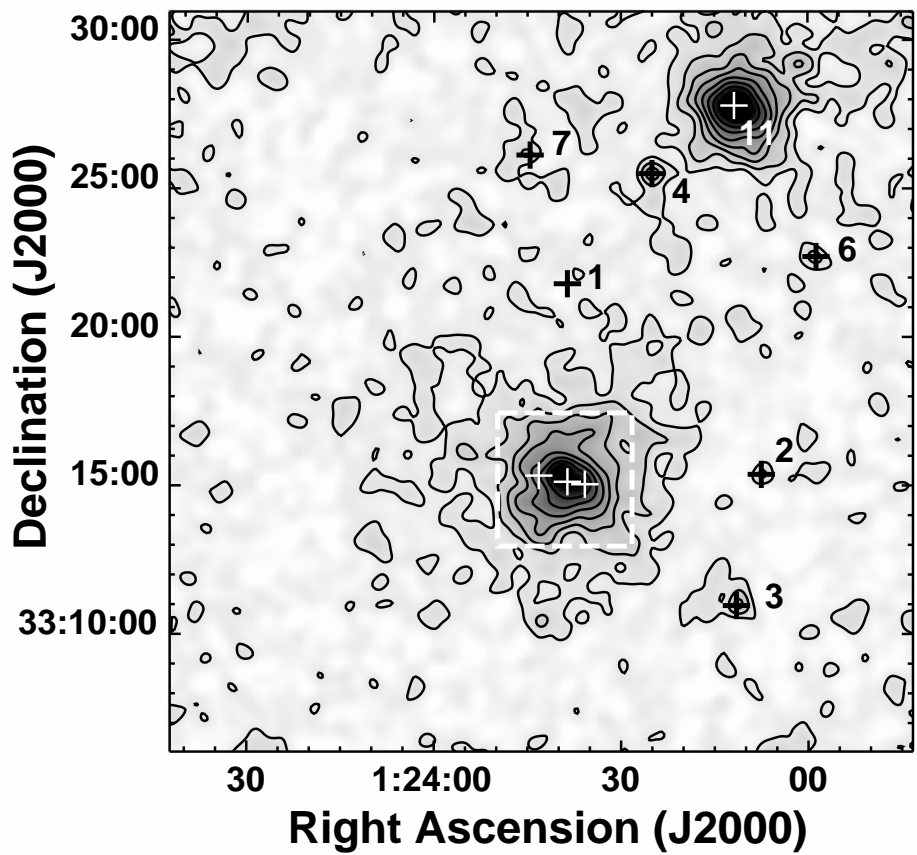,width=0.85\textwidth}}
\caption{Exposure corrected image of the center of the HRI field. The data are rebinned in $5\times 5$ arcsec pixels and smoothed with a gaussian filter of $\sigma=15"$. The X-ray contours are spaced by a factor of 1.2 with the lowest one at $6.4\times 10^{-3}$ cnts arcmin$^{-2}$ s$^{-1}$. Crosses show the position of sources detected by the wavelets algorithm (Table \ref{ngc507_source_tab}). The white dashed box is the region enlarged in Figures \ref{ngc507_centbox}, \ref{ngc507_radio} and \ref{ngc507_model}. In the figure North is up and East is left.}
\label{ngc507_smooth}
\end{figure}

To study the surface brightness distribution of NGC 507 the exposure corrected data were rebinned in $5\times 5$ arcsec pixels and then convolved with a gaussian filter of $\sigma=15"$ (Figure \ref{ngc507_smooth}). The resulting image shows the presence of an extended X-ray halo surrounding the galaxy.  The central part of the halo (the one enclosed in the dashed box) is relatively symmetric but elongated in the NE-SW direction. Two filamentary structures seem to depart from the galactic center toward NW and SE respectively. The morphology of the outer halo is very irregular due to the low S/N ratio of the HRI image.
However there appears to be more emission in the region enclosed within the two galaxies with respect to the other directions. 
The NGC 499 halo is less extended than the NGC 507 one and much more symmetric.
The positions of the sources detected by the wavelets algorithm (Table \ref{ngc507_source_tab}) and discussed in $\S$ \ref{ngc507_sources}, are marked with crosses.

The large scale distribution of hot gas was already studied by \cite{Kim95} with the ROSAT PSPC, which has a higher sensitivity than the HRI. We thus concentrate on the structure of the nuclear region (dashed box in Figure \ref{ngc507_smooth}) which is better resolved by the HRI.
The complex structure of this region is already suggested by the fact that the wavelets algorithm detects three distinct emission peaks (white crosses). 

\begin{figure*}[p]
\centerline{\hspace{0.5cm}\framebox(230,230)[br]{}\hspace{-1.5cm}\raisebox{1cm}{\parbox[b]{1cm}{\bf(a)}}\hspace{1.cm}}
\vspace{1cm}
\centerline{\psfig{figure=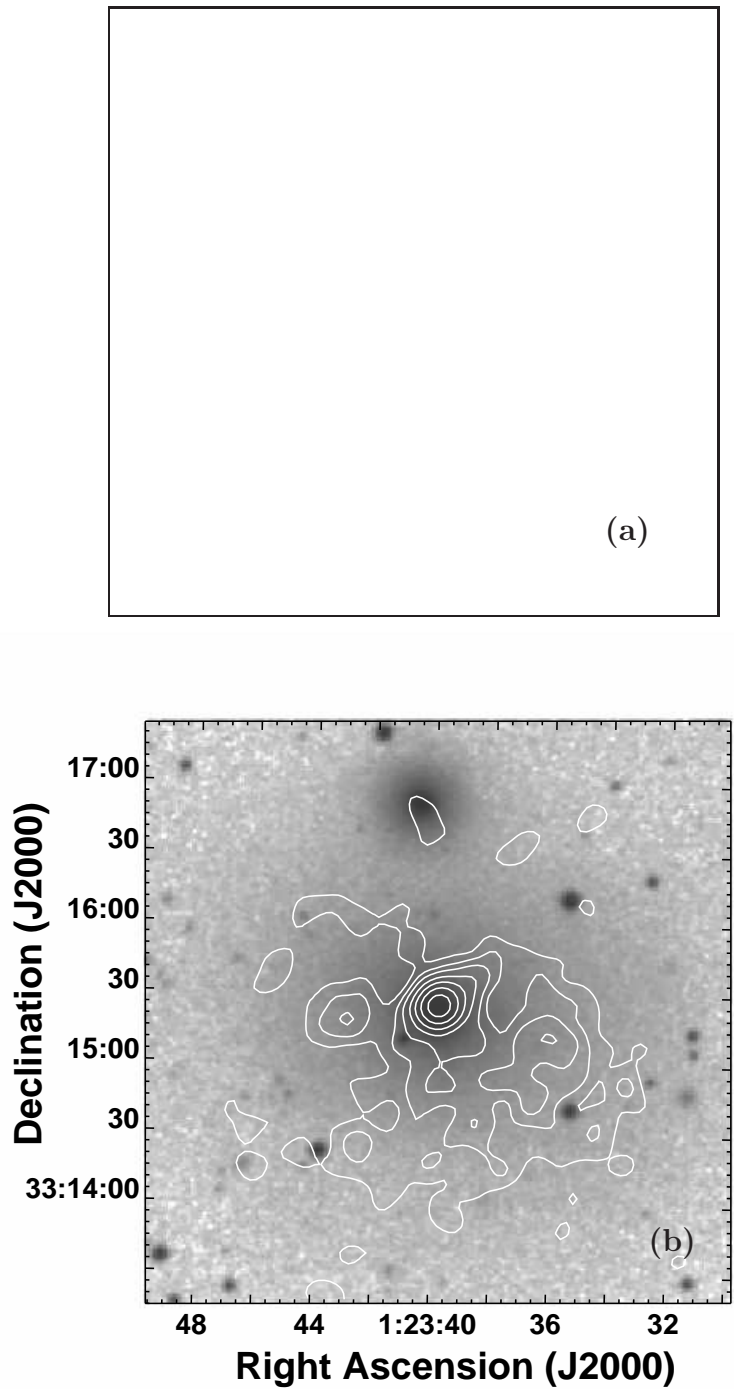,angle=0,width=0.62\textwidth}\hspace{-1.5cm}\raisebox{2cm}{\parbox[b]{1cm}{\bf(b)}}\hspace{1.5cm}}
\caption{{\bf (a)} $2.5\times 2.5$ arcsec/pixel image of the nuclear region of  NGC 507. The image is convolved with a gaussian filter of $\sigma=5"$. Colors from black to white represent logarithmic X-ray intensities from $5.7\times 10^{-3}$ to $8.1\times 10^{-2}$ cnts arcmin$^{-2}$ s$^{-1}$.
The dashed circles show the region of maximum S/N ratio for the sources detected by the wavelets algorithm.
{\bf (b)} X-ray brightness contours overlaid on the 1 arcsec/pixel 
Digital Sky Survey (DSS) image (logarithmic grayscale). Contours are spaced by a factor 1.3 with the lowest one at $1.4\times 10^{-2}$ cnts arcmin$^{-2}$ s$^{-1}$. In the figures North is up and East is left.}
\label{ngc507_centbox}
\end{figure*}

Examining this area in more detail (Figure \ref{ngc507_centbox}a), a number of interesting features can be detected:
1) a strong emission peak can be observed in the center of the X-ray halo; this peak is slightly elongated in the NW-SE direction due to the presence of two `tails': a main one extending in the NW direction and a fainter one visible on the SE side; 2) a secondary peak is visible $\sim 50"$ West of the primary peak (source No.9); 3) a region of low X-ray brightness is interposed between the primary and secondary peak; 4) a third peak is visible $\sim 40"$ East of the galaxy center (source No.5); 5) additional diffuse emission is present on the Southern side of the nucleus, making the surface brightness gradient flatter than on the Northern side.

The main peak is almost coincident with the center of the optical galaxy (Figure \ref{ngc507_centbox}b). The second and third peak, instead, are not related to any optical feature, although being within the stellar body of the galaxy. There is also some faint emission coincident with the position of the elliptical galaxy NGC 508, located 1.5 arcmin North of NGC 507, which may indicate the presence of hot gas in the core of this system.

\begin{figure}[t]
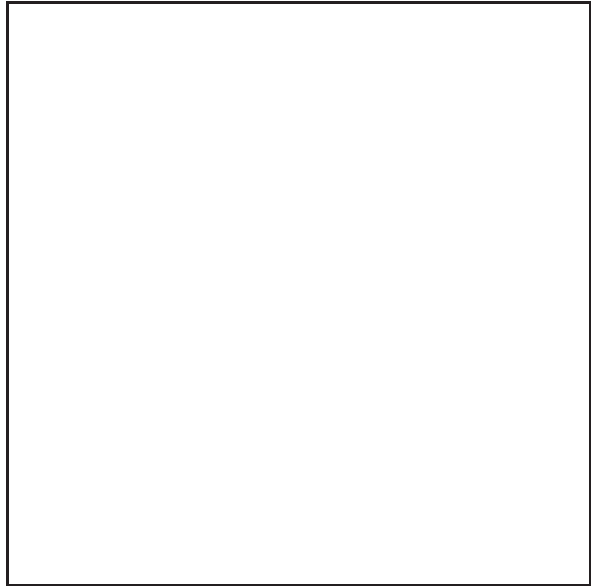

\vspace{1cm}
\centerline{\framebox(220,220)[br]{}}
\vspace{1cm}
\caption{Adaptively smoothed image of the center of the NGC 507 halo. The data are rebinned in $1\times 1$ arcsec pixels. Colors from black to yellow represent logarithmic X-ray intensities from $6.3\times 10^{-3}$ to $1.5\times 10^{-1}$ cnts arcmin$^{-2}$ s$^{-1}$. The dashed box is the region enlarged in Figures \ref{ngc507_centbox}, \ref{ngc507_radio} and \ref{ngc507_model}.}
\label{ngc507_csmooth}
\end{figure}

To reveal the presence of extended low S/N features we adaptively smoothed the HRI image with the {\it csmooth} algorithm contained in the CIAO CXC package (see $\S$ \ref{brightness}). In the resulting image (Figure \ref{ngc507_csmooth}), in addition to the three emission peaks mentioned before, we can distinguish an elongated tail extending on the North and bending Westwards out to $\sim 200"$ (63 kpc \footnote{At the the assumed distance of NGC 507, 1'=19.1 kpc}), and an X-ray tongue protruding by $\sim 160"$ (51 kpc) on the South-East side. These two features correspond to the filamentary structures visible also in Figure \ref{ngc507_smooth}.

\subsection{X-ray/Radio Comparison}
\label{ngc507_Xradio}
\begin{figure*}[p]
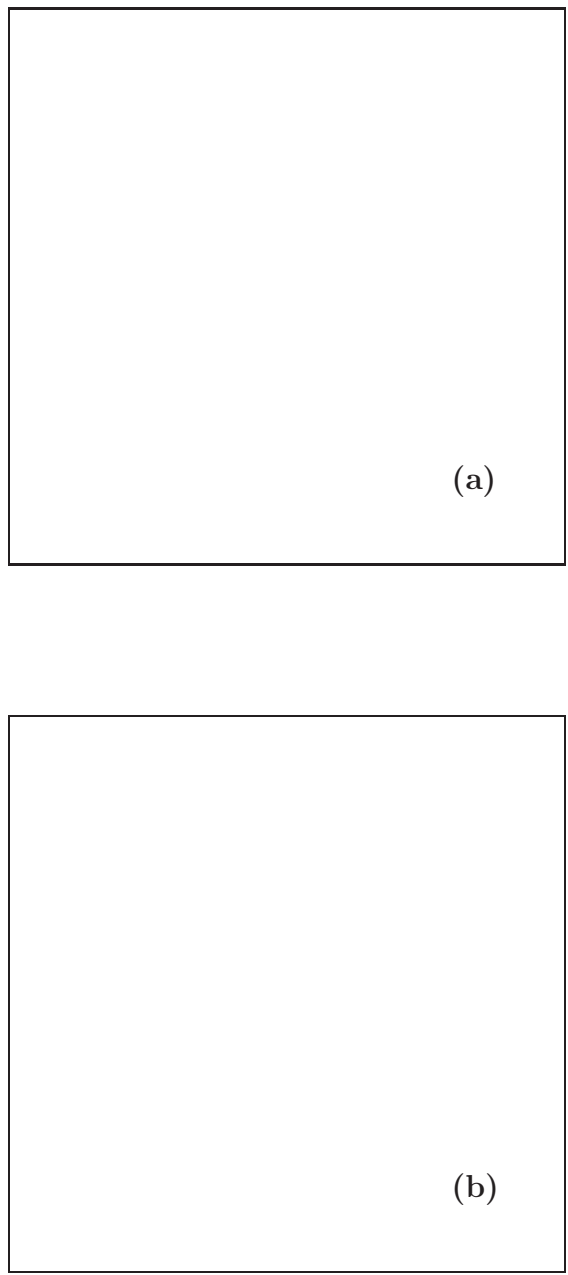

\centerline{\hspace{0.5cm}\framebox(210,210)[br]{}\hspace{-1.5cm}\raisebox{1cm}{\parbox[b]{1cm}{\bf(a)}}\hspace{1.cm}}
\vspace{2cm}
\centerline{\hspace{0.5cm}\framebox(210,210)[br]{}\hspace{-1.5cm}\raisebox{1cm}{\parbox[b]{1cm}{\bf(b)}}\hspace{1.cm}}
\vspace{1cm}
\caption{{\bf (a)} 13" FWHM radio contours superimposed on the X-ray nuclear region of NGC 507. The X-ray image is the same shown in Figure \ref{ngc507_centbox}a.
{\bf (b)} Higher resolution (3.5" FWHM) radio contours superimposed on the X-ray nuclear region of NGC 507. }
\label{ngc507_radio}
\end{figure*}

NGC 507 is known to be a weak radio source.
In Figure \ref{ngc507_radio}a,b we superimpose the radio contours at 20 cm \citep{parma86,deRuiter86} on the smoothed X-ray image. 

The nuclear radio source, visible in the high resolution map (Figure \ref{ngc507_radio}b) is coincident with the position of the central X-ray peak. However it seems to be slightly displaced ($\sim 4"$) toward the South-East edge of the peak.
The radio jets/lobes appear aligned with the direction of the secondary X-ray peaks. The western jet appears to be collimated up to 10-15 arcseconds ($\sim 4$ kpc) and then expands in a large lobe. The latter is coincident with the region of low surface brightness described in $\S$ \ref{ngc507_brightness}. The higher resolution radio contours (Figure \ref{ngc507_radio}b) reveal a very good agreement between the morphology of the Western lobe and the shape of the `cavity'. The secondary X-ray peak lies at the end of the radio lobe.

The correlation between the X-ray and radio surface brightness are less evident for the Eastern lobe. In fact this lobe is fainter than the Western one (27 mJy compared to 40 mJy at 1.4 GHz; \citealp{deRuiter86}). From the low resolution radio contours it seems that the compact X-ray peak falls in the center of the radio lobe; however, when we look at the high resolution map, the radio emission seems displaced South of the X-ray peak and falls in a region of low X-ray emission.   

\subsection{Bidimensional Halo Model}
\label{ngc507_2Dmodel}
To study the complex structure of the inner halo of NGC 507 we built a bidimensional model using the {\it Sherpa} fitting software contained in the CIAO CXC package. The model was composed by adding together two bidimensional Beta components (equation \ref{2dmodel}), representing respectively the central X-ray peak and the larger galactic halo. The 2D model was convolved for the HRI PRF and then fitted on the exposure corrected HRI image following the method outlined in $\S$ \ref{model}: we fitted the large galactic component on a $30\times 30$ arcsec pixel image and the central component on a higher resolution $5\times 5$ arcsec pixel image.

\begin{figure}[!h]
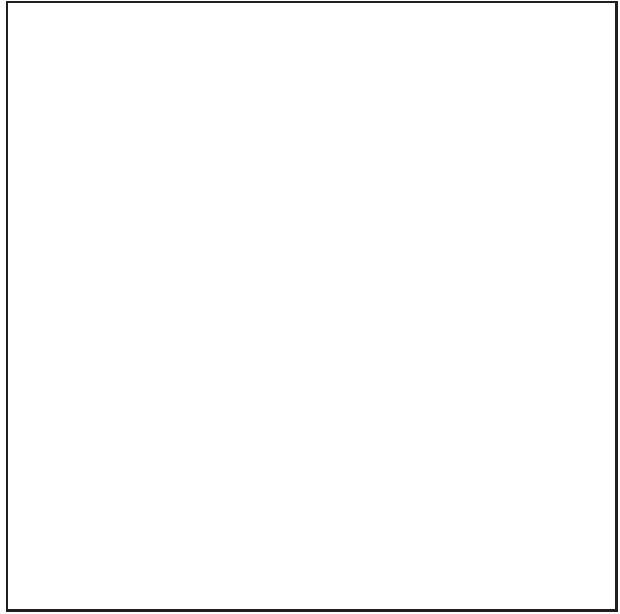

\vspace{3cm}
\centerline{\framebox(230,230)[br]{}}
\vspace{1cm}
\caption{X-ray contours of the bidimensional model superimposed on the NGC 507 central halo. The model contours are spaced by a factor 1.3 with the lowest one at $9.6\times 10^{-3}$ cnts arcmin$^{-2}$ s$^{-1}$.}
\label{ngc507_model}
\end{figure}

Table \ref{ngc507_fit_tab} shows the best-fit parameters of the bidimensional model. During the fit the ellipticity $\epsilon$ and the position angle (P.A.) $\theta$ of the two components was fixed to 0. In fact the poor S/N ratio of the HRI image at large radii makes the ellipticity and P.A. of the galactic component difficult to determine and, anyway, of poor physical meaning. For what concerns the central component leaving $\epsilon$ and $\theta$ free to vary yields $\epsilon=0.56\pm 0.02$, $\theta=137^\circ$, and results in minor changes on the slope ($\beta=0.63\pm0.01$) and the core radius ($r_0=0.32\pm 0.02$). The central component is thus elongated in the NW-SE direction, as expected from the presence of the nuclear `tails' ($\S$ \ref{ngc507_brightness}).
The galactic component centroid is displaced by $\sim 22"$ (7 kpc) South-West of the central emission peak (Figure \ref{ngc507_model}); this is in agreement with the position of source No.10 detected by the wavelets algorithm (Figure \ref{ngc507_centbox}a) which corresponds with the galactic halo centroid.
The best-values $r_0$ and $\beta$ of the galactic component are consistent within 1 $\sigma$ with those found by \cite{Kim95}; however our core radius is slightly larger due to the fact that we have separated the contribution of the central peak from the larger galactic halo.

\begin{table*}[t]
\begin{center}
\footnotesize
\caption{Best-fit parameters for the bidimensional halo model.\label{ngc507_fit_tab}}
\begin{tabular}{lcccccccc}
\\
\tableline									
\tableline									
Component & \multicolumn{2}{c}{Center Position} & $r_0$ & $\beta$ & $\epsilon$ & $\theta$ & $\chi^2_\nu$ & $\nu$ \\
& R.A. & Dec & (arcsec) & & $(1-\frac{minor~axis}{major~axis})$ & (rad) & & (d.o.f.)\\
\tableline									
Central & 01$^{\rm h}$23$^{\rm m}$39.8$^{\rm s}$ & 33$^\circ$15'26'' & 0.22$\pm $0.02 & 0.58$\pm $0.01 & 0.0\tablenotemark{(a)} & 0.0\tablenotemark{(a)} & 0.5 & 73\\
Galactic & 01$^{\rm h}$23$^{\rm m}$38.7$^{\rm s}$ & 33$^\circ$15'07'' & 49$\pm$ 9 & 0.48$\pm 0.04$ & 0.0\tablenotemark{(a)} & 0.0\tablenotemark{(a)}  & 0.9 & 776\\
\tableline									
\end{tabular}
\tablenotetext{(a)}{ The parameters $\epsilon$ and $\theta$ have no error because they were held fixed during the fit (see discussion in text).}
\tablenotetext{}{{\bf Note} - uncertainties are $1\sigma$ confidence level for 5 interesting parameters}
\end{center}
\end{table*}

\subsection{Radial Brightness Profiles}
\label{ngc507_brightness_prof}
\begin{figure}[t]
\centerline{\psfig{figure=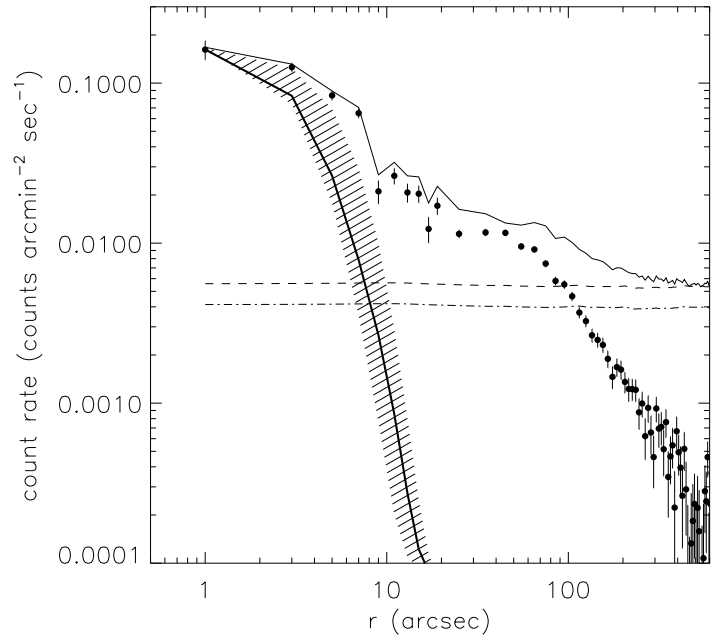,width=0.6\textwidth}}
\caption{HRI surface brightness profile of NGC 507 (thin continuous line). The dashed (dot-dashed) line represent the SMB background map profile after (before) rescaling to match PSPC counts. Filled circles show the background subtracted profile obtained using the rescaled SMB background. The PRF profile and the uncertainties due to residual aspect errors, are shown respectively as the thick continuous line and the shaded region.}
\label{ngc507_profile}
\end{figure}

To study quantitatively the hot halo of NGC 507 we derived radial profiles of the X-ray surface brightness. We binned the X-ray counts in 2 arcsec annuli, centered on the main X-ray peak, up to 20 arcsec and in 10 arcsec annuli at larger radii. This allows to both exploit the HRI resolution in the inner regions, where the X-ray emission is stronger, and to have a good S/N ratio in the outer regions. The derived profile is shown in Figure \ref{ngc507_profile} as a thin continuous line.

To derive the instrumental background level we used the background map produced by the SMB software (dot-dashed line in Figure \ref{ngc507_profile}). However, as discussed in $\S$ \ref{rad_prof}, the latter may be a poor representation of the actual background when an extended source is present in the field. We followed the same approach used for NGC 1399, comparing the HRI data with the PSPC profile obtained by \cite{Kim95}. The HRI profile was rebinned in 30 arcsec annuli to match the PSPC resolution and the SMB background was then rescaled to obtain the best agreement between the two profiles in the 100-500 arcsec range. The final adopted background (dashed line) is in close agreement with the flattening level of the HRI profile, i.e. the brightness level measured at radii $> 500$ arcsec. This result was expected since, differently from the case of NGC 1399, the shorter exposure time resulted in a lower S/N ratio in the outer HRI field. Thus the measured background is dominated by the instrumental contribution expected when no source is present. Using this background determination we derived the background-subtracted profile shown as filled circles.

\begin{figure}[t]
\centerline{\psfig{figure=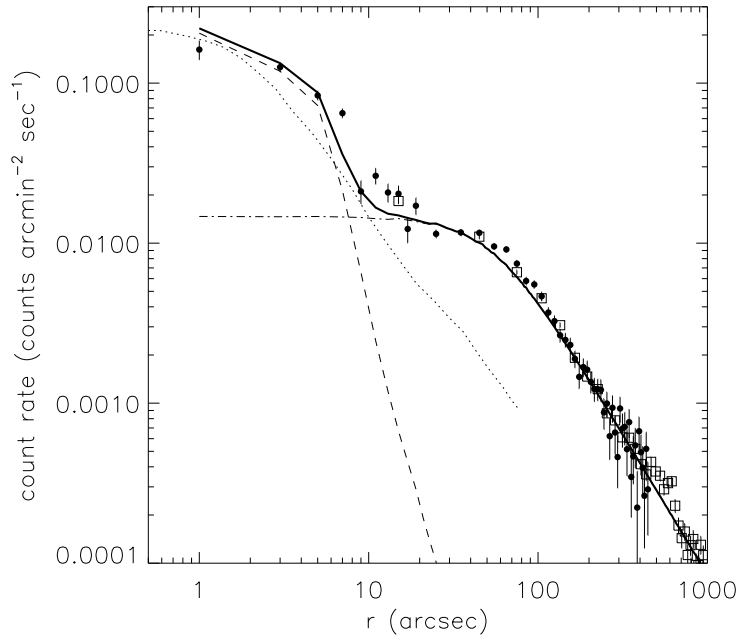,width=0.6\textwidth}}
\caption{Radial profile of the bidimensional HRI model (continuous line). The central and galactic components are represented respectively by the dashed and dot-dashed lines. HRI (PSPC) counts are shown as filled circles (empty squares). The dotted line shows the V band surface brightness profile measured by \cite{Gonzalez00}.}
\label{ngc507_model_profile}
\end{figure}

The inner 10 arcsec are dominated by the emission due to the central X-ray peak showed in Figure \ref{ngc507_centbox}a. A comparison with the HRI PRF suggests that the peak is extended. However, taking into account the PRF uncertainties due to residual aspect problems (shaded region in the Figure; \citealp{Dav96}) and the asymmetries discussed below, we find that the central peak is marginally consistent with the presence of a pointlike nuclear source. Past 20 arcsec ($\sim 6$ kpc) the X-ray profile flattens due to the presence of the galactic halo component, and then decreases with a power-law profile for $r>60"$. 

Figure \ref{ngc507_model_profile} shows the model profile ($\S$ \ref{ngc507_2Dmodel}) -- convolved with the HRI PRF -- superimposed on the HRI data.
The contribution of the two components are shown by the dashed and the dot-dashed line for the central and galactic component respectively.
The PSPC profile measured by \cite{Kim95} is in very good agreement with the HRI data and shows that the halo brightness decreases as a $r^{-1.88}$ power-law even outside the range covered by the HRI points.
For comparison we also show the optical V band surface brightness profile measured by \cite{Gonzalez00}.

\begin{figure}[t]
\centerline{\psfig{figure=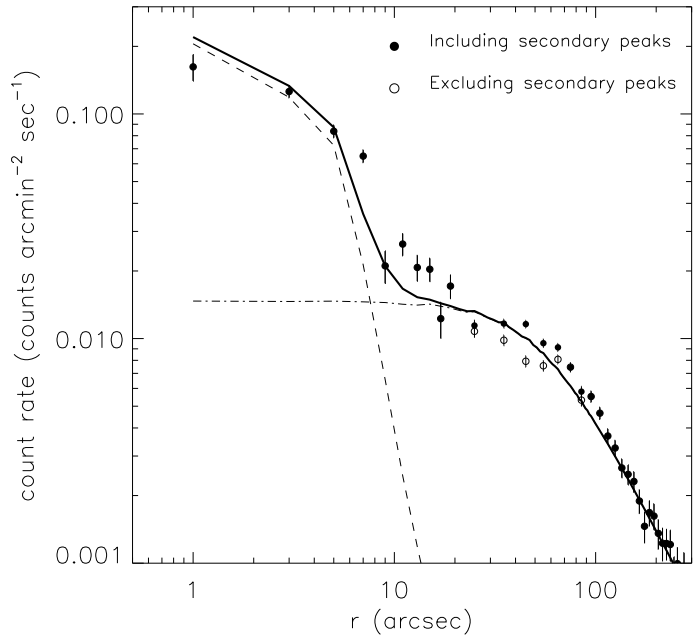,angle=0,width=0.6\textwidth}}
\caption{HRI radial profiles of the NGC 507 nuclear region including (filled circles) and excluding (empty circles) the secondary emission peaks corresponding to sources No.5 and 9 (Figure \ref{ngc507_centbox}a). The meaning of the lines is the same as in Figure \ref{ngc507_model_profile}.}
\label{ngc507_nooverdens}
\end{figure}

From Figure \ref{ngc507_model_profile} we can see that in two regions, for $10"<r<20"$ and $40"<r<100"$, the HRI counts are in excess with respect to those predicted by the bidimensional model.
To investigate these features we derived radial profiles excluding some of the structures described in $\S$ \ref{ngc507_brightness}.
Figure \ref{ngc507_nooverdens} shows that excluding  sources No.5 and 9 (Table \ref{ngc507_source_tab}), corresponding to the secondary emission peaks shown in Figure \ref{ngc507_centbox}a, significantly reduces the counts measured in the range $30"<r<70"$. 

\begin{figure*}[p]
\vspace{0.5cm}
\centerline{\psfig{figure=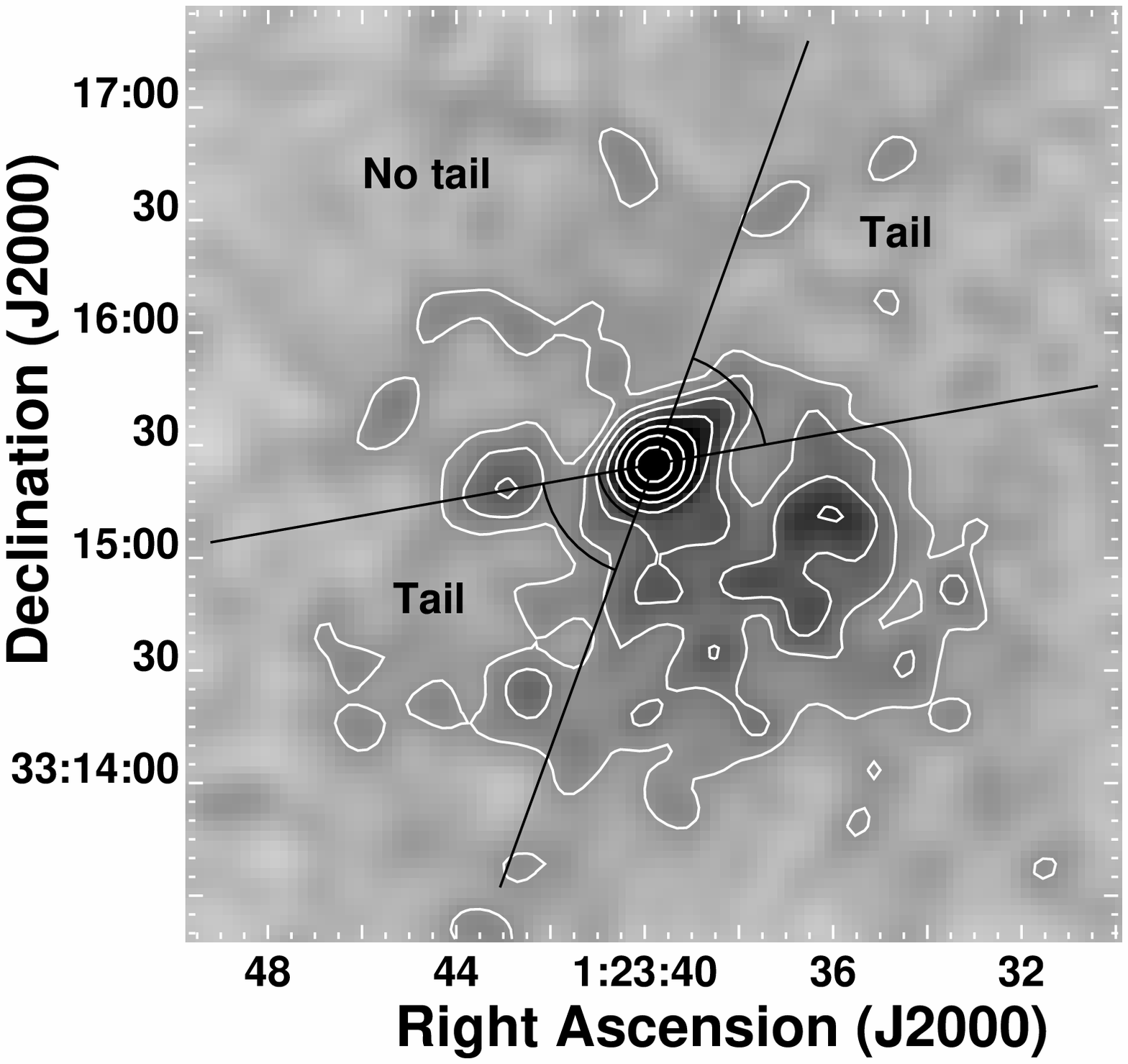,angle=0,width=0.55\textwidth}\hspace{-7.2cm}\raisebox{1.5cm}{\parbox[b]{1cm}{\bf(a)}}\hspace{7.2cm}}
\vspace{0.5cm}
\centerline{\psfig{figure=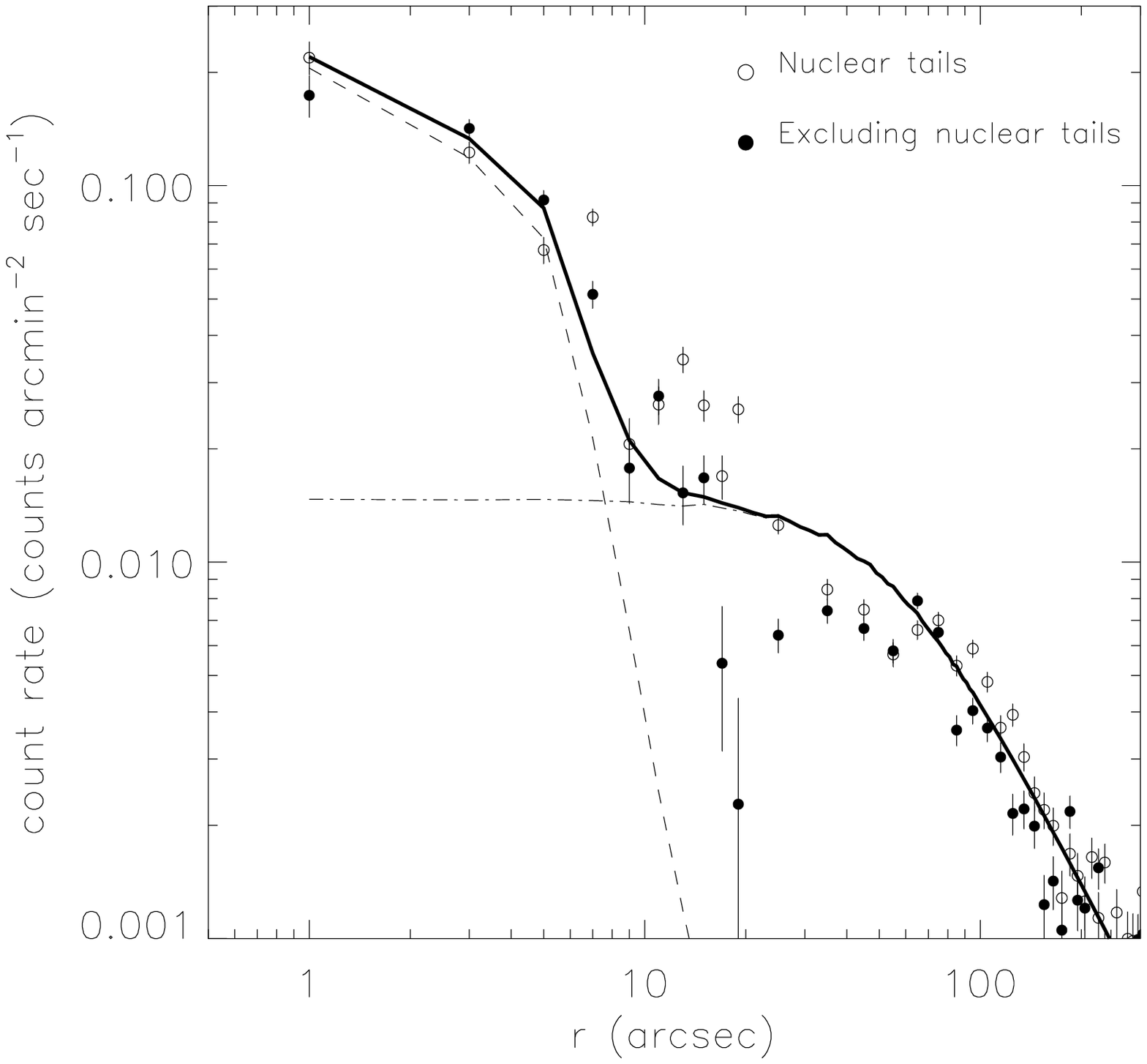,angle=0,width=0.58\textwidth}\hspace{-7.2cm}\raisebox{1.5cm}{\parbox[b]{1cm}{\bf(b)}}\hspace{7.2cm}}
\caption{{\bf (a)} X-ray surface brightness image of NGC 507 showing the NW and SE sectors used to extract the nuclear tail profiles and the NE sector excluding the nuclear tail contribution.  
{\bf (b)} HRI radial profiles of the NGC 507 central region including (empty circles) and excluding (filled circles) the contribution of the nuclear tails. 
Both profiles are extracted excluding the secondary peaks (sources No.5 and 9 in Table \ref{ngc507_source_tab}).
The meaning of the lines is the same as in Figure \ref{ngc507_model_profile}.}
\label{ngc507_notail}
\end{figure*}

To study the excess in the range $10"<r<20"$ we derived a radial brightness profile adding together the counts extracted in two sectors  ($100^\circ<$P.A.$<160^\circ$ and $280^\circ<$P.A.$<340^\circ$; `{\bf Tail}' in Figure \ref{ngc507_notail}a) aligned with the tails extending on the NW and SE side of the central X-ray peak ($\S$ \ref{ngc507_brightness}). In Figure \ref{ngc507_notail}b, this  profile is compared with the one obtained in the NE sector ($340^\circ<$P.A.$<100^\circ$; `{\bf No tail}' in Figure \ref{ngc507_notail}a) where the central component gradient is steeper. The profile in the region excluding the nuclear tails shows a better agreement with the bidimensional model out to 15"; at larger radii it falls below the model because it is missing much of the contribution of the galactic component, which is centered SW of the main peak (Figure \ref{ngc507_model}).  

\begin{figure*}[p]
\vspace{0.5cm}
\centerline{\psfig{figure=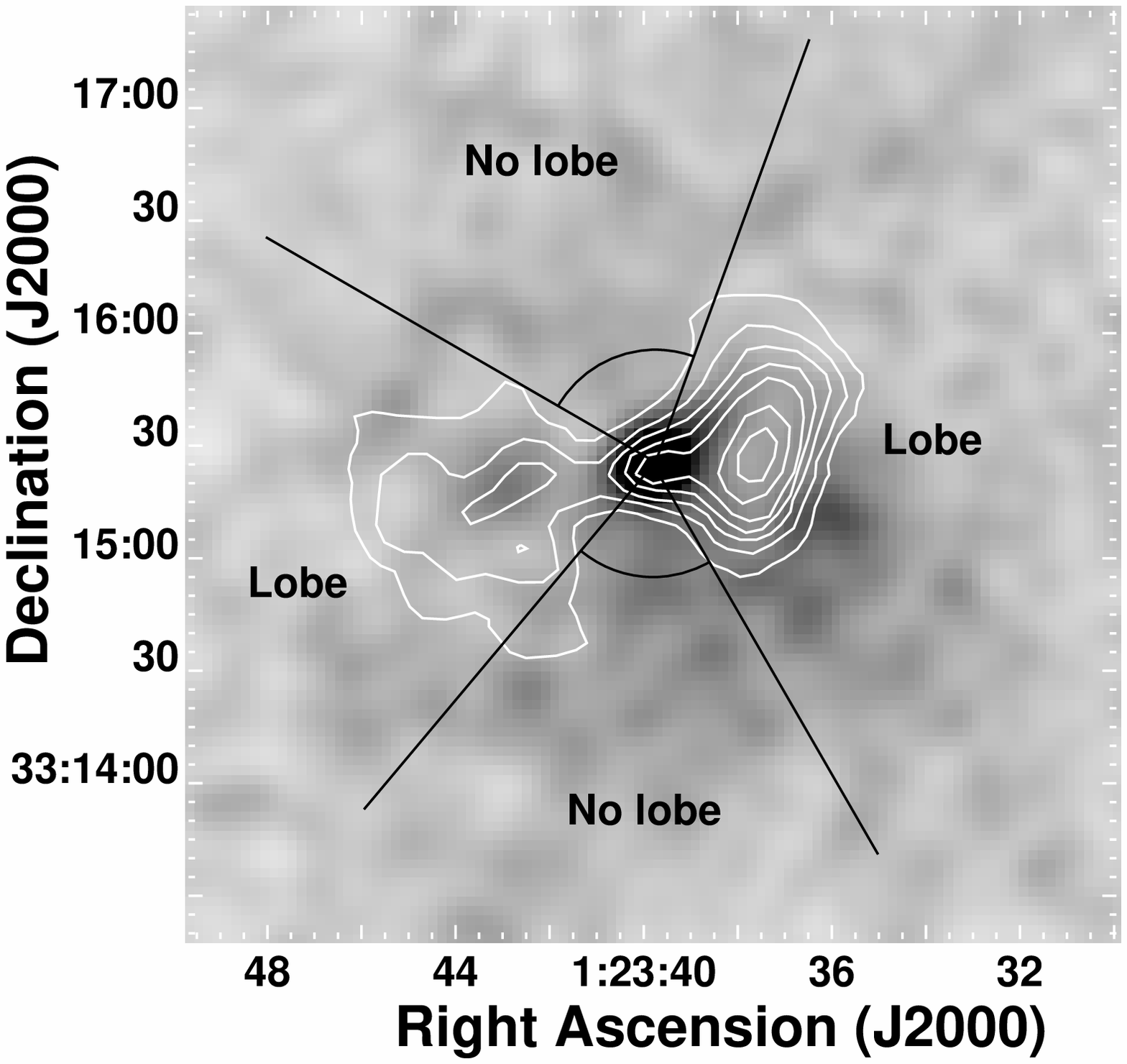,angle=0,width=0.55\textwidth}\hspace{-7.2cm}\raisebox{1.5cm}{\parbox[b]{1cm}{\bf(a)}}\hspace{7.2cm}}
\vspace{0.5cm}
\centerline{\psfig{figure=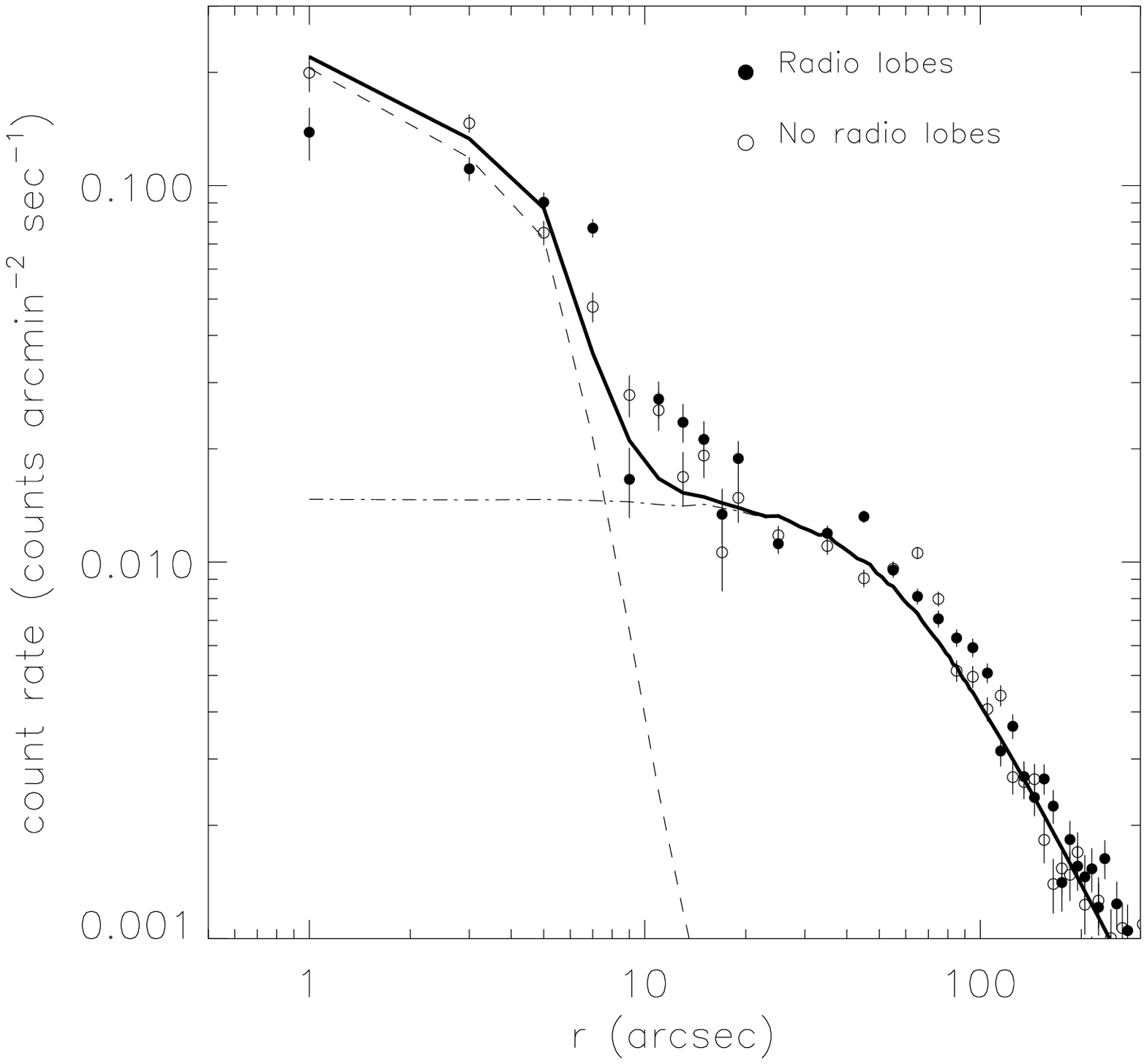,angle=0,width=0.58\textwidth}\hspace{-7.2cm}\raisebox{1.5cm}{\parbox[b]{1cm}{\bf(b)}}\hspace{7.2cm}}
\caption{{\bf (a)} X-ray surface brightness image of NGC 507 showing the East and West (North and South) sectors used to extract the X-ray profiles in the regions occupied (avoided) by the radio lobes. 
{\bf (b)} HRI radial profiles of the NGC 507 central regions occupied (filled circles) and avoided (empty circles) by the radio lobes. The meaning of the lines is the same as in Figure \ref{ngc507_model_profile}.}
\label{ngc507_nolobes}
\end{figure*}

Since most of the morphological structures seen in the central halo of NGC 507 may be due to the presence of the radio-emitting plasma, we also derived radial profiles in sectors occupied or avoided by the radio lobes (Figure \ref{ngc507_nolobes}a). The results are shown in Figure \ref{ngc507_nolobes}b. The profile avoiding the radio lobe regions shows a better agreement with the bidimensional model through all the range covered by the HRI data; conversely the profile extracted from regions occupied by the lobes shows significant deviations from the bidimensional model.

We can summarize the results of this section as follows:
 the bidimensional model is a reasonable representation of the X-ray brightness distribution. The main deviations from the model are associated to regions where the hot gas is presumably interacting with the radio emitting plasma; when such regions are ignored the radial profiles show a good agreement with the model profile.

\subsection{Density, Cooling Time and Mass Profiles}
\label{ngc507_dens}

\begin{figure}[t]
\centerline{\psfig{figure=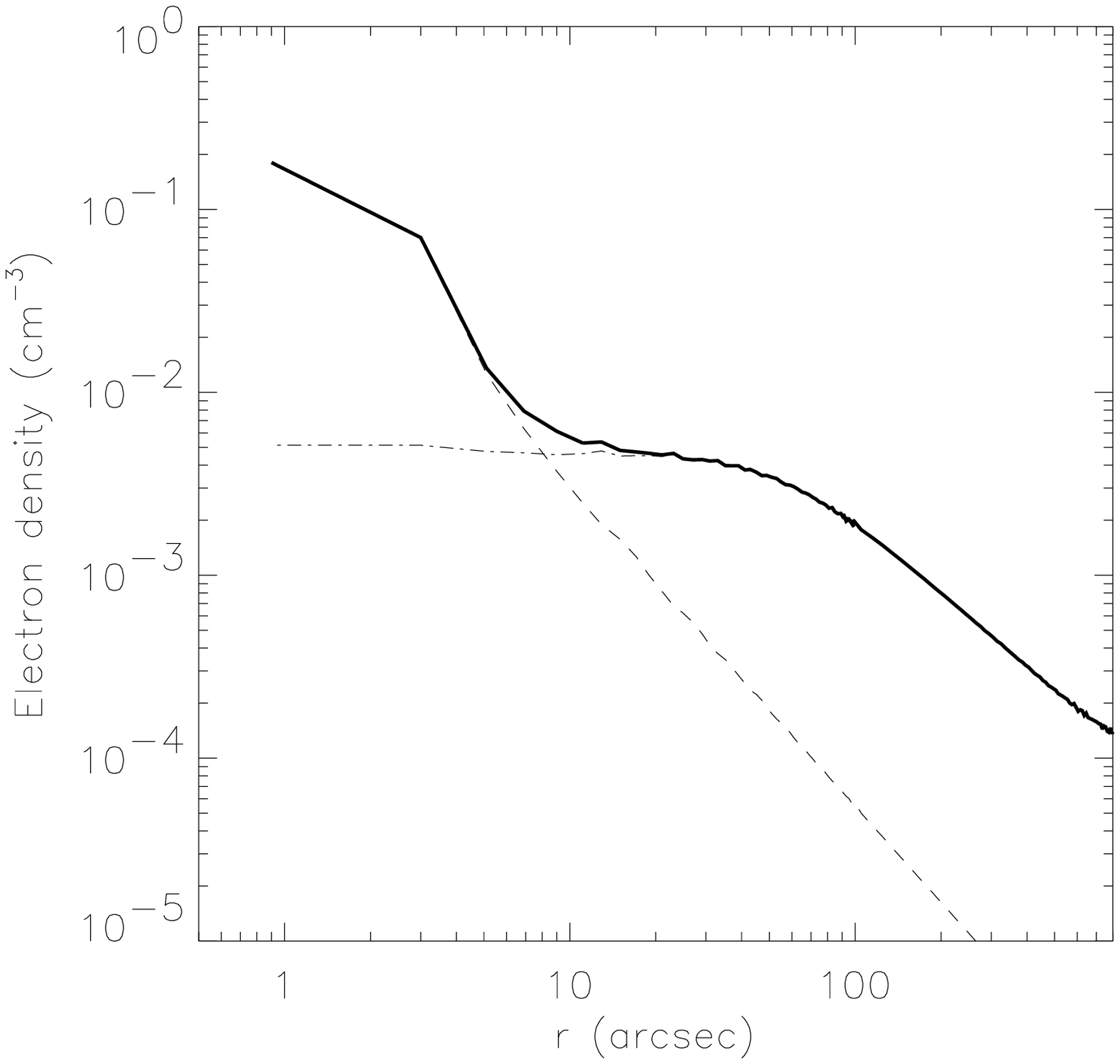,angle=0,width=0.6\textwidth}}
\caption{Deprojected density profile of the NGC 507 gaseous halo (continuous line). The dashed and dot-dashed lines show, respectively, the contribution of the central and galactic component.}
\label{ngc507_density_prof}
\end{figure}

To derive the density profile of the X-ray emitting gas we followed the method discussed in $\S$ \ref{empirical} and applied to NGC 1399 and NGC 1404 in $\S$
\ref{dens_par} and \ref{dens_par_1404} respectively: we deprojected the measured emission in concentric shells assuming an isothermal temperature profile with $kT=1.1$ keV and the cooling function given by \cite{Sar87}. As discussed in $\S$ \ref{ngc507_brightness_prof}, the bidimensional model provides a good representation of the overall X-ray emission; it also models the X-ray profile excluding the contribution of regions where the interaction with the radio-emitting plasma causes strong fluctuations in the surface brightness.
For such reason we use the model profile shown in Figure \ref{ngc507_model_profile} to derive the underlying gas density. Since the temperature profile measured by \cite{Kim95} shows the presence of a central temperature drop, we also used a lower temperature for the central component, obtaining an almost identical result.
The deprojected density is shown in Figure \ref{ngc507_density_prof}. The dashed and dot-dashed lines represent the contribution of the galactic and central component respectively.

\begin{figure}[t]
\centerline{\psfig{figure=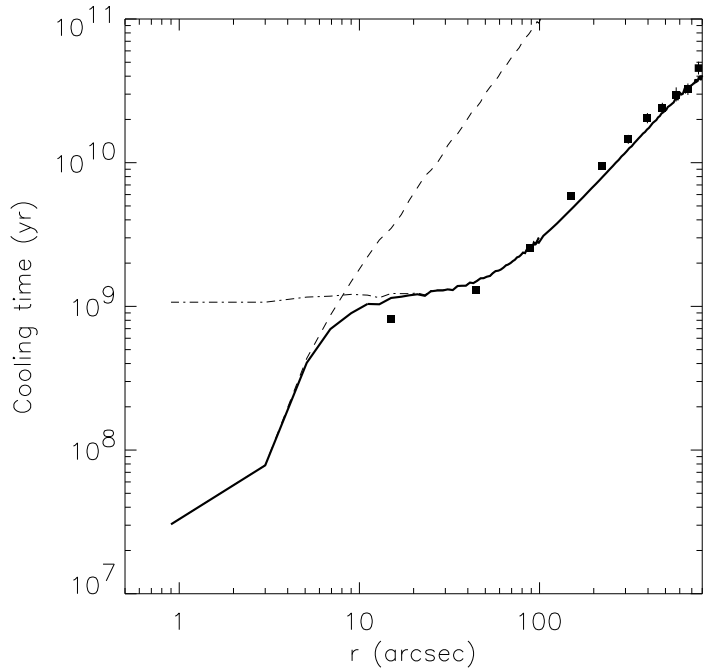,angle=0,width=0.6\textwidth}}
\caption{Cooling time profiles of NGC 507. The lines have the same meaning as in Figure \ref{ngc507_density_prof}. The cooling times derived by \cite{Kim95} are shown as filled squares.}
\label{ngc507_CT}
\end{figure}

Using this density profile we calculated the cooling time (equation \ref{CT}) of the NGC 507 halo shown in Figure \ref{ngc507_CT}. Our result is in good agreement with the \cite{Kim95} estimate at large galactocentric distances ($r>60"$) with minor differences due to their use of a non-isothermal temperature profile. At smaller radii our HRI data are able to sample the characteristics of the central component, whose cooling time is more than one order of magnitude smaller than those of the galactic component. Both components, however, have cooling times smaller than the Hubble time ($10^{10}$ yr) within the cooling radius $\simeq 250"$ (80 kpc).

\begin{figure*}[p]
\centerline{\psfig{figure=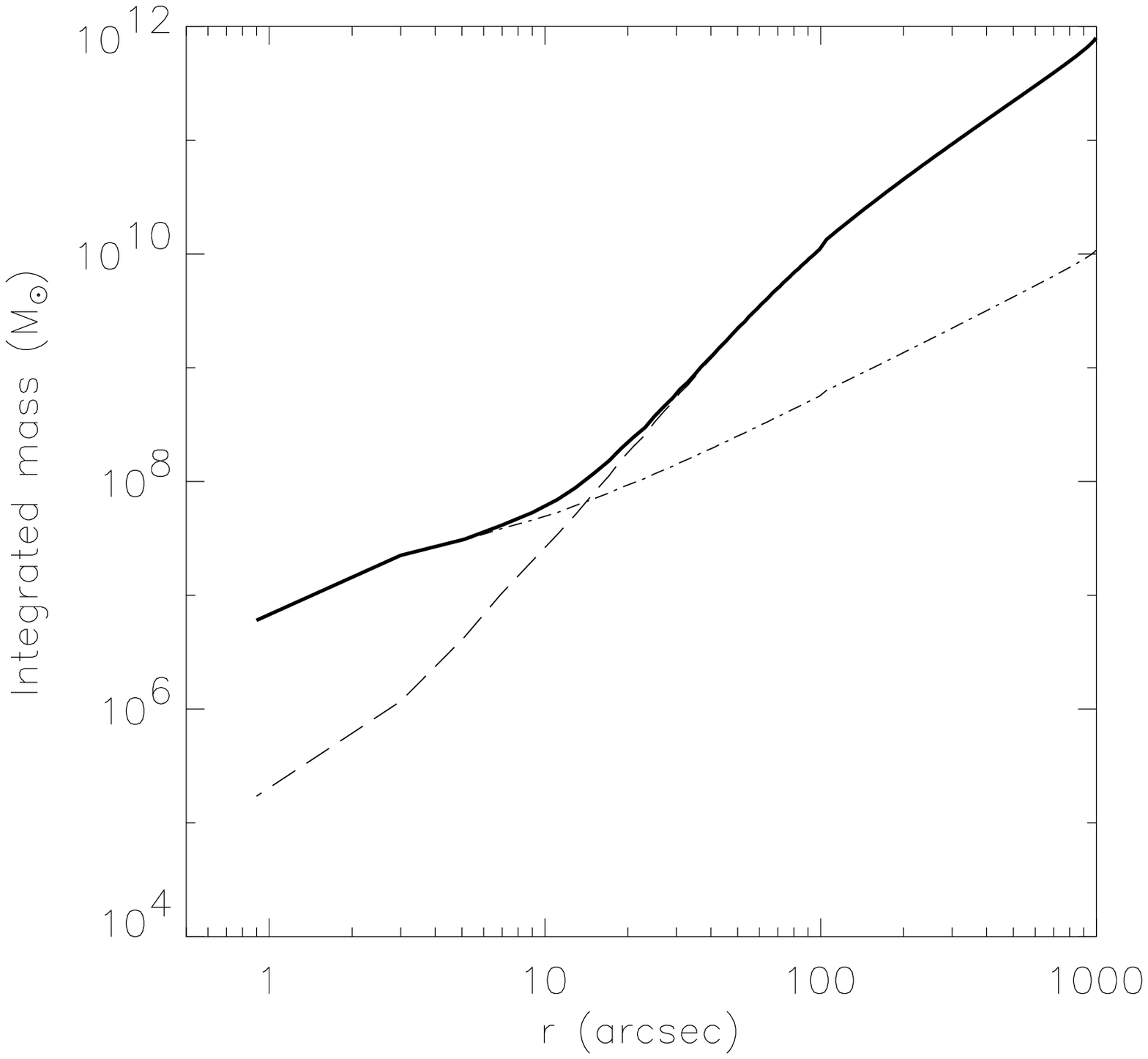,angle=0,width=0.55\textwidth}\hspace{-1.5cm}\raisebox{2cm}{\parbox[b]{1cm}{\bf(a)}}\hspace{1.5cm}}
\centerline{\psfig{figure=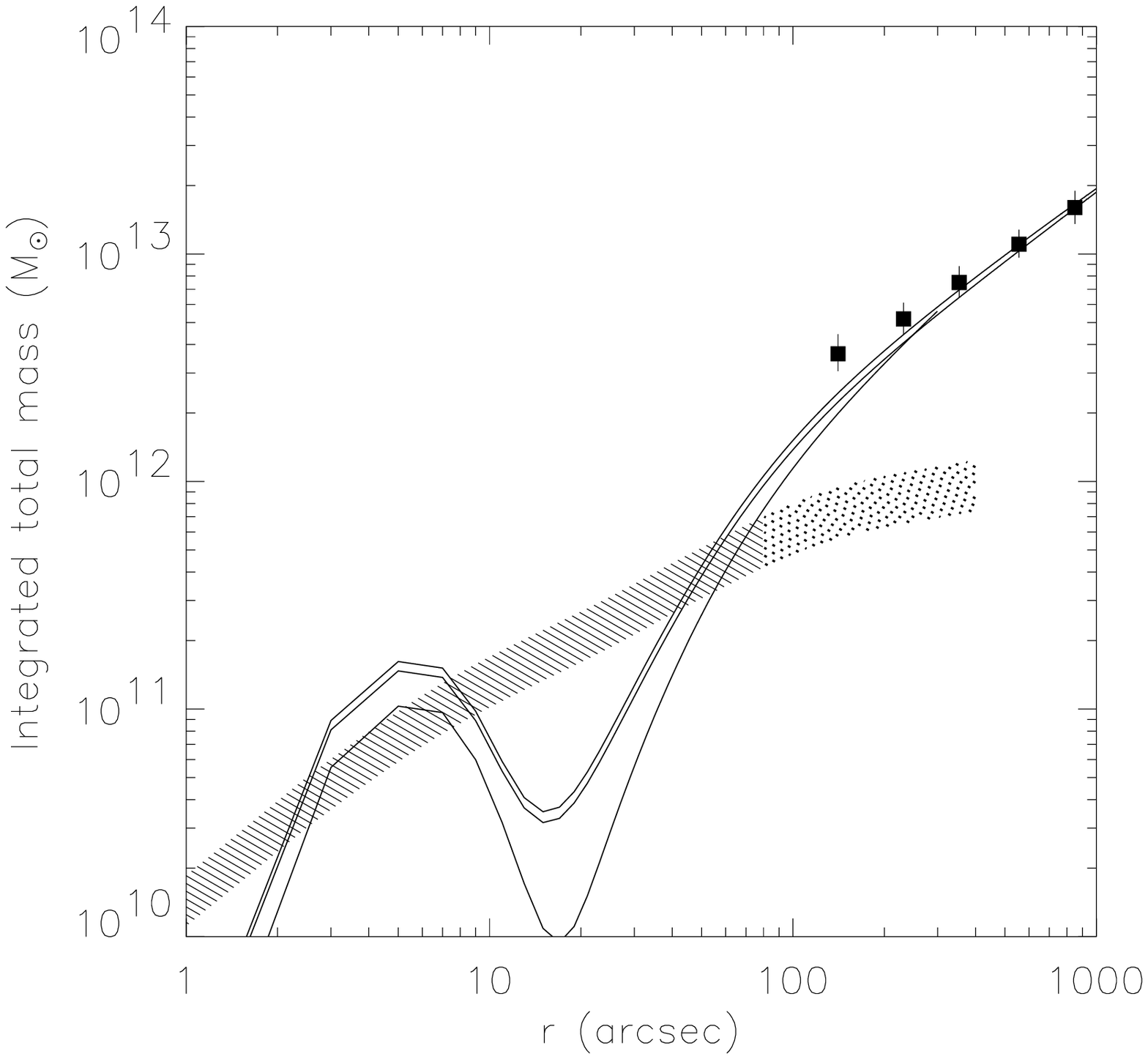,angle=0,width=0.55\textwidth}\hspace{-1.5cm}\raisebox{2cm}{\parbox[b]{1cm}{\bf(b)}}\hspace{1.5cm}}
\caption{{\bf (a)} Integrated gaseous mass within $r$. The meaning of the lines is the same as in Figure \ref{ngc507_density_prof}. {\bf (b)} Integrated total mass within $r$. The continuous lines represent the gravitating mass estimates obtained assuming the different temperature profiles described in the text. 
The total mass estimate of \cite{Kim95} is shown as filled squares.
The shaded (dotted) region shows the stellar mass contribution estimated (extrapolated) from the optical profile measured by \cite{Gonzalez00}, assuming a M/L ratio ranging from 6 to 8 M$_\odot$/L$_\odot$. Notice the unphysical behavior of the total mass profile for $r<50"$, which is discussed in the text.}
\label{ngc507_mass}
\end{figure*}

Figure \ref{ngc507_mass} shows both the integrated gaseous mass and the integrated total mass as a function of radius. The gaseous mass, calculated from the gas density profile assuming cosmic abundances (see $\S$ \ref{dens_par}), yields $\simeq 10^{12}$ M$_\odot$ within 1000" (320 kpc). The contribution of the central component, which dominates the inner 15" (5 kpc), is estimated to be $\sim 10^{10}$ M$_\odot$ at $r=1000"$, representing $\sim 1\%$ of the total gaseous mass.
 
The total mass profile was calculated using equation (\ref{mass}). We assumed three different temperature profiles: an isothermal profile at 1.1 keV, a linear fit $kT=-2.3\times 10^{-4}~r+1.21$ keV ($r$ is expressed in arcsec) to account for the temperature decline after $\sim 150"$, and a power law $kT=0.04~r^{0.5}+0.7$ keV to fit the central temperature drop. Figure \ref{ngc507_mass}b shows that the total mass profiles are weakly dependent on the assumed temperature profile, except in the inner 100" (32 kpc) where the central mass obtained accounting for the low central temperature is $\sim 50\%$ smaller.
The shaded region represents the total stellar mass calculated from the surface brightness profile measured by \cite{Gonzalez00} and assuming a M/L ratio ranging from 6 to 8 M$_\odot$/L$_\odot$. 

For $r>50"$ our mass profile is in good agreement with \cite{Kim95} and suggests the presence of dark matter, since it exceeds the stellar mass estimate extrapolated from the optical profile measured by \cite{Gonzalez00}. For r$<50"$ the X-ray mass estimate yields an unphysical result: in fact it falls below the stellar mass and its gradient is $<0$ for $10"<r<50"$, which is impossible for an integrated profile. The difficulties involved in deriving accurate mass profiles in the halo core, from X-ray measurements, have been already noticed in previous works and can be explained  by the presence of non-gravitational effects\footnote{Within a few arcsec from the main emission peak we must also consider the influence of the PRF. Even if the bidimensional model accounts for instrumental effects, an accurate deconvolution is very difficult.} \citep[see for instance][]{Brighenti97}. In our case we have seen  how the presence of the nuclear radio source strongly perturbs the central halo morphology ($\S$ \ref{ngc507_Xradio}). This is reflected in the gas density profile and thus on the total mass determination. Although we derived the total mass profile through the bidimensional model, which is less affected by X-ray/Radio interactions (see $\S$ \ref{ngc507_brightness_prof}), the reconstruction of the unperturbed gas distribution is a difficult task.
However we notice that for $2"<r<10"$, i.e. where the central component emission dominates, the total mass profile does not greatly exceed the stellar mass, suggesting that the galactic center is dominated by luminous matter. This result is in agreement with those found in other galaxies \citep[e.g.][]{Brighenti97} and will be discussed in detail in $\S$ \ref{centralpeak_discuss}.

\subsection{Discrete Sources}
\label{ngc507_sources}

\begin{figure}[p]
\centerline{\psfig{figure=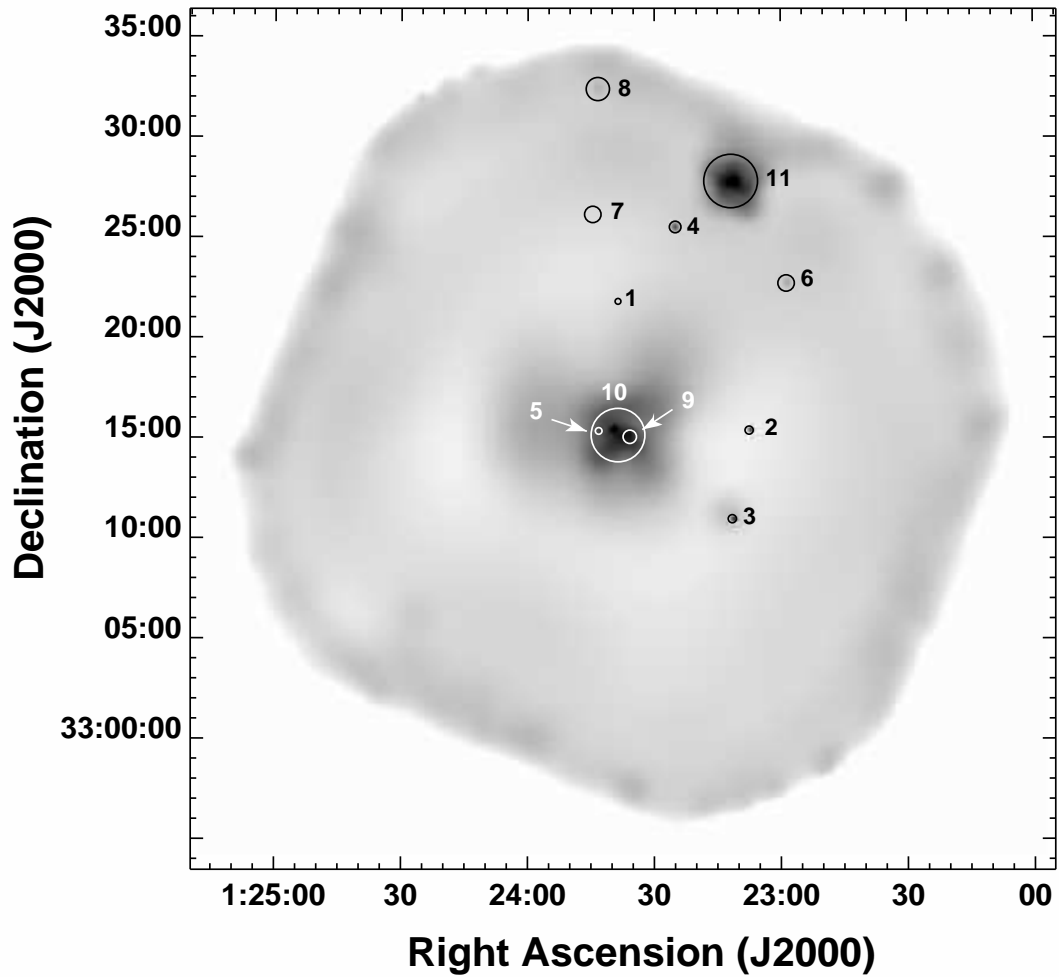,angle=0,width=0.9\textwidth}}
\caption{Discrete sources detected by the wavelets algorithm superimposed on the adaptively smoothed HRI field. The circles show the region of maximum S/N ratio for each source. The position of sources No.5 and 9 is shown in greater detail in Figure \ref{ngc507_centbox}a.}
\label{ngc507_sources_fig}
\end{figure}

To study the population of discrete sources in the HRI field we used the wavelets algorithm developed by \citet{Dam97a,Dam97b} and already described in $\S$ \ref{sources}. Using a S/N threshold of 4.2\footnote{This value differs from the one used in $\S$ \ref{sources} for NGC 1399, since it depends on the S/N ratio of the data and thus on the exposure time.} (corresponding to a contamination of 1 spurious source per field) the algorithm detected 11 sources shown in Figure \ref{ngc507_sources_fig}.
The properties of these sources, as measured by the wavelets algorithm, are reported in Table \ref{ngc507_source_tab}.
Sources No.10 and 11 are missing from the table since they are NGC 507 and NGC 499. We also notice that sources No.5 and 9 correspond to the secondary central peaks discussed in $\S$ \ref{ngc507_brightness}.

To derive the X-ray fluxes we used the PIMMS software to calculate a conversion factor between the HRI count rate and the incident flux. We assumed the same spectral models discussed in $\S$ \ref{sources}: a power law with photon index $1.96\pm0.11$ ($\mathit{f}_{HRI}^b{pow.law}$ in Table \ref{ngc507_source_tab}) following the results of \citet{Has93} for faint sources, and a thermal spectrum with $kT=0.52$ and abundance $Z\sim 0.2~Z_\odot$ ($\mathit{f}_{HRI}^{RS}$ in Table \ref{ngc507_source_tab}) as found by \cite{Kim95}. Assuming a galactic absorption of $5.3\times 10^{20}$ cm$^{-2}$ we obtain a conversion factor of 1 cps=$6.925\times 10^{-11}$ ergs s$^{-1}$ and 1 cps=$4.076\times 10^{-11}$ ergs s$^{-1}$ respectively in the 0.1-2.4 keV energy range. We notice that the energy range is different from the 0.5-2.0 keV used in $\S$ \ref{sources} since here we are mainly interested in comparing our results with those of \citet{Kim95}. Restricting our energy band to the 0.5-2.0 keV interval, as we have done in $\S$ \ref{sources}, results in a reduction of the X-ray fluxes of $\sim 55\%$ for the power law model and of $\sim 34\%$ for the thermal model. 
 
In Table \ref{ngc507_source_tab2} we cross-correlate the sources detected in the HRI field with those found by \citet{Kim95}. The fluxes that they derive from PSPC data are reported in column $\mathit{f}_{PSPC}^{RS}$.
We searched for long-term variability of the matching sources comparing these fluxes with those measured by the wavelets algorithm (Table \ref{ngc507_source_tab}). We mark as variable those sources whose flux varied by more than $3\sigma$ between the PSPC and HRI observations, i.e. between 1993 and 1995. We found that only source No.3 show significant variability in this interval.
To check if the different procedures, extraction radii and background subtraction, significantly affect the PSPC count rate determination, we extracted from the GALPIPE database (see $\S$ \ref{model}) the count rates measured by the wavelets algorithm on the second of the PSPC observations (RP600254a01) used by \citet{Kim95}. The results are in good agreement with the \citet{Kim95} determinations, except for source No.4. In this case however the proximity to the NGC 499 halo and the large PSPC PRF may compromise the wavelets accuracy.
We must notice that sources No.5 and 9, which are likely to be produced by the interaction of the radio lobes with the surrounding ISM (see $\S$ \ref{ngc507_Xradio_discuss}), have no PSPC counterpart since they are undetected due to the lower spatial resolution of this instrument.

\begin{table}[t]
\begin{centering}
\caption{Discrete Sources - Results of the Wavelets algorithm \label{ngc507_source_tab}}
\footnotesize
\begin{tabular}{cccrrrccc}
\\
\tableline
\tableline
Source & R.A.         & Dec. & 3$\sigma$ Radius & Counts 	& Max S/N & Count Rate  & $\mathit{f}_{HRI}^{pow.law}$\tablenotemark{(a)} & $\mathit{f}_{HRI}^{RS}$\tablenotemark{(a)} \\
 No.  & \multicolumn{2}{c}{(J2000)} & (arcsec)         &        &         & ($10^{-4}$ Cnts sec$^{-1}$) &    &  \\
\tableline
1  & 1:23:38.6    & +33:21:48	 & 8.7      	& 13$\pm$6  	&  4.3    & 5$\pm$2    	& 35$\pm 14$     				  & 20$\pm 8$ 	   			\\
2  & 1:23:07.5    & +33:15:23	 & 12.2		& 31$\pm$9  	&  7.0    & 11$\pm$3    & 76$\pm 21$     				  & 45$\pm 12$ 	   			\\
3  & 1:23:11.5    & +33:10:58	 & 12.2		& 37$\pm$10  	&  8.0    & 13$\pm$4   	& 90$\pm 28$     				  & 53$\pm 16$ 	   			\\
4  & 1:23:25.0    & +33:25:31	 & 17.3	  	& 47$\pm$13	&  7.6    & 17$\pm$4    & 118$\pm 28$    				  & 69$\pm 16$ 	   			\\
5  & 1:23:43.2    & +33:15:21	 & 17.3		& 33$\pm$11  	&  4.8    & 12$\pm$4   	& 83$\pm 28$     				  & 49$\pm 16$ 	   			\\
6  & 1:22:58.7    & +33:22:43	 & 24.5		& 39$\pm$12  	&  5.2    & 14$\pm$4    & 97$\pm 28$     				  & 57$\pm 16$ 	   			\\
7  & 1:23:44.6    & +33:26:08	 & 24.5		& 33$\pm$11  	&  4.6    & 13$\pm$4    & 90$\pm 28$     				  & 49$\pm 16$ 	   			\\
8  & 1:23:43.4    & +33:32:23	 & 34.6		& 46$\pm$14  	&  4.9    & 21$\pm$6    & 145$\pm 42$    				  & 65$\pm 20$ 	   			\\
9  & 1:23:35.8    & +33:15:03	 & 34.6		& 104$\pm$28  	&  6.1    & 37$\pm$10   & 256$\pm 69$    				  & 151$\pm 41$ 	   			\\
\tableline
\end{tabular}
\tablenotetext{(a)}{ Flux in the 0.1-2.4 KeV band in units of $10^{-15}$ erg s$^{-1}$ cm$^{-2}$.}
\end{centering}
\end{table}

\begin{figure}[t]
\centerline{\psfig{figure=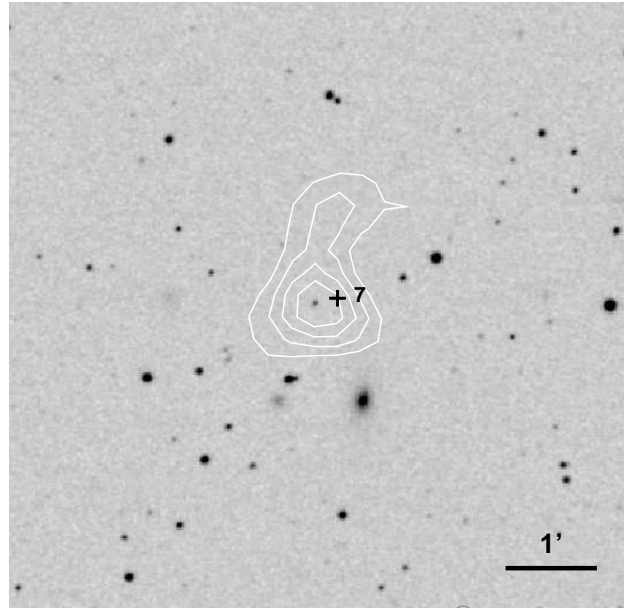,angle=0,width=0.5\textwidth}}
\caption{Radio contours superimposed on the optical DSS image of the region containing source No.7. The position of the X-ray centroid is marked by a cross.}
\label{NVSS_source7}
\end{figure}

The optical DSS image of the NGC 507/499 region was visually inspected to find optical counterparts of the X-ray sources within a $2\sigma$ radius (Table \ref{ngc507_source_tab}) from the X-ray centroid. In this case we choose a smaller size ($2\sigma$) than the one adopted for NGC 1399 since the optical image of the NGC 507 region is more crowded. The results are summarized in column ``Opt.id.'' of Table \ref{ngc507_source_tab2}. We also cross-correlated our source list with the catalogs contained in the NED database. We found that source No.7 corresponds to the radio source NVSS J012345+332554 (Figure \ref{NVSS_source7}).


\begin{table}[p]
\begin{centering}
\caption{Discrete Sources - Comparison with PSPC results and catalogues \label{ngc507_source_tab2}}
\footnotesize

\begin{tabular}{ccccccc}
\\
\tableline
\tableline
Source 	& PSPC\tablenotemark{(a)} & $\mathit{f}_{PSPC}^{RS}$\tablenotemark{(a,b)}& $\mathit{f}_{HRI}^{RS}$\tablenotemark{(b)}& ~~~Time Var.	   & Opt.id. & NED obj.\\
 No.   	&  No.         	       	&     &  											& long-term\tablenotemark{(c)} 		 					     	   &	     & \\
\tableline
1  	& -       		& -     				 & 20$\pm 8$					      & -		   & n       & - \\
2  	& 8       		& $21\pm 10$    			 & 45$\pm 12$					      & n		   & y       & - \\
3  	& 10      		& $171\pm 20$   			 & 53$\pm 16$					      & y		   & y       & - \\
4  	& 12      		& $91\pm 12$    			 & 69$\pm 16$					      & n		   & y       & - \\
5  	& -       		& -     				 & 49$\pm 16$					      & -		   & n       & - \\
6  	& 7       		& $81\pm 12$    			 & 57$\pm 16$					      & n		   & y       & - \\
7  	& 13      		& $43\pm 10$    			 & 49$\pm 16$					      & n		   & y       & NVSS J012345+332554 \\
8  	& -       		& -     				 & 65$\pm 20$					      & -		   & n       & - \\
9  	& -       		& -     				 & 151$\pm 41$  				      & -		   & n       & - \\

\tableline
\end{tabular}

\tablenotetext{(a)}{ As reported in Table 2 of \cite{Kim95}}
\tablenotetext{(b)}{ Flux in the 0.1-2.4 KeV band in units of $10^{-15}$ erg s$^{-1}$ cm$^{-2}$.}
\tablenotetext{(c)}{ Difference between PSPC and HRI fluxes: `y' means variable source ($>3\sigma$), `n' means no variability detected.}
\end{centering}\end{table}

\begin{table}[p]
\begin{centering}
\caption{PSPC-HRI Sources Comparison - Upper limits from HRI data \label{ngc507_source_tab3}}
\footnotesize

\begin{tabular}{ccccc}
\\
\tableline
\tableline
PSPC\tablenotemark{(a)} & $\mathit{f}_{PSPC}^{RS}$\tablenotemark{(a,b)} & $\mathit{f}_{HRI}^{RS}$ & $\mathit{f}_{HRI}^{\rm RS}$ & Time Var.\\
 No.		      &     &    &  $3\sigma$ upper~limit   & long-term\tablenotemark{(c)}\\
\tableline
3		       & $79\pm 15$	       & $14\pm 25$ 	& 90	   & - \\
4		       & $36\pm 10$	       & $21\pm 24$ 	& 93	   & - \\
5		       & $61\pm 14$	       & $40\pm 24$	& 111	   & - \\
6  		       & $300\pm 27$           & $116\pm 42$ 	& 243 	   & y \\
9  		       & $45\pm 11$            & $8\pm 23$	& 77 	   & - \\
11  		       & $42\pm 11$            & $14\pm 24$	& 86 	   & - \\
14		       & $38\pm 10$	       & $32\pm 25$    	& 108 	   & - \\
15		       & $32\pm 10$	       & $9\pm 21$    	& 72 	   & - \\
16		       & $28\pm 10$	       & $9\pm 22$    	& 75 	   & - \\
17		       & $51\pm 11$	       & $11\pm 22$    	& 78 	   & - \\
18		       & $46\pm 13$	       & $67\pm 23$    	& 135 	   & - \\
20		       & $60\pm 14$	       & $19\pm 21$    	& 83 	   & - \\

\tableline
\end{tabular}

\tablenotetext{(a)}{As reported in Table 2 of \cite{Kim95}}
\tablenotetext{(b)}{ Flux in the 0.1-2.4 KeV band in units of $10^{-15}$ erg s$^{-1}$ cm$^{-2}$.}
\tablenotetext{(c)}{\small Difference between PSPC flux and HRI upper limit: `y' means variable source ($>3\sigma$), `-' means no variability detected.}
\end{centering}
\end{table}

Many of the sources detected in the PSPC data by \citet{Kim95} are not detected in the HRI field. We derived upper limits on the HRI flux to find whether this result is due to the lower HRI S/N ratio or by the temporal variability of the sources. For all PSPC sources not detected in the HRI field, we extracted count rates using the radii reported in Table 2 of \citet{Kim95} rescaled by a factor 1/3 to take into account the difference between the HRI and PSPC PRF.
The background was measured locally in an annulus of radius twice the one used for the source.
The resulting fluxes (obtained with the same conversion factors discussed above) and $3\sigma$ upper limits, are shown in Table \ref{ngc507_source_tab3}. The comparison with the PSPC fluxes shows that all sources except one are below the HRI sensitivity. Source No.6 is the only one which must be significantly variable since the expected flux is higher than the $3\sigma$ upper limit indicated by the HRI data.

The implications of these results on the nature of the discrete sources found in the NGC 507/499 field will be discussed in $\S$ \ref{ngc507_sources_discuss}.

\section{DISCUSSION}

\subsection{Origin of the Central Component}
\label{ngc507_central_discuss}
The surface brightness distribution shown in Figure \ref{ngc507_centbox}a,b revealed the presence of a bright X-ray emission peak centered on the optical galaxy. The surface brightness profiles (Figure \ref{ngc507_profile} and \ref{ngc507_model_profile}) show that the peak dominates the X-ray emission within the central 10 arcsec (3 kpc). The HRI data alone do not allow to distinguish between a pointlike or extended nature of the central peak, since the comparison of the X-ray profile with the HRI PRF (Figure \ref{ngc507_profile}) and the small core radius of the central component obtained from the bidimensional fit (Table \ref{ngc507_fit_tab}) are marginally consistent with the presence of a nuclear X-ray source.  However the displacement of the radio nucleus with respect to the X-ray peak 
($\S$ \ref{ngc507_Xradio}) seems to exclude the discrete nature of this component. This result is confirmed by the {\it Chandra} X-ray profile obtained recently by \cite{Forman00}.

We notice that \cite{Canosa99}, using the ROSAT HRI data, found an upper limit on the X-ray flux of a possible nuclear source that exceeds, by more than two orders of magnitude, the emission expected from the Radio data. The extended nature of the central peak implies that the flux of the nuclear source is much smaller than though before, thus reducing the disagreement between the radio and X-ray data. 

The comparison between the optical and X-ray profile (Figure \ref{ngc507_model_profile}) seems to rule out the possibility that the central peak is due to the unresolved emission of galactic X-ray binaries. In this case the X-ray profile of the central component would be close to the optical one, since the X-ray source population should follow the stellar distribution.
The previous considerations support the gaseous nature of the central peak, even though we expect some contribution from the unresolved sources present in the galaxy center. 

\cite{Kim95} noticed that significant mass deposition should be present within the cooling radius since the cooling times of the gaseous halo are smaller than the Hubble time (Figure \ref{ngc507_CT}). Their estimate (see equation \ref{Mdot}), after rescaling to our adopted distance\footnote{The mass inflow rate is inferred from the gas density (equation \ref{Mdot}) which in turn depends on the X-ray luminosity. Since we can only measure the X-ray flux we have to rescale their estimate to our adopted distance.}, yields a mass inflow rate of $\sim 20$ M$_\odot$ yr$^{-1}$. If the central X-ray peak was due to the mass deposited by the cooling flow, we would expect to find $\sim 2\times 10^{11}$ M$_\odot$ within 250 arcsec from the galaxy center while the total mass of the central component is only $\sim 2\times 10^{9}$ (Figure \ref{ngc507_mass}a) within the same radius. An identical result holds if we assume an inhomogeneous flow where $\dot{M}\propto r$ \citep{Fabian94}. In fact, extrapolating the mass inflow down to $r=10"$ where the contribution of the central component is better defined, we obtain a deposited mass of $\sim 10^{10}$ M$_\odot$ while we observe less than $10^{8}$ M$_\odot$. This result holds also if we take into account all the uncertainties included in the calculation such as the gas filling factor and the age of the cooling flow. 

The problem of the missing deposited mass has been largely debated in literature \cite[see for instance][]{Fab89,Fabian94,rang95}. Recently the failure to observe cool gas in the center of galaxies and clusters supposed to host large cooling flows \citep{Matsu01b,Pet01,Kaa01,Tam01} has led many authors to stress the failure of the standard cooling flow models. 
In a recent study of a sample of early-type galaxies, \cite{Matsu01} noticed that the ISM temperature of X-ray compact galaxies is compatible with $\beta_{spec}\simeq 1$ where $\beta_{spec}$ is defined in equation (\ref{betaspec}). She argues that this conclusion holds also for X-ray extended galaxies (such as NGC 507) if we consider only the inner galactic halo ($r<1.5 r_e$). The lower central temperatures would be explained by kinematical heating from stellar mass losses rather than by the cooling flow scenario.

In the case of NGC 507 the central X-ray peak may be produced by the gas ejected by the stellar population. The stellar origin of the gas is supported by the solar metallicity found by \cite{Buo00b} in the halo center. If the gas is in equilibrium with the stellar component, as suggested by \cite{Matsu01}, we expect to find a central temperature close to 0.6 keV. The work of \citep{Matsumo97}, which found $\beta_{spec}\sim 0.6$, does not rule out such a possibility since it does not take into account the temperature drop within $150"$. In order to sample the inner $15"$, where the central component emission dominates, and to see if such low temperatures are present in the center of NGC 507, we will have to exploit the {\it Chandra} or XMM resolution.

\subsection{X-ray/Radio Correlations}
\label{ngc507_Xradio_discuss}
We showed in $\S$ \ref{ngc507_brightness} that the central NGC 507 halo has a very complex morphology. Figure \ref{ngc507_centbox} revealed the presence of secondary emission peaks and cavities in the X-ray emission. As discussed in $\S$ \ref{interactions}, the weak dependence of the HRI spectral response on the temperature near 1 keV \citep{clar97}, suggests a correlation between surface brightness and density fluctuations of the hot gas. The comparison of these X-ray features with the radio maps of Figures \ref{ngc507_radio}a,b supports the scenario in which the hot gas has been displaced by the pressure of the radio-emitting plasma. In fact the very good agreement between the shape of the Western radio lobe and the X-ray cavity indicates that the lobe may have produced a low-density 'bubble' in the gaseous medium, compressing the gas at the edge of the lobe. This interaction is also likely to be responsible for the N-S elongation of the lobe since the radio plasma expanded avoiding the denser Western and Southern regions. 

The presence of such interactions is not so evident in the Eastern lobe. This may be due to projection effects and/or to the fact that the Eastern lobe seems to expand in a less dense environment (the X-ray centroid of the galactic component is displaced by $\sim 20"$=6 kpc South-West of the nuclear radio source). The Radio/X-ray comparison in Figure \ref{ngc507_radio}a shows that a secondary emission peak lies in the center of the Eastern radio lobe.
As already done in $\S$ \ref{Xradio_discuss}, we used equations (\ref{B_IC}) and (\ref{B_ME}) to calculate the expected X-ray emission due to Inverse Compton scattering of Cosmic Background photons into the X-ray band, for a grid of values of the involved parameters. We used the radio flux for the Eastern lobe measured by \cite{deRuiter86} and we measured the X-ray flux density extracting counts from a 25" circle centered on the secondary emission peak. Since the X-ray flux depends on the assumed spectrum, we estimated an upper ($1.18\times 10^{-31}$ ergs s$^{-1}$cm$^{-2}$Hz$^{-1}$) and lower ($8.87\times 10^{-32}$ ergs s$^{-1}$cm$^{-2}$Hz$^{-1}$) limit using respectively power laws with photon index $\alpha_{ph}=1.8$ and $\alpha_{ph}=3$. The results are shown in Table \ref{IC_tab_ngc507}. We found that the expected X-ray emission falls an order of magnitude below the measured excess since the ratio $B_{ME}/B_{IC}$ is always $>>1$ for any choise of the parameters.

Conversely, the higher resolution radio contours shown in Figure \ref{ngc507_radio}b suggest that the bulk of the radio emission may fall south of the peak, in a low emitting region. In this case both the secondary emission peak and the SE `tongue' found in the adaptive smoothed image (Figure \ref{ngc507_csmooth}) may be produced by hot gas compressed by the radio-emitting plasma, similarly to what seen for the Western lobe.

\begin{table}[t]
\caption{Eastern lobe magnetic field strenght.\label{IC_tab_ngc507}}
\begin{center}
\begin{tabular}{cccccccc}
\\
\hline
\hline
$\eta$ & $\phi$ & $\nu_1$ & $k$ & $S_X$ & $B_{ME}$ & $B_{IC}$ & $B_{ME}/B_{IC}$\\
& (rad) & (GHz) & & (ergs s$^{-1}$cm$^{-2}$Hz$^{-1}$) & ($\mu$G) & ($\mu$G) & \\
\hline
1 & $\pi /2$    & $1.00\times 10^{-2}$ &   1 & $1.18\times 10^{-31}$ & 2.5 & 0.08 & 32\\
1 & $\pi /6$    & $1.00\times 10^{-2}$ &   1 & $1.18\times 10^{-31}$ & 3.4 & 0.11 & 29\\
1 & $\pi /2$    & $1.00\times 10^{-2}$ & 100 & $1.18\times 10^{-31}$ & 7.8 & 0.08 & 99\\
1 & $\pi /2$    & $1.00\times 10^{-4}$ &   1 & $1.18\times 10^{-31}$ & 2.5 & 0.08 & 32\\
0.1 & $\pi /2$  & $1.00\times 10^{-2}$ &   1 & $1.18\times 10^{-31}$ & 4.9 & 0.08 & 62\\
0.01 & $\pi /2$ & $1.00\times 10^{-2}$ &   1 & $1.18\times 10^{-31}$ & 9.5 & 0.08 & 120\\
1 & $\pi /2$    & $1.00\times 10^{-2}$ &   1 & $8.87\times 10^{-32}$ & 2.5 & 0.09 & 27\\
1 & $\pi /6$    & $1.00\times 10^{-2}$ &   1 & $8.87\times 10^{-32}$ & 3.4 & 0.14 & 25\\
\hline
\end{tabular}
\end{center}
\tablenotetext{}{{\bf Note} - The calculation uses equations (\ref{B_IC}) and (\ref{B_ME}) assuming the following parameters: radio spectral index $\alpha_r=-0.8$, radio flux density $S_r=0.027$ Jy at $\nu_r=1.465$ GHz, lobe diameter $\theta=60"$ and path lenght through the source $s=19$ kpc. The X-ray flux $S_X$ density is calculated at 1 keV assuming a galactic absorption of $5.3\times 10^{20}$ cm$^{-2}$; the upper syncrotron cutoff is taken to be $\nu_2=10$ GHz. }
\end{table}

The thermal confinement of the radio lobes is supported by the interactions between the radio and X-ray emitting plasma discussed above. \cite{Morganti88} studied the pressure balance between the radio lobes and the surrounding ISM, finding that the thermal pressure greatly exceeded the radio one ($P_{thermal}/P_{radio}>7$). However their estimate was based on {\it Einstein} data, which resulted in a temperature twice the ROSAT PSPC value. Taking into account the ROSAT temperature reduces the disagreement by more than 50\%; the remaining excess may be explained by the departures from the minimum pressure conditions discussed in their work. 

Finally we notice that no sharp features, such as shocks or knots, are seen in the HRI image, as expected in a weak radio source \citep{Fanti87}.

\subsection{Large-scale Halo structure}
In section \ref{ngc507_brightness} we have shown that the HRI data reveal the presence of a large halo surrounding NGC 507. Even though the irregular morphology of the halo may be affected by the low S/N ratio of the HRI data at large radii, a comparison with the PSPC image shown in Figure 4 of \cite{Kim95} suggest that some of these features are indeed real. 

The X-ray emission is more extended in the region between NGC 507 and NGC 499, creating a large halo that surrounds the two galaxies.
The total mass profile (Figure \ref{ngc507_mass}b) shows that for $r>80"$ (25 kpc) the mass distribution is dominated by dark matter. If the hot gas distribution traces the gravitational potential this dark matter is distributed preferentially on group scales rather than being associated with the dominant galaxy of the cluster. In fact, on galactic scales, the total mass is compatible with the stellar matter contribution inferred from the surface brightness profile.

Exploiting the HRI resolution to study the inner halo ($r<5'$) we found ($\S$ \ref{ngc507_brightness}) a complex morphology. We have seen in the previous section that some structures are likely to be due to the interaction with the radio lobes. However the presence of some `tails', revealed by the adaptive smoothing technique (Figure \ref{ngc507_csmooth}), requires a different explanation. 
The study of the inner halo also showed that the optical galaxy is centered on the central X-ray peak and is thus displaced with respect to the galactic halo. This result is confirmed independently by the bidimensional fit, which shows that the galactic component centroid is located $\sim 22"$ (7 kpc) South-West of the optical galaxy (Figure \ref{ngc507_model}), and by the wavelets algorithm, which centered source No.10 (associated with the NGC 507 halo) South-West of source No.9 (coincident with the central peak, see Figure \ref{ngc507_centbox}a).
These results imply that the decentering of the galactic halo with respect to the optical galaxy is not an artifact produced by the X-ray/radio interactions in the halo core which displaced South-West the X-ray centroid, since the bidimensional model fitting (and to a lesser extent the wavelets algorithm) are strongly affected by the X-ray distribution at large radii.  

We can imagine different scenarios to explain these observations: in the first one the hot halo distribution traces the gravitational potential. The latter is associated with the galaxy group and, instead of being centered on the dominant galaxy, is displaced toward NGC 499. This small displacement seems compatible with the distribution of galaxies within the NGC 507 group studied by \cite{Barton98}.

A second possibility is that tidal interactions between NGC 507 and NGC 499 are responsible for the asymmetries in the hot gas distribution. In $\S$ \ref{interactions} we discussed the possibility that such interactions may produce large density fluctuations and the formation of filamentary structures. 
The presence of `clumps' of cooler gas in the halo, suggested by \cite{Kim95} (see $\S$ \ref{ngc507_sources_discuss}) may support such scenario.

The displacement of the optical galaxy with respect to the large-scale gas distribution may be produced by the motion of the stellar body within the larger galactic halo. The central component would be centered on the optical galaxy since it may be due to gas of stellar origin in equilibrium with the stellar component ($\S$ \ref{ngc507_central_discuss}). The northern tail seen in the adaptively smoothed image (Figure \ref{ngc507_csmooth}) may represent a cooling wake produced by such motion, similar those found by \cite{Dav94} or \cite{Merrifield98}.
The smaller North-Western tail departing from the central component found in the surface brightness map and the radial brightness profiles (Figure \ref{ngc507_notail}a,b) may have a similar origin, even though we can not exclude that it simply represents a layer of gas shocked by the radio plasma on the edge of the Western radio lobe.

These considerations do not necessarily exclude each other since a combination of different effects is likely to be present in high density environments.

\subsection{Discrete Sources}
\label{ngc507_sources_discuss}

\cite{Kim95} revealed the presence of many pointlike sources in the NGC 507 halo. The co-added spectrum of these sources was found to be well fitted by a soft thermal model (kT=0.52) with low abundance (0.0-0.2), leading the authors to speculate that they could be cooling clumps in the gaseous halo. 

As discussed in $\S$ \ref{ngc507_sources} we used the wavelets algorithm to investigate the nature of discrete sources in the HRI field (\ref{ngc507_sources_fig}). After excluding NGC 507 (source No.10), NGC 499 (No.11), we cross-correlated the HRI sources (Table \ref{ngc507_source_tab}) with those detected in the PSPC data (Table 2 of \cite{Kim95}). Only 5 out of the 17 sources PSPC sources falling within the HRI field\footnote{Sources No.1,2,19,21 of \cite{Kim95} fall outside or on the edge of the HRI FOV.} are detected in the HRI data. 

We compared the fluxes measured by the two instruments, since any temporal variability would exclude the extended nature of these sources. As shown in Tables \ref{ngc507_source_tab2} and \ref{ngc507_source_tab3} two sources varied significantly between 1993 and 1995. One more object (Source No.7) is likely to be associated with a background radio source.

In conclusion $\sim 20\%$ of the sources detected by \cite{Kim95} are likely to be compact objects; this number must be considered as a lower limit since 4 additional HRI sources possess an optical counterpart candidate. This result is not
necessarily in disagreement with the soft spectrum reported by \cite{Kim95} since recently \cite{Sar01} claimed the discovery of supersoft sources in NGC 4697 (even though their sources are preferentially located near the galaxy center). The nature of the remaining sources is still uncertain. If the emission is due to accreting binaries associated with NGC 507 these systems must host massive black holes since their luminosity would exceed, by at least an order of magnitude, the Eddinghton luminosity (equation \ref{L_Edd}) for a 1.4 $M_\odot$ Neutron Star. Finally, we can not exclude the diffuse origin of many sources since the irregular structure of the galactic halo and the filamentary features seen in the halo core suggest that significant density fluctuations may be present in the hot gas distribution, similar those found for NGC 1399.

  \chapter{DISCUSSION}
\label{discussion}

\section{Summary of the Results}
In this Section we briefly summarize the main results obtained in the previous Chapters; in the following Sections we compare the different galaxies of the sample and discuss the global properties of bright Early-Types. 

\subsection{NGC 1399}
We have analyzed a deep ROSAT HRI observation centered on the cD galaxy NGC 1399, in the Fornax cluster, in conjunction with the archival ROSAT PSPC data of the same field. We found an extended and asymmetric halo extending on cluster scales ($r>90$ Kpc). The halo profile is not consistent with a simple Beta model but suggests the presence of three different components. The combined HRI and PSPC data were fitted with a multi-component bidimensional model, consisting of: (1) a `central' component (dominating for $r<50"$) centered on the optical galaxy and whose distribution follows the luminous matter profile; (2) a `galactic' component ($50"<r\leq 400"$) displaced $1'$ (5 kpc) South-West of NGC 1399 and (3) a `cluster' ellipsoidal component ($r>400"$) centered  $\sim 5.6'$ (31 kpc) North-East of the galaxy. The HRI image also revealed the existence of filamentary structures and cavities in the galactic halo, due to density fluctuations in the hot gas. The presence of these features is confirmed by a preliminary analysis of {\it Chandra} data.

The large scale surface brightness distribution can be explained if the galaxy hosts a large dark halo with different dark matter distributions on galactic and cluster scales \citep{saglia00,ikebe96}.
Alternative models, that explain the GCs abundance and its optical velocity dispersion profiles through tidal interactions with Fornax cluster galaxies (\citealp{Kiss99}; Napolitano, Arnaboldi \& Capaccioli 2001, in preparation), are compatible with our data only if such encounters result in shallower slope of the outer gas distribution. Tidal interactions may also explain the presence of the density fluctuations in the galactic halo \citep{D'Ercole00}.
Ram pressure from the cluster halo (e.g. \citealp{irw96,Dav94,fabian01}) is able to account for the decentering of the optical galaxy with respect to the galactic halo if NGC 1399 is moving subsonically (10-100 km s$^{-1}$) in the cluster potential. The displacement of the center of the cluster component with respect to the nearest Fornax galaxies may suggest that the cluster is not relaxed and may be undergoing a merger event.   

We do not detect the X-ray counterpart of the nuclear radio source \citep{kbe88} and pose an upper limit of $L_X^{3\sigma}=3.9\times 10^{39}$ erg s$^{-1}$ on its luminosity in the 0.1-2.4 keV band. In agreement with Killeen and collaborators, our data support thermal pressure confinement of the radio emitting plasma.
We found X-ray emission enhancements aligned with the radio jets that may be due to shocked gas and X-ray `holes' and enhancements coincident with the position of the radio lobes. The latter are consistent with a scenario in which the hot gas is displaced by the radio plasma pressure (e.g. \citealp{clar97}), while alternative models of Inverse Compton scattering of Cosmic Microwave Background photons fail to account for the observed excesses. 

We detected 43 discrete sources in our field. Their flux distribution is consistent at the $2\sigma$ level with the number expected from X-ray background counts. However, the spatial distribution of these sources has a significant excess peaked on the central galaxy, suggesting a population of galaxian X-ray sources in NGC 1399. This is confirmed by the {\it Chandra} data, that reveal a large number of point-like sources associated with the optical galaxy. These sources are likely to explain the hard component found by ROSAT and ASCA spectral analysis \citep{Buo99,Mat00,allen00}.

\subsection{NGC 1404}
The surface brightness distribution of the NGC 1404 halo is regular and almost consistent with the stellar distribution. The brightness profile is well represented by a single Beta model out to 90''. The similarity with the 'central' component of the NGC 1399 halo suggest that the ISM distribution of NGC 1404 is produced by the same gravitational potential that is binding the stellar population. However the agreement with the optical profile implies that 
the halo is `hotter' than the stars \citep{Fab89}, in agreement with the spectral measurements of \cite{Matsu01}.

This scenario can be explained by interactions with the surrounding environment. Indeed, for $r>90"$ the effect of ram pressure stripping becomes evident as a sharp brightness cutoff in the North-West sector and an elongated tail on the opposite side of the galaxy. 

Even though the cooling times, derived from the deprojected density profile of the NGC 1404 halo, are shorter than the Hubble time, no bright emission peak due to cool gas in the galaxy center is detected. Finally, the mass profile shows no evidence of multiple components.

\subsection{NGC 507}
The ROSAT HRI data showed, in agreement with previous studies, a bright X-ray halo surrounding NGC 507. The large-scale surface brightness distribution is irregular and more extended in the North direction. 

The halo core revealed a complex morphology: there is a main X-ray peak, coincident with the position of the optical galaxy, and several secondary peaks.
The radial brightness profile is only marginally consistent with the presence of a nuclear point source; however recent {\it Chandra} data rule out this possibility. The comparison with the optical profile seems to exclude that the main peak is due to the integrated emission of unresolved point sources. We conclude that this feature is likely produced by dense hot gas in the galaxy core.  

There are clear signs of interaction between the radio-emitting plasma and the surrounding ISM, producing density fluctuations in the hot gas. The evidence is stronger for the Western lobe than for the Eastern one. However alternative physical processes, such as IC scattering of background photons, predict X-ray emission an order of magnitude lower than observed.

Modeling the halo surface brightness with a bidimensional double Beta model, we found that the galactic halo centroid is displaced $\sim 22"$ (7 kpc) South-West of the optical galaxy. This result suggests a different origin for the central and the galactic components. Even though the cooling time of the halo within 250" (80 kpc) is shorter than the Hubble time, the mass of the gas present in the central component falls two orders of magnitude below the expected amount inferred from previous works. We propose that the central component is produced by stellar mass losses, in agreement with its high metallicity found by \cite{Buo00b}.

The displacement of the cluster halo from the optical galaxy and the filamentary structures observed in the halo core suggest that the galaxy is slowly moving within the group potential. The total mass profile derived from the bidimensional model shows that a significant amount of dark matter is required at large radii, if the gas is in hydrostatic equilibrium. The dark halo extends on cluster scales and is likely associated with the whole cluster rather than with NGC 507, since the core region within 80" is dominated by stellar matter.

We found that $\sim 20\%$ of the sources detected by \cite{Kim95} in the NGC 507 halo are due to point sources. The nature of the remaining population is unclear. If associated with NGC 507, they can be either accreting binaries hosting a massive black hole or density clumps of the X-ray halo.

\section{The Origin of the Central Emission Peak in NGC 1399 and NGC 507}
\markright{\MakeUppercase{5.2. The origin of the central emission peak}}
\label{centralpeak_discuss}

In Chapters \ref{NGC507_chapter} and \ref{NGC1399} we have shown that the X-ray halo of the two dominant galaxies analyzed in the present work, NGC 1399 and NGC 507, possess a central bright X-ray peak. These features show some interesting similarities: i) in both cases the bright peak is centered on the optical galaxy, ii) the peak is displaced several kpc (5 kpc for NGC 1399, 7 kpc for NGC 507) with respect to the galactic halo centroid, iii) the peaks cannot be due to the presence of a nuclear point source nor to an unresolved population of point sources (even though some contribution is expected) thus suggesting to originate from thermal emission from diffuse gas. 

The presence of a central emission excess over a simple Beta model, often found in galaxies and cluster of galaxies \citep[see][]{Fab89,Fabian94}, is usually attributed to the presence of a cooling flow which deposits large amounts of cooler gas in the halo core \citep{Sar88}. The higher gas density should produce a bright central peak since the thermal emissivity is $\propto n_e^2$ (see equation \ref{emiss3}). However this interpretation has been challenged by the finding that the observed excess is smaller than what is expected if all the cooling gas reaches the halo center \citep{Fabian94}. The mass deposition profiles derived by many authors have shown that $\dot{M}\propto r$ \citep{Fab89}, thus indicating that a significant amount of cooling gas must be deposited at large radii and is not able to reach the halo core. 

Our data are consistent with the hypothesis that the central peaks are not due to the gas deposited by the cooling flow. In fact the expected amount of gas deposited by the flow exceeds the mass of the central component by, at least, an order of magnitude ($\S$ \ref{inner} and $\S$ \ref{ngc507_central_discuss}). 
This result is valid also when we restrict the analysis to the region where the central peak dominates the emission (50" and 15" respectively for NGC 1399 and NGC 507) and we take into account the inhomogeneous flow scenario, 
where only part of the cooling gas reaches the halo center. 
Moreover, while in the case of NGC 1399 the disagreement is reconcilable with the cooling flow scenario if we reduce to $\sim 10^9$ yr the age of the cooling flow (but we are still ignoring stellar mass losses), this solution doesn't work for NGC 507, where the disagreement is larger.
A further indication that the central X-ray peak is not related to the cooling flow is its displacement from the galactic halo centroid. If the brightness excess was related to the deposited cooling gas we would expect the inflowing mass to be located approximately in the halo center.

These considerations lead us to two separate conclusions: the first one is that since the observed X-ray excess is not due to the presence of a cooling flow, the mass inflow must be significantly smaller than the rates predicted not only by homogeneous models (a well established fact) but by the inhomogeneous scenario ($\dot{M}\propto r$) as well for both galaxies, in agreement with the most recent spectroscopic results from XMM \citep{Matsu01b,Pet01,Kaa01,Tam01}. 

The second conclusion is that the distribution of the denser central gas must be related to the underlying stellar distribution. In fact \cite{Brighenti97} argue that the temperature drop within 4 effective radii ($r_e$) observed in many galaxies, including NGC 507 and NGC 1399, is not a natural result of cooling flow models. \cite{Matsu01} studied the spectral properties of a sample of Early-Type galaxies as a function of the distance from the halo center and found that the hot corona of X-ray compact galaxies is likely due to the gas injected into the ISM by stellar losses. The X-ray luminosities and temperatures are compatible with the expectations obtained considering stellar motions as the main source of heat, leading to a mean value of $\beta_{spec}\simeq 1$ (see equation \ref{betaspec}). X-ray extended galaxies as NGC 1399, instead, have a mean halo temperature which exceeds the value expected if the gaseous component is in equilibrium with the stellar population ($\beta_{spec}\simeq 0.5)$. 

This difference between compact and extended galaxies is reduced if we restrict the analysis to the region within $\sim 1~r_e$ from the galaxy center. In the case of NGC 1399 $\beta_{spec}\simeq 0.9$ for $r\leq r_e\simeq 45"$ \citep[$r_e$ is from][]{Faber89}, i.e. the region dominated by the central X-ray peak. Taking advantage of the HRI resolution we can further improve this result. \cite{Matsu01} could not sample the luminosity of the region within $1.5~r_e$ because of the large PSPC PRF. Our data, instead, indicate that reducing the region observed down to $r_e$ the X-ray flux is reduced to $\sim 40\%$ so that the luminosity of NGC 1399 is comparable to the energy injected in the ISM by stellar motions (see Figure 4 of \citealp{Matsu01}). 

In the case of NGC 507 the correlation between the stellar population and the central component is less obvious. In fact, if the central peak is produced by stellar mass losses, we would expect its extension to be related to the underlying stellar distribution. The extension of the central peak is $r \simeq 5$ kpc in both NGC 1399 and NGC 507 but, while this radius is comparable with the optical effective radius for the former, the latter has a broader optical distribution with $r_e\simeq 77"$\citep{Faber89}. The discrepancy is solved in part by noting that the NGC 507 galactic halo has a central density approximately twice the one of NGC 1399, which gives a surface brightness $\sim 4$ times higher (the thermal emissivity scales as $n_e^2$, see equation \ref{emiss3}). Thus the larger galactic halo luminosity prevents the observer to follow the central peak brightness, which declines as $r^{-2.5}$, out to large radii. Moreover the interaction between the ISM and the radio lobes ($\S$ \ref{ngc507_Xradio}) may have significantly altered the ISM distribution of the central component, thus making a comparison with NGC 1399 more difficult.

The comparison of the hot plasma temperature with the one expected assuming dynamical equilibrium with the stellar component is a difficult task for NGC 507. It requires to obtain a reliable spectrum from the central 15", which has become possible only very recently with the {\it Chandra} and XMM satellites and whose data are not available yet. \cite{Matsumo97} found $\beta_{spec}\sim 0.6$ for NGC 507; however they used a 1.4 arcmin extraction radius which is much larger than the central peak extension. Taking into account the central temperature drop and the fact that Matsumoto and collaborators found a $20\%$ lower temperature using the MEKA model \citep{Mewe85} instead of the classical Raymond-Smith code \citep{Ray77}, we believe that the current data are compatible with a $\beta_{spec}$ value close to one, as found for NGC 1399.  

We conclude that the central X-ray peak found in NGC 507 and NGC 1399 are likely to be produced by the material ejected by stars in the ISM. The same mechanism at work in X-ray compact galaxies, i.e. kinematical heating by stellar mass losses, may be responsible for the cooler core observed in many bright Early-Type galaxies, even though some contribution from cooling gas can not be excluded. 

\section{Dynamical Status of the Halo}
The bidimensional models used in $\S$ \ref{model} and \ref{ngc507_2Dmodel} to describe the structure of the X-ray halo surrounding NGC 1399 and NGC 507, allowed us to separate the compact central component (discussed in previous section) from the gaseous halo components extending on galactic and on cluster spatial scales. In both galaxies we found a displacement of several kpc between the X-ray centroid of these extended components and the optical galaxy. Assuming hydrostatic equilibrium we further showed that the total mass (Figures \ref{mass_prof}b and \ref{ngc507_mass}b) is dominated by stellar matter within a few effective radii from the optical centroid while, at larger distances, there is a significant contribution of dark matter. 

These two results (the displacement of the optical galaxy and the small contribution of dark matter on galactic scales) suggest that the gas distribution on large scales is mostly determined by the cluster potential rather than being associated with the dominant galaxy.
The comparison of the halo temperature outside a few effective radii and the stellar velocity dispersion yield values of $\beta_{spec}\simeq 0.5$ \citep{Matsumo97,Brown98,Brown00,Matsu01} thus supporting the need for additional heat to the one provided by stellar mass losses alone.
Sources of heat such as a central active nucleus or supernovae explosions, are not enough to account for the observational data \citep[e.g.][]{Matsumo97} and would however affect in first place the halo center. The alternative explanation proposed by \cite{Maki01} and \cite{Matsu01} is that X-ray extended galaxies are at the bottom of large potential structures corresponding to galaxy groups, which provide the additional heat by gravitational infall. This scenario is consistent with our data which show that both NGC 1399 and NGC 507 lie at the center of extended dark halos.

Nevertheless the X-ray halo presents one main difference between the two galaxies (Figure \ref{model_prof} and \ref{ngc507_model_profile}): to model the surface brightness distribution of NGC 1399 we had to introduce three different components which we associated with the central, galactic and cluster halo. Conversely the NGC 507 halo structure could be modeled with only two components. Comparing our total mass profiles to those obtained by other authors \citep{Brighenti97,Matsu98,Matsu01,Matsu01b} in bright Early-Types we find that a double structure is quite common. All these authors found a central region, within a few effective radii, dominated by stellar matter and a large dark halo extending outside the optical galaxy, in analogy with our findings. The mass profiles are thus tracing the galactic potential in the inner region and the dark extended halo elsewhere. It is however very difficult to assess the presence of a third component since this requires data of both high S/N ratio and high spatial resolution.
 
Two possibilities arise from this discussion: the first is that the galactic and cluster halo components are two separate entities reflecting the hierarchical nesting of the dark matter halos around dominant elliptical galaxies. In this case two separate components may be present in all bright Early-Types but, depending on the contribution of each component, their discovery through surface brightness analysis may be difficult, since the superposition can mimic a one-component profile. NGC 1399 may be a lucky case where the displacement of the different components and the faintness of the cluster halo makes the multi-component structure more evident. In this case we expect the galactic and cluster components to have different spectral signatures. In fact the region where the galactic component dominates the X-ray emission roughly coincides with the temperature maxima found by \cite{Kim95}. Spatially resolved spectroscopic studies with {\it Chandra} may test such a possibility.

This scenario holds if we assume that the gas distribution is tracing the gravitational potential, i.e. is in hydrostatic equilibrium. Thus an alternative possibility is that other physical mechanisms, such as tidal interactions \citep[e.g.]{D'Ercole00}, ram pressure \citep{toni01} or galactic wind stifling from the ICM \citep{Brown98,Brown00} are significantly altering the hot gas distribution. As discussed in $\S$ \ref{NGC1399_dynamical}, some of these effects (ram pressure, tidal interactions) may explain the displacement beween the optical galaxy and the galactic halo component, which is otherwise difficult to account for in the hierarchical dark halo scenario.

\section{NGC 1404: a peculiar galaxy?}
\label{ngc1404_final_discuss}
Our study of NGC 1404 was motivated by the fact that this galaxy, even though it is not a dominant cluster member, represents a very bright Early-Type with an X-ray luminosity ($L_X\sim 1.6\times 10^{41}$ erg s$^{-1}$) of the same order of magnitude of NGC 1399. This X-ray luminosity is one order of magnitude larger than other Early-Types with similar optical luminosity \citep{Brown98,Irw98b} and exceeds the one expected from kinetical heating by stellar mass losses \citep{Matsu01}. The hot halo temperature ($kT\sim 0.6$ keV) exceeds the one expected from the stellar velocity dispersion ($\sigma\sim 220-250$ Km s$^{-1}$, \citealp{D'Onofrio95,Gra98}) yielding $\beta_{spec}\sim 0.6$ \citep{Brown98,Matsu01}.
Although some authors \citep{D'Onofrio95,Gra98} noticed that NGC 1404 has a large rotation velocity ($\sim 100$ Km s$^{-1}$) and is thus likely to be a misclassified S0 galaxy, even taking into account the rotational contribution we still obtain $\beta_{spec}\leq 0.7$.

All these facts suggest that NGC 1404 is more similar to X-ray extended  galaxies, such as NGC 1399 and NGC 507, than to compact galaxies of similar optical luminosity and velocity dispersion. However the analysis presented in Chapter \ref{NGC 1404} shows a number of significant differences: i) the X-ray surface brightness of NGC 1404 is smooth and is consistent with the stellar distribution; ii) the radial brightness profile is well represented by a single Beta model and there is no sign of the presence of a multi component structure; iii) the X-ray and optical profiles are in good agreement, suggesting once more that $\beta_{spec}\sim 0.5$ (see $\S$ \ref{discussion1404}). Moreover the hot halo is characterized by a very subsolar metallicity \citep{Loew94,jones97}, in contrast with the solar abundances found in the center of many bright ellipticals \citep{Buo98,Buo00b,Mat00}.

The cooling times and total mass profiles derived in Chapter \ref{NGC 1404} are more similar to the profiles of the central components derived for NGC 1399 and NGC 507 than to the properties of the extended galactic and cluster halo, thus suggesting a similar origin. However in this case we would expect $\beta_{spec}\simeq 1$ as found for the central components of the two dominant galaxies ($\S$ \ref{centralpeak_discuss}).

These conflicting evidences suggest that NGC 1404 is a peculiar galaxy and  other physical mechanisms must be responsible for the status of the hot halo. \cite{Matsu01} noticed that environmental effects must be important since NGC 1404 is embedded in the NGC 1399 halo. We have seen that NGC 1404 possess an elongated tail which is likely to be produced by ram pressure from the surrounding ICM. \cite{toni01} simulated the effects of ram pressure stripping on galaxies similar to NGC 1404 finding that it could explain differences in the X-ray luminosity similar to those observed between NGC 1404 and other Early-Types. Anyway a comparison is difficult since their models predict a higher halo temperature and steeper brightness profiles than observed,
indicating that the balance between mass injection from stellar losses and cooling dropout is not well understood. Moreover the radial profiles shown in Figure \ref{1404profile}b also showed that ram pressure stripping seems to affect the galactic halo only at large radii ($r>3~r_e$).

\cite{Brown98,Brown00} proposed that the temperature excess, over the value calculated from stellar velocity dispersion, can be produced by supernovae heating. This would produce strong stellar winds, thus reducing the total X-ray luminosity. Conversely, in high density environment, the stifling of galactic winds by the surrounding ICM pressure would increase the total luminosity of the galaxy. This scenario conflicts with the very low metallicity observed in the NGC 1404 halo, because high supernova rates lead to higher abundances. However \cite{Loew94} notice that the presence of a metallicity gradient in the NGC 1404 halo (maybe due to the subsolar ICM abundance), may bias the ASCA results, which are affected by the low spatial resolution. This hypothesis is supported by the larger $Z\sim0.59~Z_\odot$ metallicity reported by \cite{Matsu01} within $4~r_e$.

The peculiar nature of NGC 1404 suggests the need for further studies; in fact this object represents a very good candidate to understand the effect of the environment on the physical status of the X-ray halo.

\section{Density Fluctuations in the ISM}

Most of the work done on Early-Type X-ray halos has been based on the simplified assumption of homogeneity and spherical symmetry. These assumptions were based mainly on the low spatial resolution of the X-ray satellites before ROSAT. Our HRI observations have shown the presence of significant brightness fluctuations and filamentary structures in the halo of both NGC 1399 and NGC 507. Since the ROSAT HRI spectral response is weakly dependent on the temperature \citep{clar97}, these features must reflect density fluctuations in the hot gas distribution. Excluding the structures clearly associated to the radio jets and lobes (see next section), these features have no obvious explanation. In $\S$ \ref{NGC1399_dynamical} we discussed a number of possible scenarios that could account for the density fluctuations found in the NGC 1399 halo, including tidal interactions with nearby galaxies and ram pressure stripping of the galactic corona by the ICM. 

We were not able to detect these features with comparable significance in the NGC 507 halo, because of the poorer S/N ratio of the data. In the halo center, where the S/N ratio is higher, the adaptive smoothing technique revealed the presence of filamentary structures. At larger radii the presence of `clumps' in the hot plasma was reported by \cite{Kim95} using ROSAT PSPC data. Our HRI analysis showed that $\sim 20\%$ of these sources are likely to be point sources but the origin of the remaining objects is still uncertain. We can speculate that they reflect a complex structure of the X-ray halo similar to those found in the dominant Fornax galaxy. 

The current data do not allow to distinguish between different scenarios: the central filamentary structures may represent X-ray wakes due to the motion of the galaxy within the cluster halo \citep{Dav94,Merrifield98}. However these motions alone cannot account for the distribution of these features all over the hot halo; more complex interactions, such as those proposed for NGC 1399, must be considered. One further possibility is that this complex morphology reflects the inhomogeneity of the cooling process suggested by many authors \cite[e.g.][]{Kim95,Pet01,Fab01}, thus  giving important clues to understand the failure of cooling-flow models. 

We expect that significant insight in the nature of these structures will come shortly from the analysis of {\it Chandra} and XMM data.

\section{X-ray/Radio Interaction}
Both dominant galaxies studied in this work (NGC 1399 and NGC 507) are faint radio sources. The analysis performed in $\S$ \ref{Xradio} and $\S$ \ref{ngc507_Xradio} revealed the presence of interactions between the radio emitting plasma and the ISM. 
Such interactions have been predicted by several authors \citep[e.g.][]{clar97,Reynolds01} and actually observed in many clusters of galaxies \citep{bor93,McNam00,Fab00} and in a limited number of strong radio galaxies \citep[e.g.][]{car94}. Our data extends this result to weak radio sources as well, confirming that the interaction of radio jets and lobes with the surrounding medium is a widespread phenomena.   

This interaction shows both similarities and differences between NGC 507 and NGC 1399: in both galaxies we found a low-emission region coincident with the position of the brighter radio lobe and excess emission at the edge of the lobe. We showed that the most likely explanation of these brightness variations is a change in the environmental gas density produced by the radio plasma pressure. However, while for NGC 1399 these features have been clearly revealed only by the adaptive smoothing technique, in the case of NGC 507 the effect of the radio lobes on the central halo morphology is immediately evident. Moreover the NGC 1399 radio jet and lobe morphology is only slightly disturbed while in NGC 507 the lobes look irregular and more diffuse. 

A possible interpretation of these differences can be found in the different gas density in the galactic halo core. The electron density profiles, presented in Figures \ref{density_prof} and \ref{ngc507_density_prof}, show that the NGC 507 halo density (and thus the pressure since the halo temperatures are similar) is more than twice the NGC 1399 value. This must produce a stronger interaction in NGC 507, in qualitative agreement with the observations.

Finally our data confirm the role played by ISM to confine and shape the radio plasma: in both galaxies the radio and thermal pressure are found to be in good agreement while high density regions are usually associated to `anomalies' in the radio lobe morphology.

\section{Discrete Sources Population}
The contribution of discrete sources to the total X-ray emission of Early-Type galaxies has been widely discussed in literature. However, before the launch of {\it Chandra}, most evidence was indirect, obtained through spectral analysis and by extrapolation from nearby Late-Type galaxies. Our analysis revealed a large number of sources associated with NGC 1399, within a few effective radii from the galaxy center. This result was confirmed by the high resolution {\it Chandra} observation. The brightest point sources seem to be located, in prevalence, in the globular clusters surrounding the cD galaxy, as shown by \cite{Ang01}.  

As for NGC 507, the lower quality of the data did not allow us to perform a similar analysis; hence we will have to wait for higher quality {\it Chandra} data.

  \addcontentsline{toc}{chapter}{\protect\numberline{}{CONCLUSIONS}}
\chapter*{CONCLUSIONS\markboth{CONCLUSIONS}{CONCLUSIONS}}

In this work we presented the study of three bright Early-Type galaxies hosted in groups or poor clusters of galaxies. We used ROSAT HRI and PSPC data to study the morphological structure of the X-ray halo and, when available, the higher resolution {\it Chandra} data.

Our study revealed a complex halo structure for the two dominant cluster galaxies, NGC 1399 and NGC 507, not well described by a simple Beta model. The halo center was found to be dominated by a bright X-ray peak coincident with the position of the optical galaxy. This central peak is not explained by the classical homogeneous or inhomogeneous cooling flow models. Instead our data suggest that these features are produced by stellar ejected material, kinetically heated by stellar mass losses.

The total mass distribution shows that within the effective radius the hot halo distribution is essentially tracing the galactic potential, dominated by luminous matter. At larger distances from the galaxy center the X-ray surface brightness and temperature require the presence of a large amount of dark matter. The presence of a different dynamical regime depending on the galactocentric distance was already found by many authors. Our study, however, clearly associates the central halo region with the stellar distribution and the external regions with the dark halo which extends on group and cluster spatial scales.

The NGC 1399 X-ray halo possess a more complex morphology than NGC 507. This finding requires either that the dark matter distribution has a  hierarchical structure, or that environmental effects (ram stripping from ICM, tidal interaction with nearby galaxies) are producing departures from hydrostatic equilibrium. The latter hypothesis would explain the displacement between the different halo components, which is difficult to account for otherwise.

We found significant density fluctuations in the hot gas distribution of both NGC 1399 and NGC 507. Some of these features are explained by the interaction of the radio-emitting plasma, filling the radio jets and lobes, with the surrounding ISM. This evidence indicates that Radio/X-ray interactions are a more widespread phenomena than observed before. The nature of the remaining structures is more uncertain: we speculate that they can be the result of galaxy-galaxy encounters or wakes produced by the motion of the galaxy through the ICM. Alternatively they may reflect the inhomogeneity of the cooling process invoked by many authors as an explanation for the failure of the standard cooling-flow models.
~\\
~\\
NGC 1404 represents a puzzling case. In fact, despite a large X-ray luminosity and halo temperature, which suggest a similarity with dominant cluster members, significant differences are found in the regular halo profile, small velocity dispersion and low metallicity. These conflicting evidences can be explained assuming again strong environmental effects. Even though NGC 1404 is indeed experiencing ram pressure stripping from the NGC 1399 extended halo, as demonstrated by the elongated SE `tail', this phenomena seems to affect only the outer halo regions. A more promising explanation is, instead, the stifling of galactic winds by the ICM.
~\\
~\\
We studied the population of discrete sources found in proximity of the two dominant galaxies. We were able to assess that the observed excess of sources observed in the Fornax cluster is associated with NGC 1399. This result was later confirmed by the higher resolution {\it Chandra} data.

\addcontentsline{toc}{chapter}{\protect\numberline{}{ACKNOWLEDGEMENTS}}

\chapter*{ACKNOWLEDGEMENTS
	\markboth{ACKNOWLEDGEMENTS}{ACKNOWLEDGEMENTS}}%

\thispagestyle{empty}
I wish to thank in first place my supervisor G. Peres, at the Palermo University, and G. Fabbiano, at the Harvard-Smithsonian Center for Astrophysics (C.f.A.) in Cambridge (MA - USA), for introducing and guiding me through the field of X-ray astronomy, and Dong-Woo Kim (C.f.A.) for the time dedicated to address my doubts about X-ray data reduction and interpretation.

\noindent
Moreover:\\
-- Dan Harris, at the C.f.A., for the useful discussions on the radio data interpretation;\\
-- S. Sciortino and F. Damiani, of the Palermo Observatory, for advice on the use of the GALPIPE database and of the wavelets algorithm;\\
-- Prof. N. Robba and Prof. M. Gelardi, of the Palermo University, for the useful suggestions on how to improve this thesis.\\
~\\
I'm also thankful to the Harvard-Smithsonian Center for Astrophysics, for welcoming me during my three visits to Cambridge, from 1999 to 2001.
~\\

Finally a special thank goes to my colleagues and to all the people working at the Palermo Observatory for making so pleasant these three years spent in Sicily.\\
~\\
~\\
~\\
This research has made use of:\\
- the NASA/IPAC Extragalactic Database (NED) which is operated by the Jet Propulsion Laboratory, California Institute of Technology, under contract with the National Aeronautics and Space Administration;\\
- the NASA Astrophysics Data System Bibliographic Services;\\
- data obtained from the High Energy Astrophysics Science Archive Research Center (HEASARC), provided by NASA's Goddard Space Flight Center.\\

\addcontentsline{toc}{chapter}{\protect\numberline{}{LIST OF ACRONYMS AND ABBREVIATIONS}}

\chapter*{LIST OF ACRONYMS AND ABBREVIATIONS
	\markboth{ACRONYMS AND ABBREVIATIONS}{ACRONYMS AND ABBREVIATIONS}}%

\begin{tabular}{lcl}
ACIS & - & AXAF Charged coupled Imaging Spectrometer\\ 
& & (on board of the {\it Chandra} satellite)\\ 
ADAF & - & Advection Dominated Accretion Flow\\
AGN & - & Active Galactic Nucleus\\
ASCA & - & Advanced Satellite for Cosmology and Astrophysics\\
AU & - & Astronomical Unit (1 AU = $1.49 \times 10^{13}$ cm)\\
AXAF & - & Advanced X-ray Astronomical Facility: former name of the {\it Chandra} satellite\\
BBXT & - & Broad Band X-ray Telescope\\
BETG & - & Bright Early-Type Galaxies\\
BH & - & Black Hole\\
cD & - & cD galaxy: elliptical galaxy embedded in an extended amorphous halo\\
& & (see $\S$ \ref{earlytype})\\
CIAO & - & Chandra Interactive Analysis of Observations (standard analysis tool\\
& & for {\it Chandra} data)\\
cnt & - & count\\
CXC & - & {\it Chandra} X-ray Center\\
Dec & - & Declination\\
DSS & - & Digital Sky Survey\\
EXOSAT & - & European X-ray Space Agency Telescope\\
FOV & - & Field Of View\\
FWHM & - & Full Width Half Maximum\\
HEAO & - & High Energy Astrophysical Observatory\\
HEASARC & - & High Energy Astrophysics Science Archive Research Center\\
HMXB & - & High Mass X-ray Binaries\\
HRI & - & High Resolution Camera (on board of {\it Einstein} and ROSAT satellites)\\
HST & - & Hubble Space Telescope\\
\end{tabular}


\hspace{-0.5in}
\begin{tabular}{lcl}
IPC & - & Imaging Proportional Counter (on board of the {\it Einstein} satellite)\\
IRAF & - & Image Reduction and Analysis Facility\\
IC & - & Inverse Compton\\
ICM & - & Intra Cluster Medium\\
ISM & - & Inter Stellar Medium\\
$L_\odot$ & - & Solar luminosity\\
$L_B, L_V$ & - & Optical luminosity in the $B$ ($\lambda_{eff}\sim 4300$ \AA) and $V$ ($\lambda_{eff}\sim 5500$ \AA) bands\\
LMXB & - & Low Mass X-ray Binaries\\
$M_\odot$ & - & Solar mass\\
$M_B,M_V$ & - & Absolute magnitudes in the $B$ ($\lambda_{eff}\sim 4300$ \AA) and $V$ ($\lambda_{eff}\sim 5500$ \AA)\\
& & optical bands\\
MEKA & - & Emissivity model for a thin hot plasma developed by\\
& & \cite{Mewe85}\\
$N_H$ & - & The amount of H per cm$^{-2}$ interposed between the source and the observer,\\
& & used to estimate the absorption due to the ISM.\\
NED & - & Nasa Extragalactic Database\\
NS & - & Neutron Star\\
OBI & - & OBservation Interval: temporal intervals in which each ROSAT\\
& & observation is divided\\
pc & - & parsec (1 pc = $3.1\times 10^{18}$ cm) \\
P.A. & - & Position Angle: measured counter-clockwise starting from the North\\
P.I. & - & Principal Investigator: the main scientist involved in a scientific  project.\\
PIMMS & - & Portable Interactive Multi-Mission Simulator (X-ray software)\\
PROS & - & Post Reduction Off-line Software\\
PRF & - & Point Response Function\\
PSPC & - & Position Sensitive Proportional Counter (on board of the ROSAT satellite)\\
QSO & - & Quasi Stellar Object (also quasar)\\
RA & - & Right Ascension\\
$r_e$ & - & galactic effective radius obtained from the surface brightness profile\\
& & (see $\S$ \ref{earlytype})\\
ROSAT & - & Roentgen Satellite\\
RS CVn & - &  Late-type, evolved (spectral type K) stars with high coronal\\
& & activity, found in binary systems \\
S/N & - & Signal to Noise ratio\\
SASS & - & Standard Analysis Software System (standard processing for ROSAT data)\\
SAX & - & Satellite per Astronomia X (also Beppo-SAX)\\
SED & - & Spectral Energy Distribution\\
\end{tabular}

\hspace{-0.5in}
\begin{tabular}{lcl}
SN, SNe & - & Supernova, Supernovae\\
SNR & - & Supernova Remnants\\
SSC & - & Synchrotron Self Compton\\
ULX & - & Ultra Luminous X-ray source\\
XMM & - & X-ray Multi Mirror telescope (also XMM-Newton)\\
Xspec & - & X-ray spectral analysis tool\\
yr & - & year\\
Z & - & metallicity: abundance of elements with atomic number $> 2$.\\
& & $Z_\odot$ means the solar metallicity.
\end{tabular}

  \addcontentsline{toc}{chapter}{\protect\numberline{}{BIBLIOGRAPHY}}
\def\bibname{BIBLIOGRAPHY}

\end{document}